\documentclass[preprint,12pt]{elsarticle}

\usepackage[utf8]{inputenc}
\usepackage{amssymb}
\usepackage{graphicx}
\usepackage{multirow}%
\usepackage{amsmath,amsfonts,bm}%
\usepackage{amsthm}%
\usepackage{multirow}
\usepackage{mathrsfs}%
\usepackage[title]{appendix}%
\usepackage{float}
\usepackage{enumitem}
\usepackage{placeins}
\usepackage{epsfig}
\usepackage{xcolor}%
\usepackage{textcomp}%
\usepackage{manyfoot}%
\usepackage{booktabs}%
\usepackage{array}
\usepackage{geometry}
\usepackage{subcaption}
\usepackage{algorithm}%
\usepackage{algorithmicx}%
\usepackage{algpseudocode}%
\usepackage{listings}%
\usepackage{dashrule}
\usepackage{hyperref}

\hypersetup{
    colorlinks=true,
    linkcolor=blue,
    citecolor=red,
    urlcolor=blue
}
\newtheorem{theorem}{Theorem}
\newtheorem{lemma}{Lemma}

\journal{}

\begin{document}

\begin{frontmatter}

\title{Adaptive LASSO Penalized Minimum Density Power Divergence Estimation through Least Squares Approximation: Application to Bone Mineral Density Data from the SWAN Study}

\author[inst1]{Udita Goswami}
\author[inst1]{Shuvashree Mondal*}

\address[inst1]{
Department of Mathematics and Computing, 
Indian Institute of Technology (Indian School of Mines) Dhanbad, 
Dhanbad - 826004, Jharkhand, India\\
\textit{23dr0196@iitism.ac.in, shuvasri29@iitism.ac.in*}}

\begin{abstract}
Linear mixed-effect panel data models are widely used in longitudinal biomedical, environmental, and social studies, but are often sensitive to data contamination and numerous covariates.  To address these challenges, we propose a robust variable selection approach based on a least squares approximation (LSA) of the density power divergence (DPD) objective function combined with the Adaptive LASSO penalty.  The LSA converts the nonlinear DPD objective into a computationally efficient quadratic approximation while preserving the robustness of DPD estimation.  Under suitable regularity conditions, the proposed DPD Adaptive LASSO-LSA estimator is shown to possess oracle properties, including selection consistency and asymptotic normality.  Simulation studies demonstrate improved robustness, better sparsity recovery, and significantly enhanced computational efficiency than penalized likelihood methods, while maintaining performance comparable to existing DPD-based approaches.  An application to the SWAN bone mineral density dataset illustrates the practical relevance of the proposed methodology for robust estimation and reliable identification of important covariates.
\end{abstract}


\end{frontmatter}

\section{Introduction}\label{sec1}

The growing sophistication of data acquisition systems and computational capabilities has greatly expanded the scope of panel data modeling across various disciplines.  Panel data--commonly referred to as longitudinal or cross-sectional time-series data-—comprise repeated observations on the same entities, such as individuals, regions, or commodities, over time.  This dual structure enables the simultaneous analysis of temporal variation and unit-specific effects, yielding more efficient inference compared to models based solely on cross-sectional or time-series data \cite{baltagi2005econometric, hsiao1985panel}.  Accordingly, panel data methodology has witnessed significant developments, including models incorporating random time effects and spatial dependence \cite{jang2014tests}, mixed-effects formulations for longitudinal data under flexible distributional assumptions \cite{zhao2015maximum}, and dynamic panel models accommodating threshold effects and endogeneity \cite{wan2025dynamic}.  Despite these advantages, panel data analysis poses significant methodological challenges.  Classical panel data models, including fixed-effect, random-effect, and mixed-effect formulations \cite{wallace1969use, mundlak1978pooling, laird1982random, diggle2002analysis}, are predominantly developed within likelihood-based frameworks.  While computationally convenient, such approaches are well known to be highly sensitive to outliers, leverage points, and large-scale contamination—-features that frequently arise in real-world panel datasets due to measurement errors, reporting inaccuracies, or systematic recording issues \cite{rousseeuw1990unmasking, maronna2006robust}.  Although several robust alternatives have been proposed, including least trimmed squares, MS- and MM-estimators, weighted likelihood, and related procedures \cite{rousseeuw1984least, maronna2000robust, visek2015robust, midi2018robust, beyaztas2020advantages}, much of the existing panel data literature continues to rely on likelihood-based estimation.  These limitations have stimulated growing interest in robust and penalized inference, highlighting the need for principled and scalable frameworks that jointly accommodate heterogeneity, contamination, and variable selection in panel and longitudinal data analysis.

Against this backdrop, divergence-based methods from robust statistics, and in particular the Density Power Divergence (DPD) introduced by Basu et al. \cite{basu1998robust}, offer a compelling alternative.  The DPD framework incorporates a tuning parameter that controls the trade-off between efficiency and robustness, making it especially suitable for complex panel and longitudinal data settings that are often affected by contamination.  Robust DPD-based methodologies for linear mixed-effect models and variance component estimation under contamination have been investigated by Saraceno et al. \cite{saraceno2024robust}, while Mandal et al. \cite{mandal2023robust} extended DPD to linear panel data models.  However, robust divergence-based methods for mixed-effect panel data with non-homogeneous observations and sparsity remain largely unexplored.  The present work contributes to this emerging line of research by focusing on such models under sparsity and contamination.  In addition, DPD-based methods have been successfully applied in diverse inferential frameworks, including linear models \cite{ghosh2013robust}, generalized Wald-type tests \cite{basu2016generalized}, censored data with stochastic covariates \cite{ghosh2017robust}, ordinal response models \cite{pyne2024robust}, and panel count data \cite{GOSWAMI2026116371}.  

Modern panel datasets often involve several futile covariates along with relevant ones, necessitating variable selection alongside robust estimation.  To mitigate this, we incorporate the Adaptive LASSO penalty \cite{zou2006adaptive} into the DPD framework.  Adaptive LASSO improves upon standard $\ell_1$ penalization by employing data-driven weights, thereby reducing estimation bias and achieving the oracle property introduced by Fan and Li \cite{fan2001variable}.  A substantial body of literature has investigated the asymptotic behavior and selection consistency of LASSO-type estimators \cite{knight2000asymptotics, zhao2006model, van2009conditions}, including variable selection in multiphase quantile regression \cite{ciuperca2016adaptive}.  More recent developments further incorporate correlation structures and robustness considerations \cite{wang2022regression, ghosh2024robust, maranzano2023adaptive, basu2024robust}.  However, in divergence-based frameworks, a fundamental challenge arises from the inherent nonlinearity of the DPD objective function coupled with the Adaptive LASSO penalty, which substantially complicates both computational and theoretical analysis; particularly the derivation of oracle properties in parsimonious panel data settings.

The key contribution of this paper is to address the computational and theoretical challenges of robust penalized panel data modeling through a least squares approximation (LSA) of the density power divergence (DPD) objective function.  For this, we construct an LSA by locally approximating the DPD objective around a consistent preliminary estimator, yielding an asymptotically equivalent quadratic formulation that preserves robustness through data-adaptive weighting while substantially simplifying computation.  In particular, the proposed LSA-based formulation leads to a significant reduction in computational runtime compared to the corresponding non-LSA iterative procedures, thereby considerably improving scalability and practical feasibility for large longitudinal dimensions and repeated simulation scenarios. Despite this substantial computational gain, the proposed estimator retains statistical efficiency and robustness comparable to, and often better than, its non-LSA counterpart.  Although LSA has been used in sparse estimation problems under likelihood-based frameworks \cite{wang2007unified, hui2018sparse}, its integration into a DPD-based adaptive LASSO framework for mixed-effect panel data models with non-homogeneous observations and variable selection remains unexplored, to the best of our knowledge.  The proposed approximation enables efficient implementation of adaptive $\ell_1$ penalization, substantially reduces computational burden relative to existing robust divergence-based methods, and facilitates rigorous theoretical analysis.  In particular, under the LSA formulation, we can establish the oracle properties of the proposed DPD Adaptive LASSO-LSA estimator, including consistent variable selection and asymptotic normality.  Deriving these results in a non-homogeneous mixed-effect panel data setting is technically non-trivial and has not been previously addressed, making the LSA-based formulation a principal methodological, computational, and theoretical novelty of this work.

The longitudinal bone mineral density (BMD) data from the Study of Women’s Health Across the Nation (SWAN)--a large, multi-site cohort study of midlife women initiated in 1996--pose several practical challenges, including the presence of potential outliers, complex covariate information, and within-subject correlation arising from repeated measurements.  These issues adversely affect classical estimation procedures, leading to biased inference and overfitted models.  The proposed robust and sparse estimation methodology is specifically designed to alleviate these difficulties by simultaneously mitigating the influence of anomalous observations and selecting a parsimonious subset of relevant variables.  The SWAN Bone Cohort provides repeated BMD measurements at approximately 18-month intervals, forming a rich panel dataset for examining bone loss during the menopause transition.  In the present analysis, we consider a subset of 30 women with complete lumbar spine BMD records and associated demographic and clinical covariates, enabling application of the proposed robust penalized mixed-effect panel modeling framework.

The remainder of the paper is organized as follows.  Section~\ref{sec2} introduces the model formulation and notation. Section~\ref{sec3} presents the density power divergence criterion.  Section~\ref{sec4} develops the Adaptive LASSO estimation procedure based on the LSA-DPD framework.  Section~\ref{sec5} establishes the oracle properties.  Section~\ref{sec6} reports simulation results and a real-data application using the Study of Women’s Health Across the Nation (SWAN) dataset on bone mineral density changes during menopause.  Section~\ref{sec7} concludes with a discussion and future research directions. 

\section{Model Description}\label{sec2}
\allowdisplaybreaks

We study a linear panel data model defined over a sample of $n$ independent and non-homogeneous individuals observed across $m$ time periods.  Let $y_{it}$ denote the response variable for the $i^{th}$ individual at time $t$, $x_{it} \in \mathbb{R}^k$ indicate the vector of covariates associated with fixed-effect, and $z_{it}$ represent the variables corresponding to random-effect.  The model is expressed as follows,
\begin{equation}
    y_{it} = x_{it}^\top \beta + z_{it}^\top\alpha_i + u_{it}, 
\label{eq1} 
\end{equation}

\noindent where $\boldsymbol{\beta}$ denotes a $k \times 1$ vector of fixed-effect regression coefficients, $\alpha_i$ captures a $p \times 1$ vector of individual-specific random-effects assumed to be time-invariant, and $u_{it}$ represents the random error at the individual-time level.  It is presumed that the random disturbances \( u_{it} \) are independent and identically distributed with zero mean and constant variance such that $E(u_{it} \mid x_{it}, z_{it}) = 0$, $E(u_{it}^2 \mid x_{it}, z_{it}) = \sigma_u^2$, and $E(u_{it} u_{is} \mid (x_{it}, x_{is}), (z_{it}, z_{is})) = 0 \quad \text{for } t \neq s$. 

Furthermore, if we consider $\varepsilon_{it}$ as the composite error term defined as $\varepsilon_{it} = z_{it}^\top\alpha_i + u_{it}$, the random-effect specification of the panel data model can be formally represented as follows
\begin{equation}
    y_{it} = x_{it}^\top \beta + \varepsilon_{it}. 
\label{eq2} 
\end{equation}

\noindent It is assumed that $\alpha_i \sim \mathcal{MVN}(0, \Sigma_\alpha)$ and $u_{it} \sim \mathcal{N}(0, \sigma_u^2)$ independently for all $i$ and $t$.  

To better capture the panel structure, we now arrange the responses, covariates, random-effects, and error terms for each individual across all time periods.  Consider the response vector for the $i^{th}$ individual, $\boldsymbol{y}_i = \begin{pmatrix} y_{i1}, y_{i2}, \ldots, y_{im} \end{pmatrix}^\top \in \mathbb{R}^{m \times 1}$ which collects the time-series outcomes over $m$ periods.  Stacking across all individuals produces a single column vector of dimension $nm \times 1$, thereby capturing all observations in the panel.  The corresponding fixed-effect covariates for individual $i$ are arranged as $x_i = \begin{pmatrix} x_{i1},  x_{i2}, \ldots, x_{im} \end{pmatrix}^\top \in \mathbb{R}^{m \times k}$.  In parallel, the random-effect covariates are collected in $z_i = \begin{pmatrix} z_{i1}, z_{i2}, \ldots, z_{im} \end{pmatrix}^\top \in \mathbb{R}^{m \times p}$.  Finally, $u_i = \begin{pmatrix} u_{i1}, u_{i2}, \ldots, u_{im} \end{pmatrix}^\top \in \mathbb{R}^{m \times 1}$ stores the random error terms for the $i^{th}$ individual.  The composite error term $\varepsilon_i = \begin{pmatrix} \varepsilon_{i1}, \varepsilon_{i2}, \ldots, \varepsilon_{im} \end{pmatrix}^\top$ inherits the following variability structure,
\begin{align}
    \boldsymbol{\Omega}_i = \operatorname{Cov}(\varepsilon_{i}) = z_i\Sigma_\alpha z_i ^{\top} + \sigma^2_u \boldsymbol{I}_m,
\label{eq3}
\end{align}
\noindent where \( \boldsymbol{I}_m \) denotes the identity matrix of dimension \( m \).

These results suggest that although observations from various individuals are uncorrelated, repeated observations from the same individual exhibit correlation through the common individual-specific factor $\alpha_i$.  This simplification presumes that the influence of unobserved heterogeneity remains consistent over time for each individual and allows for a more tractable model formulation while still accounting for the essential individual-level variation.  

In real-world panel datasets, outliers are common and may arise for various reasons, such as measurement errors, data entry mistakes, structural breaks, or unobserved heterogeneity across individuals.  To mitigate such irregularities, we adopt an M-estimation framework based on the DPD criterion rather than relying on classical least-squares or likelihood methods.  While the working assumption is that errors are normally distributed, the framework remains flexible enough to accommodate substantial departures from this assumption, giving rise to a contaminated model.  The resulting estimator naturally downweights the influence of outlying observations without requiring prior identification of outliers and still guarantees convergence towards the true parameter.  The subsequent section focuses on robust point estimation in the presence of these outliers.

\section{Minimum Density Power Divergence Estimator }\label{sec3}
\allowdisplaybreaks

Basu et al. \cite{basu1998robust} proposed a new class of divergence measure, called the density power divergence, to quantify the difference between two probability densities.  In this context, we consider a family of models ${F_{\theta} : \theta \in \boldsymbol{\Theta}}$, each characterized by a density function $f_{\theta}$.  Assume that the true data-generating distribution is $G$, having the corresponding density $g$.  The density power divergence between the model density $f_{\theta}$ and the true density $g$ is then defined as 
\begin{equation}
d_\gamma(f_\theta, g) =
\begin{cases}
\displaystyle \int \left[ f_\theta^{1+\gamma}(y) - \left(1 + \frac{1}{\gamma}\right) f_\theta^\gamma(y) g(y) + \frac{1}{\gamma} g^{1+\gamma}(y) \right] dy, & \text{if } \gamma > 0, \\
\displaystyle \int g(y) \log\left( \frac{g(y)}{f_\theta(y)} \right) dy, & \text{if } \gamma \rightarrow 0, \label{eq4}
\end{cases}
\end{equation}

\noindent where the tuning parameter $\gamma$ balances the trade-off between robustness and efficiency of the parameter estimation.  For $\gamma$ approaches zero from the right, the DPD converges to the well-known Kullback-Leibler (KL) divergence.    

In the context of the DPD framework, we now describe the mixed-effect model defined in \eqref{eq2} for estimation purposes.  Let us define the parameter vector as $\boldsymbol{\theta} = (\boldsymbol{\beta}^\top, vech(\Sigma_{\alpha})^\top, \sigma^{2}_u)^\top,$ which includes the regression coefficients together with the variance components of $ vech(\Sigma_{\alpha})^\top$ and $\sigma^{2}_u$.  To characterize the assumed distribution of the response vectors, we express the conditional density through the transformation $z_i \alpha_i + u_i = y_i - x_i \boldsymbol{\beta},$ as follows:
\begin{align}
    f_{\boldsymbol{\theta}}(y_i \mid x_i, z_i) 
    &= \frac{1}{(2\pi)^{m} \left| \boldsymbol{\Omega} \right|^{\frac{1}{2}}} \exp\left( -\frac{1}{2} \left[ (y_i - x_i \boldsymbol{\beta})^\top \boldsymbol{\Omega}_i^{-1} (y_i - x_i \boldsymbol{\beta}) \right] \right). \label{eq5}
\end{align}

Specifically, the observed data $Y_1, \ldots, Y_n$ are independent but non-homogeneous; where $g_1, \ldots, g_n$ are the true data-generating densities of $Y_1, \ldots, Y_n$.  Let $F_{i,\boldsymbol{\theta}}$ denote the distribution of the assumed conditional density $f_{\boldsymbol{\theta}}(y_i \mid x_i, z_i)$.  Here, we consider that the true distribution $G$ belongs to the model so that $G_i = F_{i,\boldsymbol{\theta_0}}$ for $\boldsymbol{\theta_0} = (\boldsymbol{\beta_0}^\top, vech(\Sigma_{\alpha0})^\top, \sigma^{2}_{u0})^\top$.  The parameter vector $\boldsymbol{\theta}$ is then estimated by minimizing the DPD between the empirical estimator of the true data-generating density and the assumed model density.  In our setting, the divergence between the empirical density corresponding to $i^{th}$ observation and the assumed model density is defined as
\begin{equation}
    d_\gamma^{(i)} = \int f_{\boldsymbol{\theta}}^{1+\gamma}(y \mid x_i, z_i) \, dy - \left(1 + \frac{1}{\gamma} \right) f_{\boldsymbol{\theta}}^\gamma(y_i \mid x_i, z_i) + c_i, 
\label{eq6}
\end{equation}
where $c_i$ is a constant corresponding to $i^{th}$ observation, and it is independent of $\boldsymbol{\theta}$.  Accordingly, parameter estimation in this setup involves minimizing the following objective function,
\begin{align}
    &\frac{1}{n} H_n(\boldsymbol{\theta}) = \frac{1}{n}\sum_{i=1}^{n}\left(\int f_{\boldsymbol{\theta}}^{1+\gamma}(y \mid x_i, z_i) \, dy - \left(1 + \frac{1}{\gamma} \right) f_{\boldsymbol{\theta}}^\gamma(y_i \mid x_i, z_i)\right) = \frac{1}{n}\sum_{i=1}^{n}W_{\boldsymbol{\theta}}(y_i|x_i, z_i). \label{eq7}
\end{align}
Differentiating \( H_n(\boldsymbol{\theta}) \) with respect to \( \boldsymbol{\theta} \), we obtain the estimating equation under independent and non-identically distributed observations,
\begin{equation}
    \frac{\partial H_n(\boldsymbol{\theta})}{\partial \boldsymbol{\theta}}  = 0.
\label{eq8}
\end{equation}
This can be rewritten explicitly as
\begin{equation}
    \sum_{i=1}^n \Big[f_{\boldsymbol{\theta}}^\gamma(y_i \mid x_i, z_i) u_{\boldsymbol{\theta}}(y_i \mid x_i, z_i) - \int f_{\boldsymbol{\theta}}^{1+\gamma}(y \mid x_i, z_i) u_{\boldsymbol{\theta}}(y \mid x_i, z_i)\,dy \Big]  = 0, 
\label{eq9}
\end{equation}
\noindent where \( u_{\boldsymbol{\theta}}(y_i \mid x_i, z_i) = \frac{\partial}{\partial \boldsymbol{\theta}}  \log f_{\boldsymbol{\theta}}(y_i \mid x_i, z_i) \) is the score function for the \( i^{\text{th}} \) observation.  By solving \eqref{eq9}, we obtain the corresponding minimum density power divergence (MDPD) estimator, denoted by $\tilde{\boldsymbol{\theta}} = (\tilde{\boldsymbol{\beta}}^\top, vech(\tilde{\Sigma}_{\alpha})^\top, \tilde{\sigma}^{2}_u)^\top$.

However, with large $k$ and a complex structure of $\boldsymbol{\Omega}$, the resulting formulation is computationally demanding, which poses practical challenges.  To overcome these difficulties, we adopt a quadratic approximation of the DPD objective, enabling a more tractable and computationally efficient estimation procedure.  The subsequent developments build upon this approximation to establish a feasible sparse estimation methodology.  Therefore, expanding \( H_n(\boldsymbol{\theta}) \) around \( \boldsymbol{\tilde{\theta}} \) using a Taylor series, we get
\begin{equation}
    \frac{1}{n} H_n(\boldsymbol{\theta}) \approx \frac{1}{n} H_n(\boldsymbol{\tilde{\theta}}) + \frac{1}{n} \frac{\partial H_n(\tilde{\boldsymbol{\theta})}}{\partial \boldsymbol{\theta}}(\boldsymbol{\theta} - \boldsymbol{\tilde{\theta}}) + \frac{1}{2} (\boldsymbol{\theta} - \boldsymbol{\tilde{\theta}})^\top \left( \frac{1}{n} \frac{\partial^2 H_n(\boldsymbol{\tilde{\theta}})}{\partial \boldsymbol{\theta} \partial \boldsymbol{\theta}^\top} \right)(\boldsymbol{\theta} - \boldsymbol{\tilde{\theta}}). 
\label{eq10}
\end{equation}
In this case, $\tilde{\boldsymbol{\theta}}$ being a minimizer of $H_n(\cdot)$ simplifies \eqref{eq10} to the subsequent form
\begin{equation}
    \frac{1}{n} H_n(\boldsymbol{\theta}) \approx \frac{1}{n} H_n(\boldsymbol{\tilde{\theta}}) + \frac{1}{2} (\boldsymbol{\theta} - \boldsymbol{\tilde{\theta}})^\top \left( \frac{1}{n} \frac{\partial^2 H_n(\boldsymbol{\tilde{\theta}})}{\partial\boldsymbol{\theta} \partial\boldsymbol{\theta}^\top} \right)(\boldsymbol{\theta} - \boldsymbol{\tilde{\theta}}).
\label{eq11}
\end{equation}
This indicates that, in a neighborhood of $\tilde{\boldsymbol{\theta}}$, the original DPD loss function is closely approximated by the least squares-type objective function.

\vspace{0.1cm}

As the term $\frac{1}{n} H_n(\boldsymbol{\tilde{\theta}})$ becomes zero, the objective function \eqref{eq11} reduces to $(\boldsymbol{\theta} - \boldsymbol{\tilde{\theta}})^\top \left(\frac{1}{n} \frac{\partial^2 H_n(\boldsymbol{\tilde{\theta}})}{\partial\boldsymbol{\theta} \partial\boldsymbol{\theta}^\top}\right) (\boldsymbol{\theta} - \boldsymbol{\tilde{\theta}})$.  This simplification motivates us to consider the LSA of $H_n(\boldsymbol{\theta})$, defined as
\begin{equation}
    H_A(\boldsymbol{\theta}) = \frac{1}{2} (\boldsymbol{\theta} - \boldsymbol{\tilde{\theta}})^\top \left \{\frac{1}{n} R(\boldsymbol{\tilde{\theta}})\right \}(\boldsymbol{\theta} - \boldsymbol{\tilde{\theta}}), 
\label{eq12}
\end{equation}
where $R(\boldsymbol{\tilde{\theta}}) = \frac{\partial^2 H_n(\boldsymbol{\tilde{\theta}})}{\partial\boldsymbol{\theta} \partial\boldsymbol{\theta}^\top}$.  In cases where the objective function lacks adequate smoothness, the LSA method is useful provided a consistent estimator like $\boldsymbol{\tilde{\theta}}$ is available.

\section{Variable Selection with Adaptive LASSO penalty}\label{sec4}
\allowdisplaybreaks

In panel data models with mixed-effects, the underlying structure is often assumed to be sparse, meaning only a subset of the regression coefficients \( \boldsymbol{\beta} \) are non-zero.  This implies that several predictors \( x_{i} \) have no significant influence on the response variable \( y_{i} \).  In such settings, efficient variable selection becomes crucial.  The Least Absolute Shrinkage and Selection Operator (LASSO), introduced by Tibshirani \cite{tibshirani1996regression}, is a widely used method that simultaneously performs estimation and variable selection by exploiting the \(\ell_1\)-norm of the coefficient vector.  The corresponding mathematical formulation leads to the shrinkage of some coefficient estimates to exactly zero, effectively excluding irrelevant predictors from the model.  

Suppose the true coefficient vector in the linear mixed-effect model is denoted by \( \boldsymbol{\beta}_0 = (\beta_{01}, \ldots, \beta_{0k})^\top \).  We define the true active set of relevant predictors as \( P_0 = \{ j : \beta_{0j} \neq 0 \} \), and assume \( |P_0| = p_0 < k \), which represents the sparsity assumption.  An estimator \( \hat{\boldsymbol{\beta}} \) is said to possess variable selection consistency if it correctly identifies the set of non-zero coefficients with probability tending to one as the sample size increases.  Although LASSO is commonly employed for simultaneous estimation and selection, it is known to introduce bias in the coefficient estimates.  Moreover, its ability to consistently select the true model often relies on restrictive conditions, which may not hold in practice.  In addition, the conventional LASSO imposes the same amount of shrinkage on all regression coefficients, for which the resulting estimator lacks efficiency and may lead to inconsistency in results.  To circumvent these limitations, Fan and Li \cite{fan2001variable} introduced the Smoothly Clipped Absolute Deviation (SCAD) penalty, which possesses attractive oracle properties.  A penalized estimator is said to possess the oracle property if it can consistently identify the true set of relevant variables and is asymptotically equivalent to the ideal (oracle) estimator that would be obtained when estimation is performed using only the true signal variables without penalization.  However, the non-convex nature of SCAD can lead to computational challenges, particularly when dealing with sparse data.  As a more computationally feasible alternative, Zou and Hui \cite{zou2006adaptive} proposed the Adaptive LASSO, which enhances performance by incorporating data-dependent weights in the penalty function.  The key idea is that the efficiency of the estimator can be improved by applying heavier shrinkage to coefficients that are truly zero while imposing lighter penalties on the non-zero ones.  

Building on this, and utilizing the least squares approximation as outlined in \eqref{eq12}, we formulate an objective function that integrates the Adaptive LASSO penalty.  The corresponding estimator $\hat{\boldsymbol{\theta}} = (\hat{\boldsymbol{\beta}}^\top, vech(\hat{\Sigma}_{\alpha})^\top, \hat{\sigma}^{2}_u)^\top$ is obtained by minimizing this penalized objective,
\begin{equation}
    Q(\boldsymbol{\theta}) = H_A(\boldsymbol{\theta}) + \lambda\sum_{j=1}^{k}w_j \left|\boldsymbol{\beta_j}\right|.
\label{eq13} 
\end{equation}
Here, 
\begin{equation}
    w_j = \frac{1}{\left|\boldsymbol{\tilde{\beta}_j}\right| + \delta_nI(\boldsymbol{\tilde{\beta}_j}=0)} 
\label{eq14} 
\end{equation}
denotes the pre-defined adaptive weights, where $\boldsymbol{\tilde{\beta}_j}$ represents a consistent initial estimate of $\boldsymbol{\beta_j}$ for $j = 1, \ldots, k$, and \( \delta_n \) is a small positive constant introduced to avoid division by zero.  We shall refer to $\hat{\boldsymbol{\theta}}$ as the DPD Adaptive LASSO-LSA estimator of $\boldsymbol{\theta}$.  This formulation retains the accuracy of variable selection while enhancing computational efficiency.

\vspace{0.1cm}

Note that Ghosh and Basu \cite{ghosh2013robust} have established that MDPD estimator $\boldsymbol{\tilde{\theta}}$ is both consistent and robust.  The stability of $\hat{\boldsymbol{\theta}}$ hinges on these properties of $\boldsymbol{\tilde{\theta}}$.

\section{Asymptotic Properties}\label{sec5}
\allowdisplaybreaks

We encounter substantial theoretical challenges in obtaining a closed-form expression for the DPD Adaptive LASSO-LSA estimator, which hinders a direct analysis of its finite-sample properties. To overcome this limitation, we turn our attention to its large-sample properties, adopting asymptotic frameworks that align with the underlying structure of the data.  In many real-world scenarios, the number of time periods $m$ remains constant, prompting us to explore the large-sample behavior of the model as the number of cross-sectional units $n$ increases.  Nevertheless, the same methodological approach applies when $m$ grows with $n$ fixed, or when both dimensions tend to infinity.  In this setting, we now rewrite the true parameter vector as
\begin{equation*}
    \boldsymbol{\theta_0} = (\beta_{10}, \ldots, \beta_{k0}, vech(\Sigma_{\alpha0})^\top, \sigma^{2}_{u0})^\top = (\boldsymbol{\beta_{10}}^\top, \boldsymbol{\beta_{20}}^\top, vech(\Sigma_{\alpha0})^\top, \sigma^{2}_{u0})^\top,
\end{equation*}
where, without loss of generality, it is assumed that $\boldsymbol{\beta_{10}}$ denotes the $r-$ dimensional vector of truly non-zero fixed-effects while \( \boldsymbol{\beta_{20}} = \mathbf{0}_{k-r}, \ r \leq k \).  

\vspace{0.1cm}

Based on this, we have modified the regulatory conditions suggested by \cite{ghosh2013robust} and \cite{hui2018sparse} according to our context.
\begin{itemize}
    \item[R1)] The support $\chi = \{ y \mid f_{\boldsymbol{\theta}}(y \mid x_i , z_i) > 0 \}$ remains the same across all $i$ and $\boldsymbol{\theta}$.
    
    \item[R2)] There exists an open neighborhood $\omega \subseteq \Theta$ containing the true parameter $\boldsymbol{\theta_0}$, such that for almost every $y \in \chi$ and for every $\boldsymbol{\theta} \in \Theta$, the function $f_{\boldsymbol{\theta}}(y \mid x_i, z_i)$ is thrice differentiable with respect to $\boldsymbol{\theta}$ and the third partial derivatives are continuous with respect to $\boldsymbol{\theta}$ for all $i$.

    \item[R3)] There exists a constant $\nu_1 > 0$ such that
    \[
    \nu_1 < \min_{\substack{j = 1, \ldots, r}} \{|\beta_{j0}| : \beta_{j0} \neq 0 \} < \infty.
    \]
    
    \item[R4)] For all $i$, the integral $\int f^{1+\gamma}_{\boldsymbol{\theta}}(y \mid x_i, z_i ) \, dy$ admits third-order derivatives with respect to $\boldsymbol{\theta}$, and differentiation under the integral sign is valid.

    \item[R5)] There exists an open neighborhood $\Theta^* \subseteq \Theta$ containing the true parameter $\boldsymbol{\theta_0}$, such that for every $\boldsymbol{\theta} \in \Theta^*$, there exist integrable functions $S_{uvw}(r_i)$ satisfying
    \[
    \left| \frac{\partial^3}{\partial \theta_u \, \partial \theta_v \, \partial \theta_w} W_{\boldsymbol{\theta}}(y_i|x_i, z_i) \right| < S_{uvw}(r_i),
    \]
    for all $u, v$ and $w$, where $r_i$ denotes the observed data for the $i^{th}$ individual, including responses $y_i$ and covariates $x_i$ and $z_i$. Furthermore, the functions satisfy
    \[
    E \left[ S_{uvw}^2(r_i) \right] < \infty.
    \]

    \item[R6)] The regularization parameter $\lambda$ is chosen such that $\sqrt{n} \lambda \to 0$ and $n \lambda \to \infty$ as $n \to \infty$.
\end{itemize}

\noindent For simplicity, we denote $f_{\boldsymbol{i\theta}} = f_{\boldsymbol{\theta}}(y_i|x_i, z_i)$.  Then, as $n\rightarrow\infty$, the second-order derivative can be obtained as follows.
\begin{align*}
    &\frac{1}{n} \left. \frac{\partial^2 H_n(\boldsymbol{\theta})}{\partial \boldsymbol{\theta} \partial \boldsymbol{\theta}^\top} \right|_{\boldsymbol{\theta} = \boldsymbol{\theta_0}}
    = \frac{1}{n} \sum_{i=1}^{n} \left. \frac{\partial^2}{\partial {\theta} \partial {\theta}^\top} W_{\boldsymbol{\theta}}(y_i|x_i, z_i) \right|_{\boldsymbol{\theta} = \boldsymbol{\theta_0}} \\
    &= \frac{1}{n} \sum_{i=1}^{n} \left. \frac{\partial}{\partial \boldsymbol{\theta}} \left( \frac{\partial}{\partial \boldsymbol{\theta}} W_{\boldsymbol{\theta}}(y_i|x_i, z_i) \right) \right|_{\boldsymbol{\theta} = \boldsymbol{\theta_0}} \\   
    &\xrightarrow{p} \lim_{n \to \infty} \frac{1}{n} \sum_{i=1}^{n} E \left[\left. \frac{\partial}{\partial \boldsymbol{\theta}} \left(\int u_{\boldsymbol{i\theta}} f_{\boldsymbol{i\theta}}^{1+\gamma} - u_{\boldsymbol{i\theta}} f_{\boldsymbol{i\theta}}^\gamma \right) \right|_{\boldsymbol{\theta} = \boldsymbol{\theta_0}} \right] \nonumber \\
    &\xrightarrow{p} \lim_{n \to \infty} \frac{1}{n} \sum_{i=1}^{n} E \left[ \int \left\{ -I_{\boldsymbol{i\theta_0}} f_{\boldsymbol{i\theta_0}}^{1+\gamma} + (1 + \gamma) u_{\boldsymbol{i\theta_0}} u_{\boldsymbol{i\theta_0}}^\top f_{\boldsymbol{i\theta_0}}^{1+\gamma} \right\} - I_{\boldsymbol{i\theta_0}} f_{\boldsymbol{i\theta_0}}^\gamma - \gamma u_{\boldsymbol{i\theta_0}} u_{\boldsymbol{i\theta_0}}^\top f_{\boldsymbol{i\theta_0}}^\gamma \right] \nonumber \\
    &\xrightarrow{p} \lim_{n \to \infty} \frac{1}{n} \sum_{i=1}^{n} \left[ -\int I_{\boldsymbol{i\theta_0}} f_{\boldsymbol{i\theta_0}}^{1+\gamma} + (1 + \gamma) \int u_{\boldsymbol{i\theta_0}} u_{\boldsymbol{i\theta_0}}^\top f_{\boldsymbol{i\theta_0}}^{1+\gamma} - \int I_{\boldsymbol{i\theta_0}} f_{\boldsymbol{i\theta_0}}^\gamma g - \gamma \int u_{\boldsymbol{i\theta_0}} u_{\boldsymbol{i\theta_0}}^\top f_{\boldsymbol{i\theta_0}}^\gamma g \right] \nonumber \\
    &\xrightarrow{p} \lim_{n \to \infty} \frac{1}{n} \sum_{i=1}^{n} \left[\int u_{\boldsymbol{i\theta_0}} u_{\boldsymbol{i\theta_0}}^\top f_{\boldsymbol{i\theta_0}}^{\gamma+1} \right]. 
\end{align*}

\noindent Now, by applying the weak law of large numbers (WLLN), we can establish
\begin{align*}
    &\frac{1}{n} \sum_{i=1}^{n} \left. \frac{\partial^2}{\partial {\theta} \partial {\theta}^\top} W_{\boldsymbol{\theta}}(y_i|x_i, z_i) \right|_{\boldsymbol{\theta} = \boldsymbol{\theta_0}} \\
    &\xrightarrow{p} \lim_{n \to \infty} E \left[ \frac{1}{n} \sum_{i=1}^{n} \frac{\partial^2}{\partial {\theta} \partial {\theta}^\top} W_{\boldsymbol{\theta_0}}(y_i|x_i, z_i) \right] \\
    &   \xrightarrow{p} \lim_{n \to \infty} \frac{1}{n} \sum_{i=1}^n B_{i}(\boldsymbol{\theta_0}) = B(\boldsymbol{\theta_0}),
\end{align*}
where $B(\boldsymbol{\theta_0})$ is assumed to be a positive-definite matrix.  Hence, 
\begin{equation}
    \frac{1}{n}R(\tilde{\boldsymbol{\theta}}) = \frac{1}{n} \frac{\partial^2 H_n(\boldsymbol{\tilde{\theta}})}{\partial \boldsymbol{\theta} \partial \boldsymbol{\theta}^\top} \xrightarrow{p} B(\boldsymbol{\theta_0}).
\label{eq15}
\end{equation}

\noindent Further, we can deduce
\begin{align*}
    &\left. \left(\frac{1}{\sqrt{n}} \frac{\partial H_n(\boldsymbol{\theta})}{\partial \boldsymbol{\theta}} \right) \left(\frac{1}{\sqrt{n}} \frac{\partial H_n(\boldsymbol{\theta})}{\partial \boldsymbol{\theta}}  \right) ^\top\right|_{\boldsymbol{\theta} = \boldsymbol{\theta_0}} \\
    &= \left.\frac{1}{n} \sum_{i = 1}^{n} \left( \frac{\partial}{\partial \boldsymbol{\theta}} W_{\boldsymbol{\theta}}(y_i|x_i, z_i) \right) \left( \frac{\partial}{\partial \boldsymbol{\theta}} W_{\boldsymbol{\theta}}(y_i|x_i, z_i) \right)^\top \right|_{\boldsymbol{\theta} = \boldsymbol{\theta_0}} \\             &\xrightarrow{p} \lim_{n \to \infty} \left. E \left[ \frac{1}{n} \sum_{i = 1}^{n} \left( \frac{\partial}{\partial \boldsymbol{\theta}} W_{\boldsymbol{\theta}}(y_i|x_i, z_i) \right) \left( \frac{\partial}{\partial \boldsymbol{\theta}} W_{\boldsymbol{\theta}}(y_i|x_i, z_i) \right)^\top \right] \right|_{\boldsymbol{\theta} = \boldsymbol{\theta_0}} \nonumber \\
    &\xrightarrow{p} \lim_{n \to \infty} \left. E \left[ \frac{1}{n} \sum_{i = 1}^{n} \left( \int u_{\boldsymbol{i\theta}} f_{\boldsymbol{i\theta}}^{1+\gamma} - u_{\boldsymbol{i\theta}} f_{\boldsymbol{i\theta}}^\gamma \right) \left( \int u_{\boldsymbol{i\theta}} f_{\boldsymbol{i\theta}}^{1+\gamma} - u_{\boldsymbol{i\theta}} f_{\boldsymbol{i\theta}}^\gamma \right)^\top \right] \right|_{\boldsymbol{\theta} = \boldsymbol{\theta_0}} \nonumber \\
    &\xrightarrow{p} \lim_{n \to \infty} \frac{1}{n} \sum_{i = 1}^{n} E \left[ u_{\boldsymbol{i\theta_0}} u^{\top}_{\boldsymbol{i\theta_0}} f_{\boldsymbol{i\theta_0}}^{2\gamma} - \left( \int u_{\boldsymbol{i\theta_0}} f_{\boldsymbol{i\theta_0}}^{\gamma + 1}\right)\left( \int u_{\boldsymbol{i\theta_0}} f_{\boldsymbol{i\theta_0}}^{\gamma + 1}\right)^\top \right] \nonumber \\
    &\xrightarrow{p} \lim_{n \to \infty} \frac{1}{n} \sum_{i = 1}^{n} \left[ \int u_{\boldsymbol{i\theta_0}} u^{\top}_{\boldsymbol{i\theta_0}} f_{\boldsymbol{i\theta_0}}^{2\gamma}g - \left( \int u_{\boldsymbol{i\theta_0}} f_{\boldsymbol{i\theta_0}}^{\gamma + 1}\right)\left( \int u_{\boldsymbol{i\theta_0}} f_{\boldsymbol{i\theta_0}}^{\gamma + 1}\right)^\top \right] \nonumber \\
    &\xrightarrow{p} \lim_{n \to \infty} \frac{1}{n} \sum_{i = 1}^{n} \left[ \int u_{\boldsymbol{i\theta_0}} u^{\top}_{\boldsymbol{i\theta_0}} f_{\boldsymbol{i\theta_0}}^{2\gamma + 1} - \xi_{\boldsymbol{i\theta_0}} \xi_{\boldsymbol{i\theta_0}}^\top \right] \nonumber \\
    & \xrightarrow{p} \lim_{n \to \infty} \frac{1}{n} \sum_{i=1}^n A_{i}(\boldsymbol{\theta_0}) = A(\boldsymbol{\theta_0}). \nonumber
\end{align*}
Therefore, we get
\begin{equation}
     \left(\frac{1}{\sqrt{n}} \frac{\partial}{\partial \boldsymbol{\theta}} H_n(\tilde{\boldsymbol{\theta})} \right) \left(\frac{1}{\sqrt{n}}\frac{\partial}{\partial \boldsymbol{\theta}} H_n(\tilde{\boldsymbol{\theta})} \right)^\top \xrightarrow{p} A(\boldsymbol{\theta_0}). 
\label{eq16}
\end{equation}
where $A(\boldsymbol{\theta_0})$ is also assumed to be a positive-definite matrix.

\vspace{0.1cm}

Under the aforementioned regulatory conditions, we will now establish the following results to assess the asymptotic properties of the DPD Adaptive LASSO-LSA estimator. 

\begin{lemma}[$\sqrt{n}$-consistency of the DPD Adaptive LASSO-LSA estimator]\label{lem1}
Suppose that the regularity conditions R1) - R6) hold. Then, as $n \to \infty$, the DPD Adaptive LASSO-LSA estimator satisfies
\[
\bigl\lVert \boldsymbol{\hat{\theta}}-\boldsymbol{\theta}_0 \bigr\rVert
= O_p\!\left(\frac{1}{\sqrt{n}}\right).
\]
\end{lemma}

Lemma illustrates the $\sqrt{n}$-consistency of the DPD Adaptive LASSO-LSA estimator.  This, in turn, enables us to evaluate the behavior of the adaptive weights.  Next, we will demonstrate the key result concerning the oracle property of the proposed estimator.  Denote, $\boldsymbol{\theta_{10}} = (\boldsymbol{\beta_{10}}^\top, vech(\Sigma_{\alpha0})^\top, \sigma^{2}_{u0})^\top$ and $\boldsymbol{\theta_{20}} = \boldsymbol{\beta_{20}}.$ Here, $\boldsymbol{\hat{\theta}_1}$ and $\boldsymbol{\hat{\theta}_2}$, derived from $\boldsymbol{\hat{\theta}}$, are the corresponding estimators of $\boldsymbol{\theta_{10}}$ and $\boldsymbol{\theta_{20}}$, respectively.\\

\begin{theorem}[Oracle properties of the DPD Adaptive LASSO-LSA estimator]\label{thm1}
Suppose that the regularity conditions R1) - R6) hold. As $n \to \infty$, the DPD Adaptive LASSO-LSA estimator satisfies the following oracle properties:

\begin{enumerate}
\item[a)] \textbf{Asymptotic normality.}  
\[
\sqrt{n}\bigl(\boldsymbol{\hat{\theta}}_{1}-\boldsymbol{\theta}_{10}\bigr)
\xrightarrow{d}
N\!\left(
\mathbf{0},
\, B_1^{-1}(\boldsymbol{\theta}_0)\,
A_1(\boldsymbol{\theta}_0)\,
B_1^{-1}(\boldsymbol{\theta}_0)
\right),
\]
where $A_1(\boldsymbol{\theta}_0)$ and $B_1(\boldsymbol{\theta}_0)$ denote the $r\times r$ submatrices of
$A(\boldsymbol{\theta}_0)$ and $B(\boldsymbol{\theta}_0)$, respectively, corresponding to the nonzero components $\boldsymbol{\theta}_{10}$.

\item[b)] \textbf{Selection consistency.}  
\[
P\bigl(\boldsymbol{\hat{\theta}}_{2}=\boldsymbol{0}\bigr)\xrightarrow{}1 .
\]
\end{enumerate}
\end{theorem}

According to part a) of Theorem  1, the estimator corresponding to the truly non-zero coefficients converges in distribution to a multivariate normal, with a covariance structure given by $B^{-1}_1(\boldsymbol{\theta_0})A_1(\boldsymbol{\theta_0})B^{-1}_1(\boldsymbol{\theta_0})$.  Meanwhile, part b) of Theorem 1 guarantees that with high probability, the DPD Adaptive LASSO-LSA estimator correctly identifies only the non-zero fixed-effect parameters as the sample size grows.  The proofs of all these results are relegated to the \textbf{Appendix}.

\section{Numerical Experiments}\label{sec6}
\allowdisplaybreaks

This section carries comprehensive numerical experiments, encompassing both simulation studies and real data analysis, that demonstrate the proposed method's robustness, accuracy, and practical relevance.

\subsection{Simulation Study}

To evaluate the finite-sample performance of the proposed DPD Adaptive LASSO-LSA procedure, we conduct an extensive Monte Carlo simulation study based on the mixed-effect panel-data model in \eqref{eq1}.  The study is carried out across different combinations of sample sizes and time periods to examine the stability and robustness of the estimator in multiple-dimensional settings.

\noindent Specifically, three sample sizes are considered, $n \in \{60,\ 120,\ 300\}$ and each individual is observed over $m \in \{10,\ 15,\ 20\}$ time points.  The total number of covariates is fixed at $k = 15,$ among which the first nine covariates are relevant and the remaining six are irrelevant.  The true regression coefficient vector is taken as
\[
\boldsymbol{\beta}_0 =
(3.0,\ 2.5,\ 2.0,\ 1.8,\ 3.5,\ 1.2,\ 1.0,\ 0.8,\ 4.1,\ 0,\ 0,\ 0,\ 0,\ 0,\ 0)^\top.
\]
The robust tuning parameter of the density power divergence criterion is considered for $\gamma \in \{0.2,\ 0.5,\ 0.8\}$.  For each individual $i = 1,2,\ldots,n$ and time point $t = 1,2,\ldots,m$, the covariates are generated as follows.

\vspace{0.1cm}

\noindent The first three covariates are generated from Gaussian distributions with time-varying means,
\[
\begin{aligned}
X_{1,it} &\sim N(2 + 0.10t,\ 1),\\
X_{2,it} &\sim N(1 + 0.08t,\ 1),\\
X_{3,it} &\sim N(0.5 + 0.05t,\ 1).
\end{aligned}
\]

\noindent The next three covariates are constructed as moderately correlated linear combinations of the previous variables,
\[
\begin{aligned}
X_{4,it} &= 0.35X_{1,it} + 0.25X_{2,it} + \varepsilon_{4,it},\\
X_{5,it} &= 0.30X_{2,it} + 0.30X_{3,it} + \varepsilon_{5,it},\\
X_{6,it} &= 0.25X_{1,it} + 0.20X_{3,it} + \varepsilon_{6,it},
\end{aligned}
\]
where
\[
\varepsilon_{4,it},\varepsilon_{5,it},\varepsilon_{6,it}
\stackrel{\text{i.i.d.}}{\sim} N(0,1).
\]

\noindent To incorporate temporal variation in the model, the next three covariates are generated using sinusoidal and trend-based structures,
\[
\begin{aligned}
X_{7,it} &\sim N(\sin(t/2),\ 1),\\
X_{8,it} &\sim N(\cos(t/3),\ 1),\\
X_{9,it} &\sim N(0.1t,\ 1).
\end{aligned}
\]

\noindent The tenth covariate is generated as a correlated linear combination of the previous time-varying predictors,
\[
X_{10,it}
=
0.30X_{7,it}
+
0.20X_{8,it}
+
\varepsilon_{10,it},
\]
where
\[
\varepsilon_{10,it}\sim N(0,1).
\]

\noindent The remaining five covariates are generated independently from the standard normal distribution,
\[
X_{11,it},X_{12,it},X_{13,it},X_{14,it},X_{15,it}
\stackrel{\text{i.i.d.}}{\sim} N(0,1),
\]
and are treated as irrelevant variables. 

\vspace{0.1cm}

The random-effect design variable is generated from the exponential distribution,
\[
Z_{it}\sim \text{exp}(1).
\]

\noindent The individual-specific random effects are generated as $\alpha_i \sim N(0,\sigma_\alpha^2),$ while the random error terms are generated from $u_{it}\sim N(0,\sigma_u^2)$.  Here, the variance components for the model are set to $\sigma_{u0}^2 = 1.0$ and $\sigma_{\alpha0}^2 = 1.2$.  Using these quantities, the response variable is generated from the mixed-effect panel-data model defined in \eqref{eq1}.

To assess the robustness of the proposed estimator, both pure and contaminated data schemes are considered.  Under the pure scheme, the response and covariates are generated directly from the distributions described above without any contamination.  Under the contaminated scheme, two types of outliers are introduced.
\begin{enumerate}
    
\item \textbf{Response contamination ($y$-outliers):} 
 In $\epsilon \% \in \{5\%,\ 10\%,\ 15\%\}$  of the sample , the response variable is contaminated using additive noise generated from the inverse Gaussian distribution, $\text{IG}(4.5,\ 0.8)$.

\item \textbf{Covariate contamination ($X$-outliers):}
In  $\epsilon \% \in \{5\%,\ 10\%,\ 15\%\}$ of the sample, the covariate observations are perturbed by additive noise generated from the Cauchy distribution, $\text{C}(4.8,\ 0.9)$.

\end{enumerate}
These contamination mechanisms are introduced to investigate the robustness of the proposed estimation procedure under heavy-tailed and asymmetric perturbations.

Subsequently, the consistent estimator of both the fixed-effect parameter vector ($\boldsymbol{\beta}$) and the variance component ($\boldsymbol{\Omega}$) is obtained using the Minimum Density Power Divergence Estimator (MDPDE).  The corresponding robust objective function is minimized using the exponentially weighted gradient descent (EWGD) optimization procedure.  To ensure stable convergence of the EWGD algorithm and to avoid numerical singularity during optimization, suitable initial values are required.  For this, the regression coefficients are perturbed slightly around the true parameter vector,
\[
\boldsymbol{\beta}^{(0)}
=
\boldsymbol{\beta}_0
+
\boldsymbol{\eta},
\]
where the perturbation vector $\boldsymbol{\eta}$ consists of small Gaussian noises with very small variance.  Here, the initial variance component values are taken to be $\sigma_{u}^{2{(0)}} = 1.4$ and $\sigma_{\alpha}^{2{(0)}} = 1.2$.  Using these consistent MDPD estimates $\boldsymbol{\tilde{\beta}}$, $\tilde{\sigma_{u}}^2$, and $\tilde{\sigma_{\alpha}}^2$, the adaptive weights $w_j$ are constructed for all $j=1 \ldots k$ and the Hessian matrix associated with the robust objective function is computed. 

Thereafter, to select the optimal tuning parameter, we employ the Extended Regularized Information Criterion (ERIC) proposed by Hui et al.~\cite{hui2015tuning} and further applied to sparse estimation in multivariate longitudinal mixed models by \cite{hui2018sparse}.  The detailed algorithm for ERIC is provided in the supplementary material.

Following this, the penalized objective function is minimized using the Fast Iterative Shrinkage Thresholding Algorithm (FISTA) introduced by Beck and Teboulle \cite{beck2009fast}.  FISTA is particularly well-suited for penalized optimization problems due to its computational efficiency and accelerated convergence properties.  The steps for FISTA are provided in the following algorithm.  Once the estimates of the regression coefficients are obtained, the variance parameters are estimated using a closed-form expression derived from the quadratic objective function. 

\begin{table}[h!]
\renewcommand{\arraystretch}{1.2}
\setlength{\tabcolsep}{4pt}
\begin{tabular}{p{0.95\textwidth}}
\hline
\textbf{Algorithm: Fast Iterative Shrinkage Thresholding} \\
\hline

\begin{minipage}[!htbp]{0.95\textwidth}
\footnotesize
\vspace{0.3cm}

\begin{itemize}[leftmargin=1.2em] 

\item \textbf{Initialization:}
\[
\boldsymbol{\beta}^{(0)}
=
\widetilde{\boldsymbol{\beta}},
\qquad
\mathbf{h}^{(0)}
=
\boldsymbol{\beta}^{(0)},
\qquad
t_0 = 1.
\]

\item \textbf{Iteration:}

\begin{itemize}[leftmargin=1em]

\item[\textbf{-}] Set $k=0$.

\item[\textbf{-}] Compute the gradient of the objective function at the current iterate.

\item[\textbf{-}] Perform the gradient update:
\[
\mathbf{z}^{(k)}
=
\mathbf{h}^{(k)}
-
\eta
\nabla Q(\mathbf{h}^{(k)}),
\]
where $\eta$ denotes the step size.

\item[\textbf{-}] Apply the soft-thresholding operator:
\[
\beta_j^{(k+1)}
=
\text{sign}(z_j^{(k)})
\left(
|z_j^{(k)}|
-
\eta\lambda w_j
\right)_+,
\qquad
j=1,\ldots,k.
\]

\item[\textbf{-}] Update the acceleration parameter:
\[
t_{k+1}
=
\frac{1+\sqrt{1+4t_k^2}}{2}.
\]

\item[\textbf{-}] Compute the accelerated iterate:
\[
\mathbf{h}^{(k+1)}
=
\boldsymbol{\beta}^{(k+1)}
+
\frac{t_k-1}{t_{k+1}}
\left(
\boldsymbol{\beta}^{(k+1)}
-
\boldsymbol{\beta}^{(k)}
\right).
\]

\end{itemize}

\item \textbf{Update:}

\begin{itemize}[leftmargin=1em]

\item[\textbf{-}] Set $k=k+1$ and update the parameter vector using the accelerated iterate.

\item[\textbf{-}]  Return $\boldsymbol{\beta}^{(k+1)}$ as the optimal estimate. 

\end{itemize}

\item \textbf{Repeat:} Until convergence. \\

\end{itemize}

\end{minipage}
\\

\hline
\end{tabular}
\end{table}

Consequently, a Monte Carlo simulation study is conducted using $D=500$ independently generated datasets for each combination of sample size and time period.  The complete estimation procedure is repeated for each generated dataset under both pure and contaminated data settings.

The performance of the proposed estimator is evaluated using several estimation and variable-selection criteria.  Let
\[
\mathcal{S}_0
=
\left\{
j : \beta_{0j} \neq 0
\right\}
\]
denote the support set of the true parameter vector, and let
\[
\hat{\mathcal{S}}
=
\left\{
j : \hat{\beta}_j \neq 0
\right\}
\]
represent the support set of the estimated parameter vector.  The mean squared error corresponding to the estimated coefficients of truly relevant or non-zero covariates is defined as
\[
\text{MSES}
=
\frac{1}{\left|\mathcal{S}_0\right|}
\sum_{j \in \mathcal{S}_0}
\left(\boldsymbol{\hat{\beta}}_{j} - \boldsymbol{\beta}_{j0}\right)^2.
\]
Similarly, the mean squared error associated with the estimated coefficients of truly irrelevant or zero covariates is computed as
\[
\text{MSEN}
=
\frac{1}{\left|\mathcal{S}_0^c\right|}
\sum_{j \in \mathcal{S}_0^c}
\boldsymbol{\hat{\beta}}_{j}^2.
\]
The average absolute bias corresponding to the estimated coefficients of significant covariates is evaluated as
\[
\text{Bias}(\boldsymbol{\hat{\beta}}_\mathcal{S})
=
\frac{1}{\left|\mathcal{S}_0\right|}
\sum_{j \in \mathcal{S}_0}
\left|\boldsymbol{\hat{\beta}}_{j} - \boldsymbol{\beta}_{j0}\right|.
\]

\noindent Likewise, the average absolute bias associated with the estimated coefficients of insignificant covariates is computed as
\[
\text{Bias}(\boldsymbol{\hat{\beta}}_{\mathcal{S}^c})
=
\frac{1}{\left|\mathcal{S}_0^c\right|}
\sum_{j \in \mathcal{S}_0^c}
\left|\boldsymbol{\hat{\beta}}_{j}\right|.
\]
Here, $\hat{\boldsymbol{\beta}_\mathcal{S}}$ and $\boldsymbol{\hat{\beta}}_{\mathcal{S}^c}$  denote the subvectors of $\boldsymbol{\hat{\beta}}$ corresponding to the relevant and irrelevant covariates, respectively.  In addition, the model size (MS), which measures the number of selected variables, is defined as
\[
\text{MS}(\boldsymbol{\hat{\beta}})
=
\left|
\hat{\mathcal{S}}
\right|.
\]
The true positive proportion (TP), representing the proportion of significant variables correctly identified by the estimator, is computed as
\[
\text{TP}(\boldsymbol{\hat{\beta}})
=
\frac{
\left|
\hat{\mathcal{S}}
\cap
\mathcal{S}_0
\right|
}{
\left|
\mathcal{S}_0
\right|
}.
\]
Analogously, the true negative proportion (TN), which measures the proportion of insignificant variables correctly excluded from the model, is defined as
\[
\text{TN}(\boldsymbol{\hat{\beta}})
=
\frac{
\left|
\hat{\mathcal{S}}^{c}
\cap
\mathcal{S}^{c}_0
\right|
}{
\left|
\mathcal{S}^{c}_0
\right|
}.
\]
The estimation accuracy of the variance component is assessed through the absolute estimation error, defined as
\[
EE(\hat{\sigma}^2)
=
\left|
\hat{\sigma}^2
-
\sigma_0^2
\right|,
\]
where $\sigma_0^2$ represents the overall true variability in the dataset, derived as
\[
\sigma_0^2
=
\frac{1}{n}
\sum_{i=1}^{n}
\left\{
\text{tr}(\Omega_{0i})
\right\}.
\]
and $\hat{\sigma}^2$ denotes its estimated value.  

Finally, the computational efficiency of the proposed optimization procedure is examined by measuring the average execution time across repeated runs of the FISTA algorithm.  In order to assess the overall performance of the proposed estimator, the DPD Adaptive LASSO-LSA (DPD-Ad-LASSO-LSA) estimator is compared with several penalized estimators, namely DPD-LASSO-LSA, DPD-Ad-LASSO, and DPD-LASSO corresponding to tuning parameter values $\gamma = 0.2,\ 0.5,\ 0.8$.  These competing approaches are further assessed within the classical maximum likelihood framework by incorporating both LASSO and Adaptive LASSO penalties under LSA as well as non-LSA formulations.  However, it is observed that MLE-based methods generally require smaller regularization than the corresponding robust DPD-based estimators to achieve stable variable selection performance.  The penalty parameter values used in the simulation study are selected separately for each estimation procedure.  The selected values of the tuning parameter $\lambda$ are reported in \hyperref[tab1]{Table \ref{tab1}}.

\begin{table}[!htbp]
\centering
\caption{Penalty parameter values used for different estimation procedures.}
\label{tab1}
\vspace{0.1cm}

\resizebox{0.50\textwidth}{!}{%
\begin{tabular}{lc}
\toprule
\textbf{Method} & \textbf{Penalty Parameter} ($\boldsymbol{\lambda}$) \\
\midrule

MLE-LASSO & $0.6 \times 10^{-2}$ \\
DPD-LASSO & $1.8 \times 10^{-2}$ \\

MLE-Adaptive-LASSO & $1.1 \times 10^{-2}$ \\
DPD-Adaptive-LASSO & $2.6 \times 10^{-2}$ \\

MLE-LASSO-LSA & $2.9 \times 10^{-4}$ \\
DPD-LASSO-LSA & $4.3 \times 10^{-4}$ \\

MLE-Adaptive-LASSO-LSA & $3.7 \times 10^{-4}$ \\
DPD-Adaptive-LASSO-LSA & $5.2 \times 10^{-4}$ \\

\bottomrule
\end{tabular}
}
\end{table}

Further, for each competing method, the corresponding execution time is recorded to compare the computational efficiency of the estimation procedures.  The estimation performance of the proposed methods is also evaluated by examining the empirical bias of both relevant and irrelevant regression coefficients across different sample sizes, contamination schemes, and tuning parameter values.  These results assess the finite-sample stability, sparsity recovery, and robustness of the competing procedures.  Since the complete set of numerical results is extensive, only selected representative cases are presented in the main manuscript, while the detailed bias results are provided in \textcolor{red}{Table 1-Table 6 of the supplementary material}.

\begin{figure}[h!]
\centering

\subfloat[n = 60, m = 10]{
\includegraphics[width=0.48\textwidth]{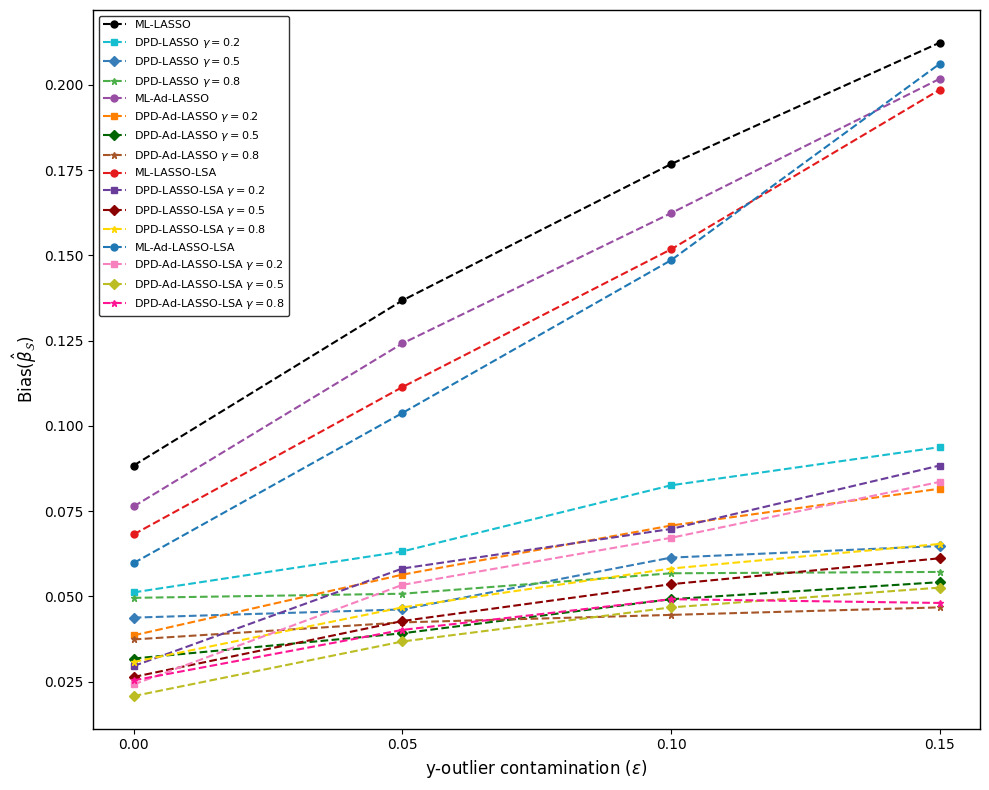}
}
\hfill
\subfloat[n = 300, m = 20]{
\includegraphics[width=0.48\textwidth]{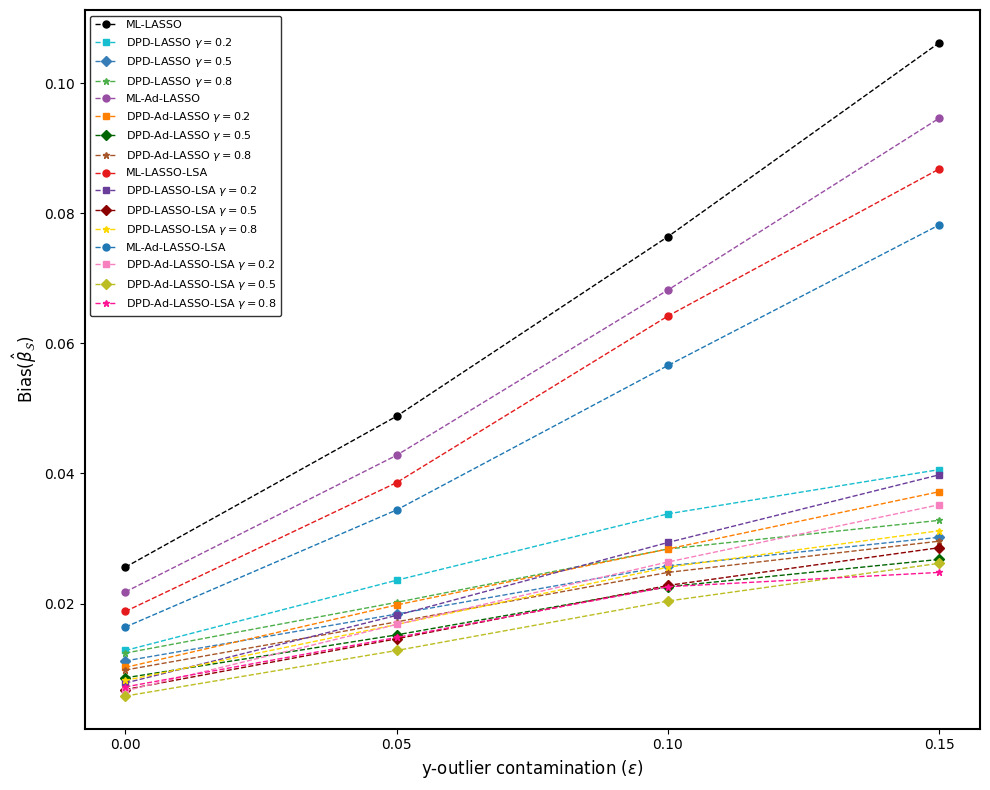}
}

\vspace{0.3cm}

\subfloat[n = 60, m = 10]{
\includegraphics[width=0.48\textwidth]{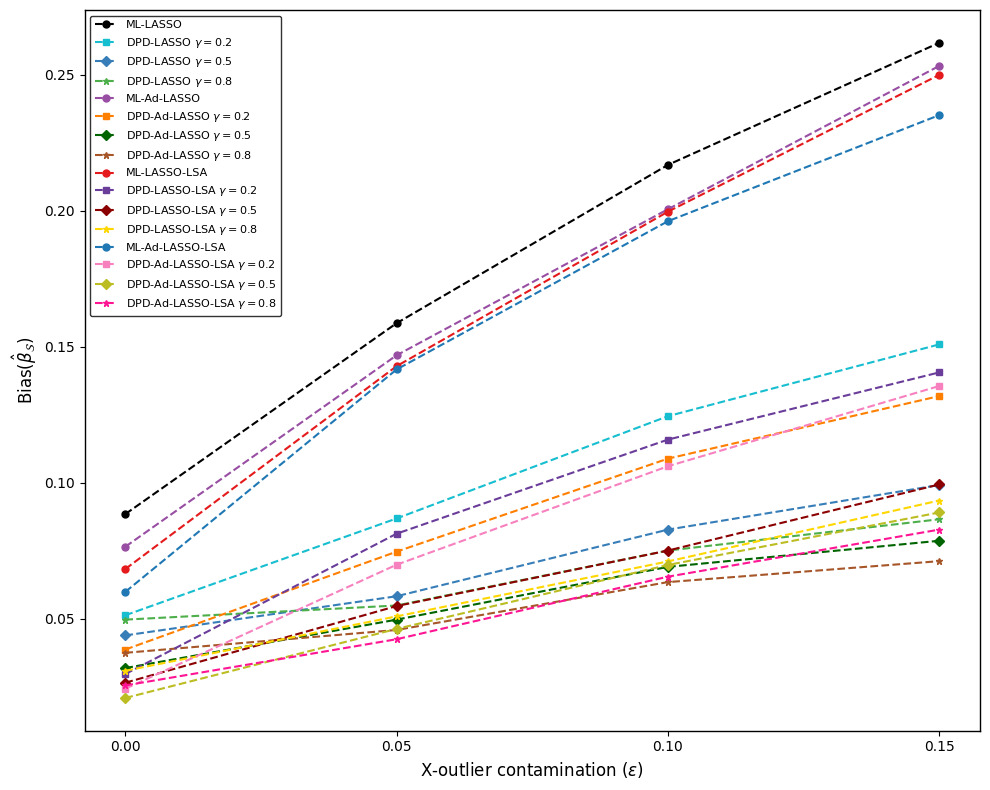}
}
\hfill
\subfloat[n = 300, m = 20]{
\includegraphics[width=0.48\textwidth]{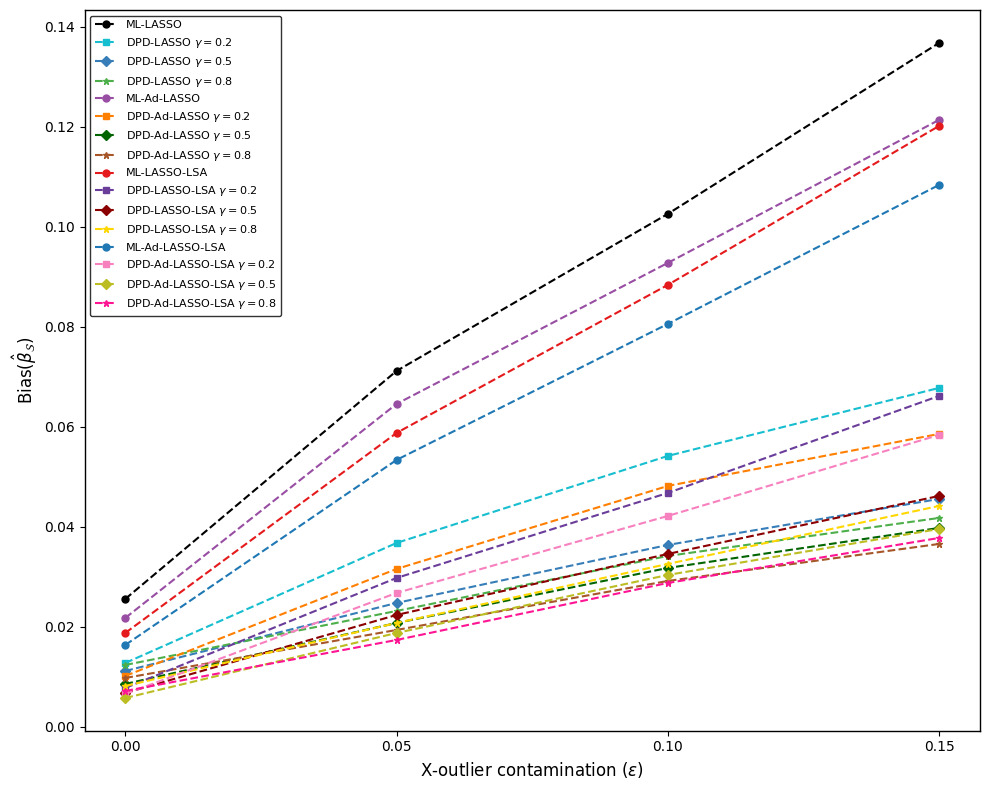}
}

\caption{Bias$(\boldsymbol{\hat{\beta}}_{\mathcal{S}})$ of various estimation procedures under different sample sizes $n$ for selected combinations of sample size $(n)$ and number of time points $(m)$.}

\vspace{0.1cm}

\begin{minipage}{\textwidth}
\footnotesize
\raggedright
(a), (b): Bias$(\boldsymbol{\hat{\beta}}_{\mathcal{S}})$ vs $y$-outlier contamination. \\
(c), (d): Bias$(\boldsymbol{\hat{\beta}}_{\mathcal{S}})$ vs $X$-outlier contamination.
\end{minipage}

\label{fig1}
\end{figure}

\begin{figure}[!htbp]
\centering

\subfloat[m = 10, 15\% y-outliers]{
\includegraphics[width=0.48\textwidth]{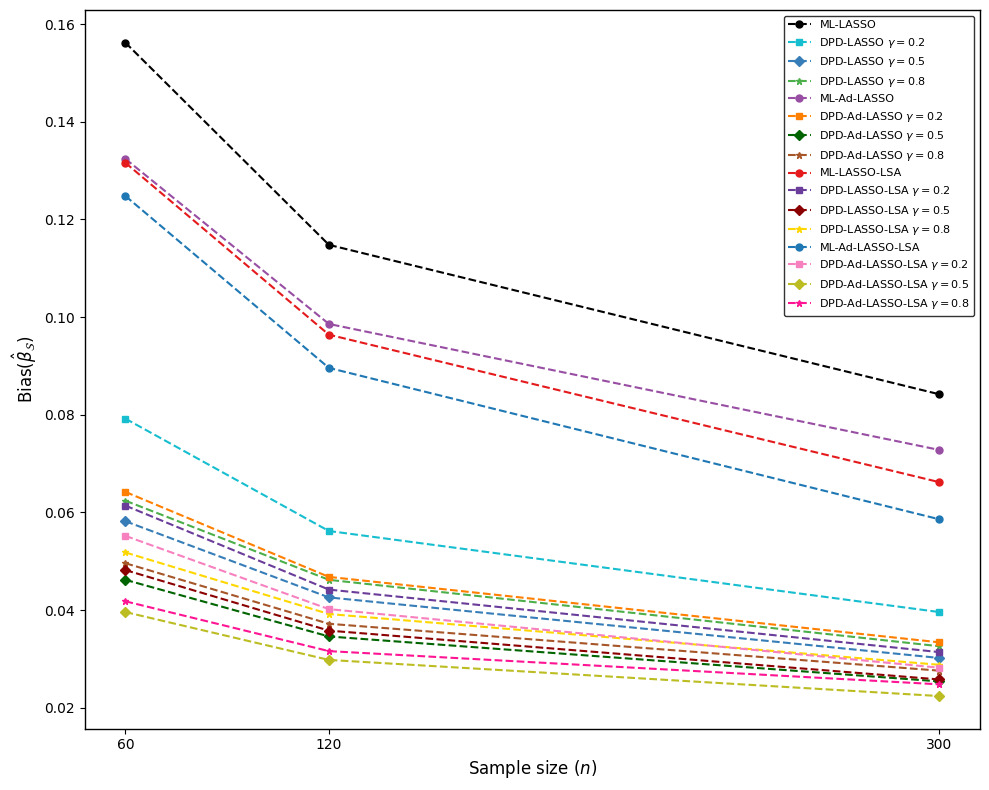}
}
\hfill
\subfloat[m = 20, 15\% y-outliers]{
\includegraphics[width=0.48\textwidth]{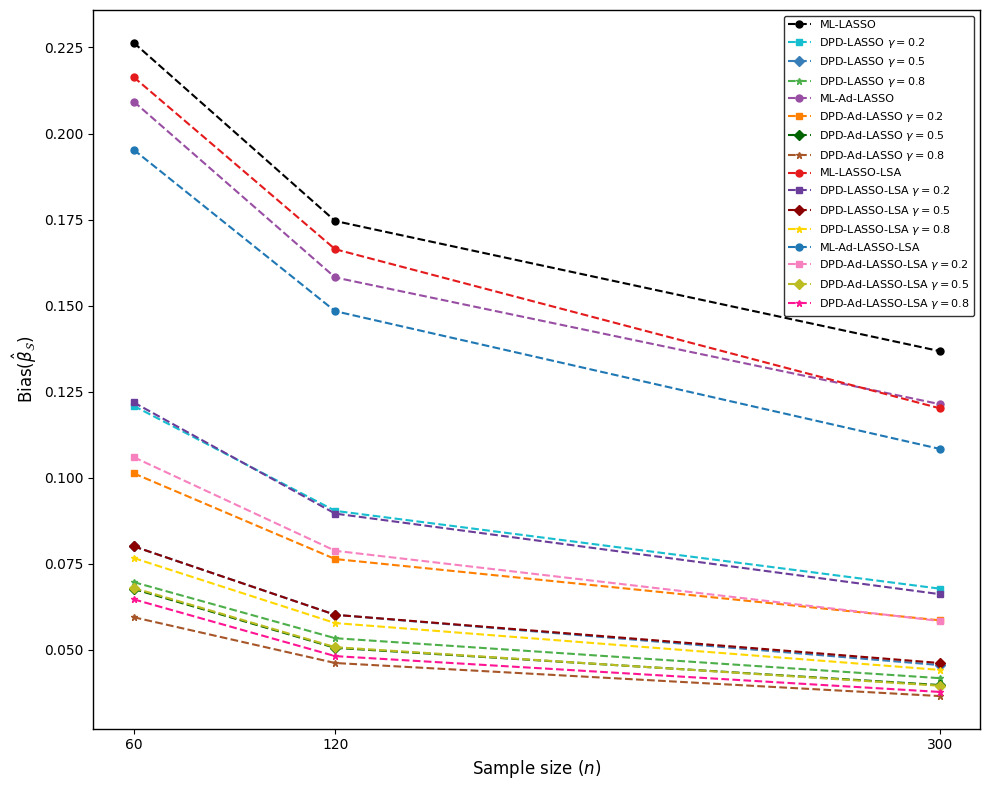}
}

\caption{Bias$(\boldsymbol{\hat{\beta}}_{\mathcal{S}})$ of various estimation procedures with increasing sample size $(n)$ under $10\%$ $y$-outlier contamination for different values of time points $(m)$.}
\label{fig2}
\end{figure}

\begin{figure}[!htbp]
\centering

\subfloat[m = 10, 15\% X-outliers]{
\includegraphics[width=0.48\textwidth]{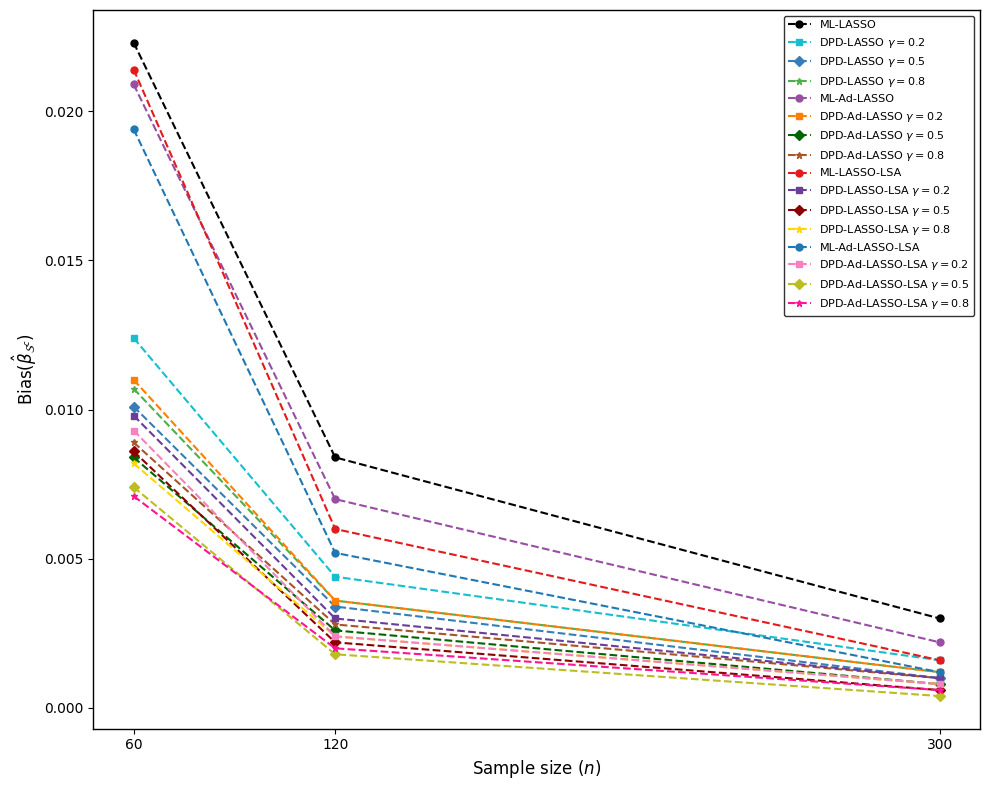}
}
\hfill
\subfloat[m = 20, 15\% X-outliers]{
\includegraphics[width=0.48\textwidth]{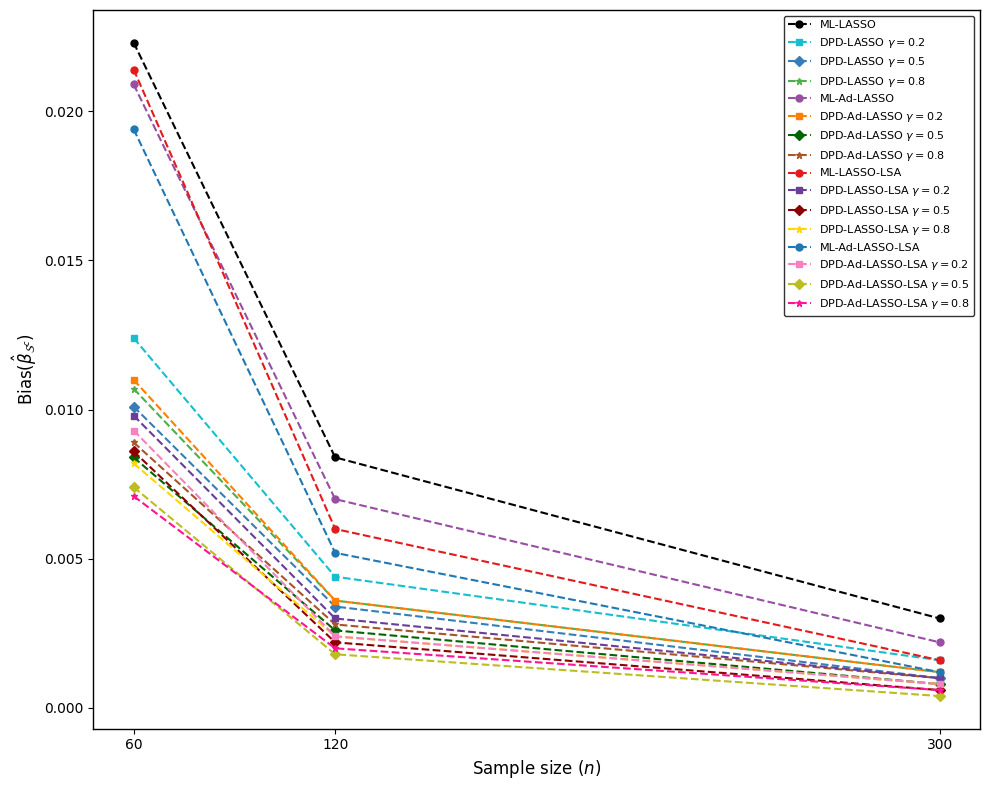}
}

\caption{Bias$(\boldsymbol{\hat{\beta}}_{\mathcal{S}^c})$ of various estimation procedures with increasing sample size $(n)$ under $15\%$ $X$-outlier contamination for different values of time points $(m)$.}
\label{fig3}
\end{figure}

\hyperref[fig1]{Figure \ref{fig1}} \textcolor{blue}{(a)-(d)} illustrate the behavior of Bias$(\boldsymbol{\hat{\beta}}_{\mathcal{S}})$ for the regression coefficients of relevant variables under different contamination schemes and selected combinations of sample size $(n)$ and time points $(m)$, while \hyperref[fig2]{Figure \ref{fig2}} \textcolor{blue}{(a)-(b)} further demonstrate its variation with increasing sample size under a fixed contamination level of $15\% \ y\text{-outliers}$ for $m=10$ and $m=20$, respectively.  The figures portray that the empirical bias decreases steadily as both $n$ and $m$ increase, indicating improved finite-sample stability and convergence behavior of all estimators.  Under both pure and contaminated settings, the DPD-based procedures consistently outperform the corresponding ML-based methods, with the superiority becoming more pronounced at higher contamination levels.  From \textcolor{red}{Table 1-Table 3 of the supplementary material}, it is evident that in case of true non-zero coefficients, although the bias increases for all methods as contamination severity rises, the increase is substantially larger for the classical likelihood-based procedures, particularly under $X$-outlier contamination, whereas the DPD-based estimators remain comparatively stable even under $15\%$ contamination, demonstrating strong robustness against both response and covariate outliers.  Moreover, the adaptive penalization schemes consistently improve estimation accuracy over the ordinary LASSO procedures in terms of bias.  Further, the incorporation of the least squares approximation in DPD-LASSO and DPD-Ad-LASSO depicts that the estimation performance is not compromised relative to its non-LSA counterparts in terms of bias (refer to \textcolor{red}{Table 1-Table 3 of the supplementary material}) for the cases investigated in the present study.

It can be inferred from \textcolor{red}{Table 4-Table 6 of the supplementary material} that the bias associated with insignificant coefficients decreases substantially with increasing sample size and longitudinal dimension, indicating improved sparsity recovery and variable selection consistency, with the reduction being more pronounced for the robust DPD-based procedures, particularly for larger tuning parameters $(\gamma=0.5,0.8)$.  Although both $y$- and $X$-outliers inflate the bias of irrelevant variables, the deterioration is considerably larger for the ML-based methods, especially under leverage-point contamination.  In contrast, the DPD-based estimators maintain comparatively stable and much smaller bias values, with the DPD-Ad-LASSO-LSA estimator consistently achieving the smallest Bias$(\boldsymbol{\hat{\beta}}_{\mathcal{S}^c})$, thereby demonstrating superior oracle-like behavior and more accurate elimination of irrelevant variables across all settings considered.   \hyperref[fig3]{Figure \ref{fig3}} \textcolor{blue}{(a)-(b)} illustrate the same behavior of Bias$(\boldsymbol{\hat{\beta}}_{\mathcal{S}^c})$ for the regression coefficients of irrelevant variables under the representative settings $(m=10, 15\% \ X\text{-outliers})$ and $(m=20, 15\% \ X\text{-outliers})$, respectively.  

In addition to the empirical bias study, the overall performance of the competing estimators is examined using measures such as MS$(\boldsymbol{\hat{\beta}})$, TP$(\boldsymbol{\hat{\beta}})$, TN$(\boldsymbol{\hat{\beta}})$, MSES$(\boldsymbol{\hat{\beta}})$, MSEN$(\boldsymbol{\hat{\beta}})$, and EE$(\hat{\sigma}^2)$ under both pure and contaminated settings.  For brevity, only selected representative results corresponding to $n=60$ are discussed in the main manuscript, while the complete numerical results for $n=120$ and $n=300$ are provided in the supplementary material.

The results presented in \hyperref[tab2]{Table \ref{tab2}}-\hyperref[tab8]{Table \ref{tab8}} summarize the performance of the competing penalized estimators for $n=60$ under pure data as well as different levels of $y$- and $X$-outlier contamination.  Across all settings, the estimation performance improves with increasing number of time points $(m)$, leading to smaller values of MSES$(\boldsymbol{\hat{\beta}})$, MSEN$(\boldsymbol{\hat{\beta}})$, and EE$(\hat{\sigma}^2)$ together with improved TP and TN rates.  The ML-based procedures exhibit substantial deterioration under contaminated settings, particularly in the presence of leverage-point contamination, where large estimation errors and weaker sparsity recovery are observed.  In contrast, the DPD-based estimators remain considerably more stable and robust across all contamination schemes, with the improvement becoming more pronounced for moderate and larger robustness tuning parameters $(\gamma=0.5, 0.8)$.  The adaptive penalization schemes consistently outperform the ordinary LASSO procedures by achieving improved sparsity recovery, higher TN probabilities, and lower estimation errors, indicating more accurate identification and elimination of irrelevant variables.  A key feature of this proposed framework is the incorporation of the least squares approximation (LSA), which yields a substantial and consistent reduction in computational runtime compared to non-LSA counterparts, thereby making the estimation procedure considerably faster.  Despite this significant computational gain, the LSA-based estimators preserve, not compromise, statistical performance in terms of estimation accuracy and sparsity recovery.  Compared with the DPD-LASSO and DPD-LASSO-LSA procedures, the adaptive version provides improved variable selection consistency and smaller estimation errors due to its adaptive weighting mechanism.  Overall, the proposed DPD-Ad-LASSO-LSA procedure, especially for $\gamma=0.5$ and $\gamma=0.8$, provides the best balance among robustness, sparsity recovery, estimation accuracy, and computational efficiency, making it the preferred estimator among all competing methods.

The detailed performance measures for the moderate and large sample size settings $(n=120,300)$ under pure data as well as different levels of $y$- and $X$-outlier contamination, are presented in \textcolor{red}{Table 7-Table 13 and Table 14-Table 20 of the supplementary material}, respectively.  Throughout all considered settings, estimation accuracy, sparsity recovery, and variance estimation improve consistently with increasing sample size and longitudinal dimension $(m)$, as reflected through progressively smaller values of MS$(\boldsymbol{\hat{\beta}})$, MSES$(\boldsymbol{\hat{\beta}})$, MSEN$(\boldsymbol{\hat{\beta}})$, and EE$(\hat{\sigma}^2)$ together with increasingly higher TP and TN rates.  The results further demonstrate that the oracle-like behavior of the proposed robust adaptive procedures becomes stronger as the sample size increases from $n=60$ to $n=300$ and the longitudinal dimension increases from $m=10$ to $m=20$, leading to near-perfect identification of relevant and irrelevant variables in many scenarios. The robustness advantage of the DPD-based procedures over the ML-based counterparts becomes even more evident for larger sample sizes and severe contamination settings, particularly under leverage-point contamination.  Among all competing estimators, the proposed DPD-Ad-LASSO-LSA procedure consistently delivers comparable performance in terms of robustness, sparsity recovery, estimation accuracy, and variable selection consistency, while achieving the highest computational efficiency.  Consequently, the DPD Adaptive LASSO-LSA (DPD-Ad-LASSO-LSA) estimator emerges as the most practically attractive and computationally efficient procedure among all competing methods for sparse panel data models, owing to its excellent balance among robustness, oracle property, estimation efficiency, sparsity recovery, and computational scalability.

\begin{table}[!htbp]
\centering
\caption{Performance measures of different methods for $n=60$ with no outliers.}
\label{tab2}
\vspace{0.1cm}

\resizebox{0.98\textwidth}{!}{%
\begin{tabular}{clccccccc}
\toprule
\multirow{2}{*}{$\boldsymbol{m}$} & \multirow{2}{*}{\textbf{Method}} 
& \textbf{MS$(\boldsymbol{\hat{\beta}})$} & \textbf{TP$(\boldsymbol{\hat{\beta}})$} & \textbf{TN$(\boldsymbol{\hat{\beta}})$} 
& \textbf{MSES$(\boldsymbol{\hat{\beta}})$} & \textbf{MSEN$(\boldsymbol{\hat{\beta}})$} & \textbf{EE$(\hat{\sigma}^2)$} & \textbf{Runtime} \\
& & & & & \textbf{$(10^{-3})$} & \textbf{$(10^{-5})$} & \textbf{$(10^{-2})$} & \textbf{$(10^{-2})$} \\
\midrule

\multirow{16}{*}{10}
& ML-LASSO                 & 10.48 & 0.93 & 0.74 & 10.86 & 18.92 & 14.85 & 138.42 \\
& DPD-LASSO $\gamma=0.2$   & 10.67 & 0.98 & 0.76 & 3.28 & 11.74 & 10.12 & 196.84 \\
& DPD-LASSO $\gamma=0.5$   & 10.82 & 1.00 & 0.79 & 2.84 & 8.96  & 8.67  & 182.36 \\
& DPD-LASSO $\gamma=0.8$   & 11.04 & 1.00 & 0.75 & 3.51 & 10.21 & 8.21  & 208.57 \\

& ML-Ad-LASSO              & 9.68 & 0.95 & 0.84 & 9.42 & 15.84 & 13.06 & 164.28 \\
& DPD-Ad-LASSO $\gamma=0.2$ & 9.86 & 1.00 & 0.87 & 2.54 & 8.12  & 9.06  & 254.63 \\
\bottomrule
\end{tabular}
}
\end{table}

\begin{table}[!htbp]
\centering
\caption{Performance measures of different methods for $n=60$ with $5\%$ y-outliers.}
\label{tab3}
\vspace{0.1cm}

\resizebox{0.98\textwidth}{!}{%
\begin{tabular}{clccccccc}
\toprule
\multirow{2}{*}{$\boldsymbol{m}$} & \multirow{2}{*}{\textbf{Method}} 
& \textbf{MS$(\boldsymbol{\hat{\beta}})$} & \textbf{TP$(\boldsymbol{\hat{\beta}})$} & \textbf{TN$(\boldsymbol{\hat{\beta}})$} 
& \textbf{MSES$(\boldsymbol{\hat{\beta}})$} & \textbf{MSEN$(\boldsymbol{\hat{\beta}})$} & \textbf{EE$(\hat{\sigma}^2)$} & \textbf{Runtime} \\
& & & & & \textbf{$(10^{-3})$} & \textbf{$(10^{-5})$} & \textbf{$(10^{-2})$} & \textbf{$(10^{-2})$} \\
\midrule

\multirow{16}{*}{10}
& ML-LASSO                 & 11.18 & 0.89 & 0.66 & 18.42 & 34.76 & 43.58 & 156.84 \\
& DPD-LASSO $\gamma=0.2$   & 10.74 & 0.97 & 0.74 & 5.12 & 15.84 & 18.72 & 214.63 \\
& DPD-LASSO $\gamma=0.5$   & 10.46 & 1.00 & 0.81 & 3.06 & 9.42  & 11.36 & 198.42 \\
& DPD-LASSO $\gamma=0.8$   & 10.58 & 0.99 & 0.79 & 3.58 & 10.86 & 10.48 & 226.75 \\

& ML-Ad-LASSO              & 10.36 & 0.92 & 0.71 & 15.64 & 29.82 & 38.14 & 182.57 \\
& DPD-Ad-LASSO $\gamma=0.2$ & 9.74 & 0.99 & 0.86 & 4.02 & 11.28 & 15.06 & 276.48 \\
\bottomrule
\end{tabular}
}
\end{table}

\begin{table}[!htbp]
\centering
\caption{Performance measures of different methods for $n=60$ with $10\%$ y-outliers.}
\label{tab4}
\vspace{0.1cm}

\resizebox{0.98\textwidth}{!}{%
\begin{tabular}{clccccccc}
\toprule
\multirow{2}{*}{$\boldsymbol{m}$} & \multirow{2}{*}{\textbf{Method}} 
& \textbf{MS$(\boldsymbol{\hat{\beta}})$} & \textbf{TP$(\boldsymbol{\hat{\beta}})$} & \textbf{TN$(\boldsymbol{\hat{\beta}})$} 
& \textbf{MSES$(\boldsymbol{\hat{\beta}})$} & \textbf{MSEN$(\boldsymbol{\hat{\beta}})$} & \textbf{EE$(\hat{\sigma}^2)$} & \textbf{Runtime} \\
& & & & & \textbf{$(10^{-3})$} & \textbf{$(10^{-5})$} & \textbf{$(10^{-2})$} & \textbf{$(10^{-2})$} \\
\midrule

\multirow{16}{*}{10}
& ML-LASSO                 & 11.84 & 0.84 & 0.58 & 28.74 & 54.18 & 68.42 & 164.82 \\
& DPD-LASSO $\gamma=0.2$   & 10.96 & 0.95 & 0.70 & 8.26  & 24.84 & 28.16 & 226.48 \\
& DPD-LASSO $\gamma=0.5$   & 10.58 & 0.99 & 0.79 & 4.72  & 13.86 & 15.42 & 208.74 \\
& DPD-LASSO $\gamma=0.8$   & 10.44 & 0.99 & 0.82 & 4.08  & 11.74 & 12.68 & 238.16 \\

& ML-Ad-LASSO              & 10.92 & 0.88 & 0.66 & 24.16 & 46.82 & 59.36 & 191.45 \\
& DPD-Ad-LASSO $\gamma=0.2$ & 9.96 & 0.98 & 0.82 & 6.12 & 17.28 & 20.84 & 292.74 \\
\bottomrule
\end{tabular}
}
\end{table}

\begin{table}[!htbp]
\centering
\caption{Performance measures of different methods for $n=60$ with $15\%$ y-outliers.}
\label{tab5}
\vspace{0.1cm}

\resizebox{0.98\textwidth}{!}{%
\begin{tabular}{clccccccc}
\toprule
\multirow{2}{*}{$\boldsymbol{m}$} & \multirow{2}{*}{\textbf{Method}} 
& \textbf{MS$(\boldsymbol{\hat{\beta}})$} & \textbf{TP$(\boldsymbol{\hat{\beta}})$} & \textbf{TN$(\boldsymbol{\hat{\beta}})$} 
& \textbf{MSES$(\boldsymbol{\hat{\beta}})$} & \textbf{MSEN$(\boldsymbol{\hat{\beta}})$} & \textbf{EE$(\hat{\sigma}^2)$} & \textbf{Runtime} \\
& & & & & \textbf{$(10^{-3})$} & \textbf{$(10^{-5})$} & \textbf{$(10^{-2})$} & \textbf{$(10^{-2})$} \\
\midrule

\multirow{16}{*}{10}
& ML-LASSO                 & 12.36 & 0.79 & 0.52 & 36.82 & 68.74 & 84.16 & 172.48 \\
& DPD-LASSO $\gamma=0.2$   & 11.18 & 0.92 & 0.66 & 11.24 & 31.86 & 36.42 & 236.84 \\
& DPD-LASSO $\gamma=0.5$   & 10.62 & 0.98 & 0.78 & 5.86  & 16.42 & 18.74 & 218.63 \\
& DPD-LASSO $\gamma=0.8$   & 10.28 & 0.99 & 0.84 & 4.62  & 12.68 & 13.94 & 246.35 \\

& ML-Ad-LASSO              & 11.24 & 0.84 & 0.61 & 31.74 & 58.42 & 73.28 & 198.26 \\
& DPD-Ad-LASSO $\gamma=0.2$ & 10.08 & 0.96 & 0.79 & 8.42 & 22.86 & 26.74 & 304.58 \\
\bottomrule
\end{tabular}
}
\end{table}

\begin{table}[!htbp]
\centering
\caption{Performance measures of different methods for $n=60$ with $5\%$ X-outliers.}
\label{tab6}
\vspace{0.1cm}

\resizebox{0.98\textwidth}{!}{%
\begin{tabular}{clccccccc}
\toprule
\multirow{2}{*}{$\boldsymbol{m}$} & \multirow{2}{*}{\textbf{Method}} 
& \textbf{MS$(\boldsymbol{\hat{\beta}})$} & \textbf{TP$(\boldsymbol{\hat{\beta}})$} & \textbf{TN$(\boldsymbol{\hat{\beta}})$} 
& \textbf{MSES$(\boldsymbol{\hat{\beta}})$} & \textbf{MSEN$(\boldsymbol{\hat{\beta}})$} & \textbf{EE$(\hat{\sigma}^2)$} & \textbf{Runtime} \\
& & & & & \textbf{$(10^{-3})$} & \textbf{$(10^{-5})$} & \textbf{$(10^{-2})$} & \textbf{$(10^{-2})$} \\
\midrule

\multirow{16}{*}{10}
& ML-LASSO                 & 11.86 & 0.83 & 0.47 & 29.84 & 51.36 & 68.72 & 154.36 \\
& DPD-LASSO $\gamma=0.2$   & 10.92 & 0.91 & 0.60 & 9.62  & 25.18 & 33.46 & 218.47 \\
& DPD-LASSO $\gamma=0.5$   & 10.46 & 0.96 & 0.72 & 4.54  & 13.86 & 17.42 & 202.64 \\
& DPD-LASSO $\gamma=0.8$   & 10.34 & 0.96 & 0.74 & 4.06  & 12.94 & 16.28 & 229.58 \\

& ML-Ad-LASSO              & 11.18 & 0.87 & 0.54 & 24.96 & 43.84 & 59.84 & 182.75 \\
& DPD-Ad-LASSO $\gamma=0.2$ & 10.22 & 0.94 & 0.72 & 7.64 & 20.24 & 27.36 & 278.36 \\
\bottomrule
\end{tabular}
}
\end{table}

\begin{table}[!htbp]
\centering
\caption{Performance measures of different methods for $n=60$ with $10\%$ X-outliers.}
\label{tab7}
\vspace{0.1cm}

\resizebox{0.98\textwidth}{!}{%
\begin{tabular}{clccccccc}
\toprule
\multirow{2}{*}{$\boldsymbol{m}$} & \multirow{2}{*}{\textbf{Method}} 
& \textbf{MS$(\boldsymbol{\hat{\beta}})$} & \textbf{TP$(\boldsymbol{\hat{\beta}})$} & \textbf{TN$(\boldsymbol{\hat{\beta}})$} 
& \textbf{MSES$(\boldsymbol{\hat{\beta}})$} & \textbf{MSEN$(\boldsymbol{\hat{\beta}})$} & \textbf{EE$(\hat{\sigma}^2)$} & \textbf{Runtime} \\
& & & & & \textbf{$(10^{-3})$} & \textbf{$(10^{-5})$} & \textbf{$(10^{-2})$} & \textbf{$(10^{-2})$} \\
\midrule

\multirow{16}{*}{10}
& ML-LASSO                 & 12.48 & 0.76 & 0.38 & 42.86 & 76.42 & 102.58 & 168.74 \\
& DPD-LASSO $\gamma=0.2$   & 11.24 & 0.86 & 0.54 & 15.48 & 39.84 & 52.36 & 236.42 \\
& DPD-LASSO $\gamma=0.5$   & 10.62 & 0.93 & 0.68 & 6.84  & 18.62 & 24.18 & 218.35 \\
& DPD-LASSO $\gamma=0.8$   & 10.42 & 0.95 & 0.72 & 5.62  & 15.84 & 20.46 & 248.57 \\

& ML-Ad-LASSO              & 11.74 & 0.81 & 0.46 & 36.28 & 64.72 & 88.34 & 198.26 \\
& DPD-Ad-LASSO $\gamma=0.2$ & 10.46 & 0.90 & 0.66 & 11.84 & 30.48 & 40.62 & 304.84 \\
\bottomrule
\end{tabular}
}
\end{table}

\begin{table}[!htbp]
\centering
\caption{Performance measures of different methods for $n=60$ with $15\%$ X-outliers.}
\label{tab8}
\vspace{0.1cm}

\resizebox{0.98\textwidth}{!}{%
\begin{tabular}{clccccccc}
\toprule
\multirow{2}{*}{$\boldsymbol{m}$} & \multirow{2}{*}{\textbf{Method}} 
& \textbf{MS$(\boldsymbol{\hat{\beta}})$} & \textbf{TP$(\boldsymbol{\hat{\beta}})$} & \textbf{TN$(\boldsymbol{\hat{\beta}})$} 
& \textbf{MSES$(\boldsymbol{\hat{\beta}})$} & \textbf{MSEN$(\boldsymbol{\hat{\beta}})$} & \textbf{EE$(\hat{\sigma}^2)$} & \textbf{Runtime} \\
& & & & & \textbf{$(10^{-3})$} & \textbf{$(10^{-5})$} & \textbf{$(10^{-2})$} & \textbf{$(10^{-2})$} \\
\midrule

\multirow{16}{*}{10}
& ML-LASSO                 & 13.06 & 0.68 & 0.28 & 58.42 & 108.76 & 146.84 & 182.36 \\
& DPD-LASSO $\gamma=0.2$   & 11.86 & 0.81 & 0.46 & 22.74 & 56.42 & 74.18 & 252.48 \\
& DPD-LASSO $\gamma=0.5$   & 10.94 & 0.90 & 0.63 & 9.82  & 26.74 & 34.62 & 236.74 \\
& DPD-LASSO $\gamma=0.8$   & 10.58 & 0.93 & 0.69 & 7.48  & 20.86 & 27.18 & 268.53 \\

& ML-Ad-LASSO              & 12.18 & 0.74 & 0.36 & 49.84 & 92.36 & 126.42 & 214.68 \\
& DPD-Ad-LASSO $\gamma=0.2$ & 10.84 & 0.86 & 0.58 & 17.36 & 44.82 & 58.74 & 328.42 \\
\bottomrule
\end{tabular}
}
\end{table}

\subsection{Real Data Analysis}

For the real data analysis, we have utilized the panel data on bone mineral density (BMD) of women during the menopause transition (MT) stage from the Study of Women’s Health Across the Nation (SWAN).  SWAN is a large-scale, multi-site, community-based cohort study designed to investigate the biological and psychosocial changes that occur in midlife women.  Since osteoporotic fractures in postmenopausal women often result from bone loss that begins before menopause, the SWAN data are particularly suitable for examining bone loss progression and its determinants during the MT stage.  The study recruited women aged 42–52 years with an intact uterus and at least one ovary, who were not on hormone therapy and had menstruated within the last three months before screening.  Participants self-identified with one of five ethnic groups and were enrolled across seven U.S. sites (Boston, Detroit, Los Angeles, Oakland, Pittsburgh, Chicago, and Newark).  The Bone Cohort comprised 2,413 women from five sites (excluding Chicago and Newark, where bone data were not available).  Initiated in 1996, the cohort has undergone one baseline and fifteen follow-up visits at approximately 18-month intervals, providing a rich panel dataset for analyzing changes in bone health over time.

In the present analysis, we have considered a subset of $30$ women from the five Bone Cohort sites of the Study of Women’s Health Across the Nation (SWAN), for whom detailed repeated measurements of bone mineral density (BMD) and associated demographic, clinical, and biochemical covariates were available.  We employ this comprehensive longitudinal dataset within a robust statistical framework, applying the proposed DPD Adaptive LASSO-LSA (DPD-Ad-LASSO-LSA) estimator to the linear panel mixed-effect model given in \eqref{eq1}.  In this model, $y_{it}$ denotes the bone mineral density of the lumbar spine for the $i^{th}$ respondent at the $t^{th}$ visit date, $X_{it}$ represents the corresponding fixed-effect covariates, and $u_{it}$ is the noise term.  A random intercept term, $\alpha$, is incorporated to account for the unobserved heterogeneity among the respondents. 

In our model, the fixed effects ($X_{it}$) include clinical, biochemical, and lifestyle covariates such as: \textit{Age}, \textit{Race}, \textit{Site}, \textit{Education Years}, \textit{Income Category}, \textit{Marital Status}, \textit{Parity}, \textit{Height (cm)}, \textit{Weight (kg)}, \textit{BMI}, \textit{Menopausal Status}, \textit{Estradiol (pg/ml)}, \textit{FSH (mIU/ml)}, \textit{Cigarette Units per week}, \textit{Alcohol Units per week}, \textit{Physical Activity (min/week)}, \textit{Number of fractures}, \textit{Calcium intake (mg/day)}, \textit{Vitamin D (ng/ml)}, \textit{Systolic BP (mmHg)}, \textit{Low-density lipoprotein (LDL) (mg/dl)}, \textit{High-density lipoprotein (HDL) (mg/dl)}, \textit{Triglycerides (mg/dl)}, \textit{Glucose (mg/dl)}, \textit{C-Reactive Protein (CRP) (mg/L)}, \textit{Center for Epidemiologic Studies Depression Scale (CESD) Score}, \textit{Bone Turnover Marker}, along with binary indicators such as \textit{Diabetes} and \textit{Hypertension status}.  These covariates collectively capture both physiological and behavioral factors that influence bone mineral density.  By distinguishing between population-level influences and individual bone density patterns, this framework provides critical insights into the progression of bone loss during menopause.  The resulting findings facilitate the early detection of osteoporosis risk and promote sustainable health and an improved quality of life for midlife women.

In large-scale biomedical longitudinal datasets, such as the bone mineral density (BMD) measurements obtained from the SWAN Bone Cohort, the presence of outliers is inevitable due to measurement errors, missing clinical records, or unexpected physiological fluctuations. Since the true outliers are not known a priori, an effective detection mechanism is essential before performing robust estimation.  Accordingly, we employ the distance covariance measure proposed by Wang and Li \cite{wang2017outlier} for identifying anomalous observations, through which a total of 54 outlying data points were detected in our dataset.

We next estimate the parameters under both the classical likelihood and the robust DPD $\gamma = (0.2, 0.5, 0.8)$ frameworks, incorporating LASSO, Adaptive LASSO, and their corresponding LSA versions.  To begin with, the penalty parameter $\lambda$ is selected using ERIC, and the optimal value is obtained as $\lambda = 0.065$.  The initial values of the model coefficients are obtained from the ordinary least squares estimates derived from the original model.  The numerical values of these initial estimates are obtained as (0.3526, 0.0849, 0.0163, 0.0082, 0.0253, 0.0411, 0.0158, 0.0827, -0.0958, 0.2453, 0.4517, 0.1225, 0.1102, 0.0111, 0.0084, 0.3746, 0.0537, -0.2815, -0.2724, 0.0045, 0.0032, -0.0074, -0.1215, -0.3438, -0.0630, 0.0093, 0.0754, -0.3176, -0.4262), with the fixed intercept term estimated as $1.8483$.  

\begin{figure}[!htbp]
\centering
\includegraphics[width=0.85\textwidth]{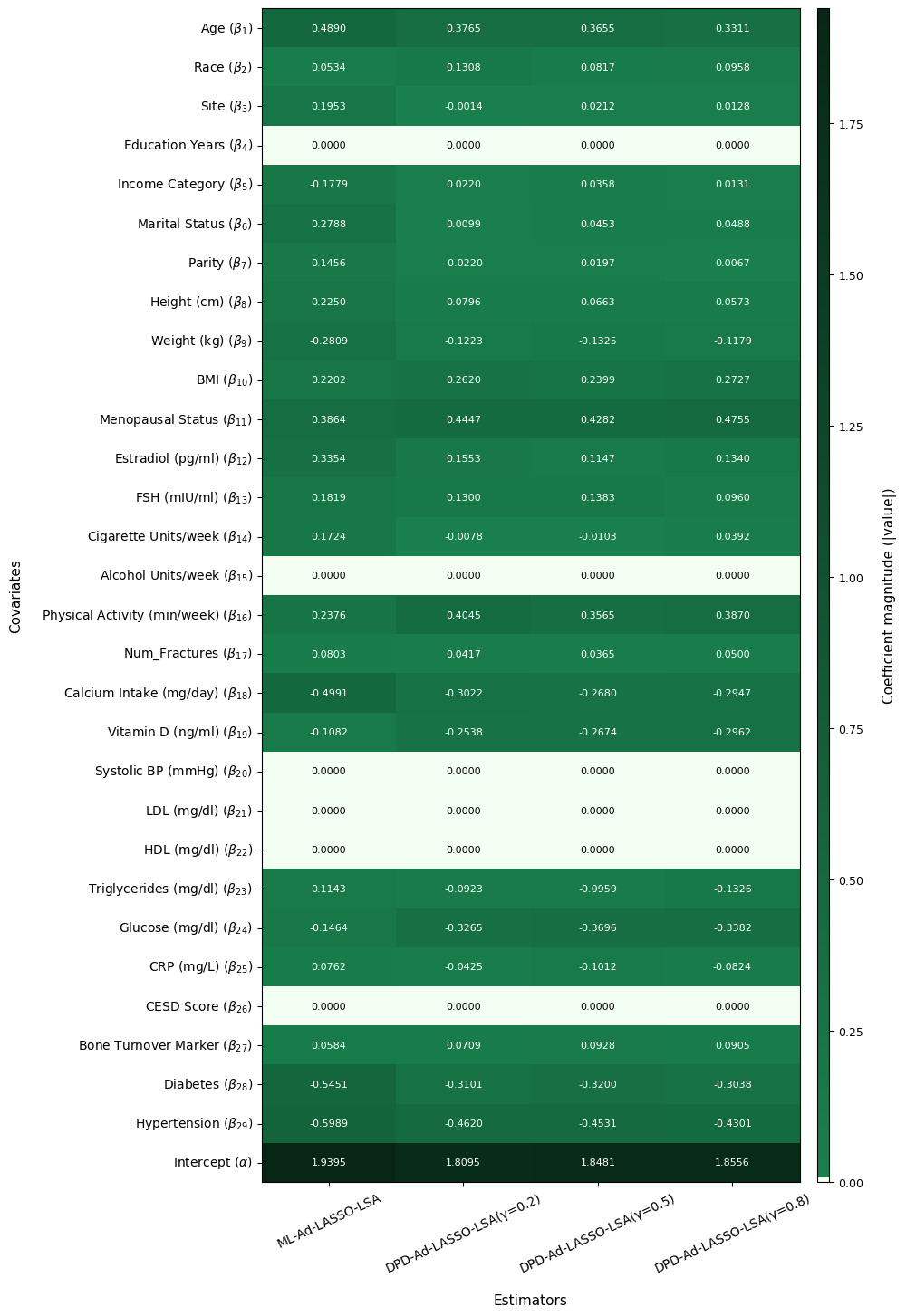}
\caption{Estimated values of model coefficients.}
\label{fig4} 
\end{figure}

An overview of the estimated coefficient values derived from the Adaptive-LASSO-LSA procedures under both the ML and DPD frameworks is presented in \hyperref[fig4]{Figure \ref{fig4}}.  The corresponding estimated coefficient values obtained through the remaining competing procedures are reported in \textcolor{red}{Table 21-Table 23 of the supplementary material}.  Among the predictors, Age, BMI, Menopausal Status, Estradiol, Physical Activity, Calcium Intake, Vitamin D, Diabetes, and Hypertension display relatively large coefficient magnitudes, indicating a strong association with Lumbar spine BMD.  

\noindent Moderate effects are observed for Site, Marital Status, Parity, Height, Weight, FSH, Cigarette Units/week, Number of Fractures, Triglycerides, Glucose, and CRP, suggesting that these factors contribute meaningfully but less dominantly to the model.  On the other hand, variables such as Education Years, Income Category, Race, Alcohol Units/week, Systolic BP, LDL, HDL, CESD Score, and Bone Turnover Marker exhibit very small or near-zero coefficients across all estimators, indicating a minimal or statistically insignificant influence on Lumbar spine BMD.  

\begin{figure}[h]
\centering
\includegraphics[width=1.00\textwidth]{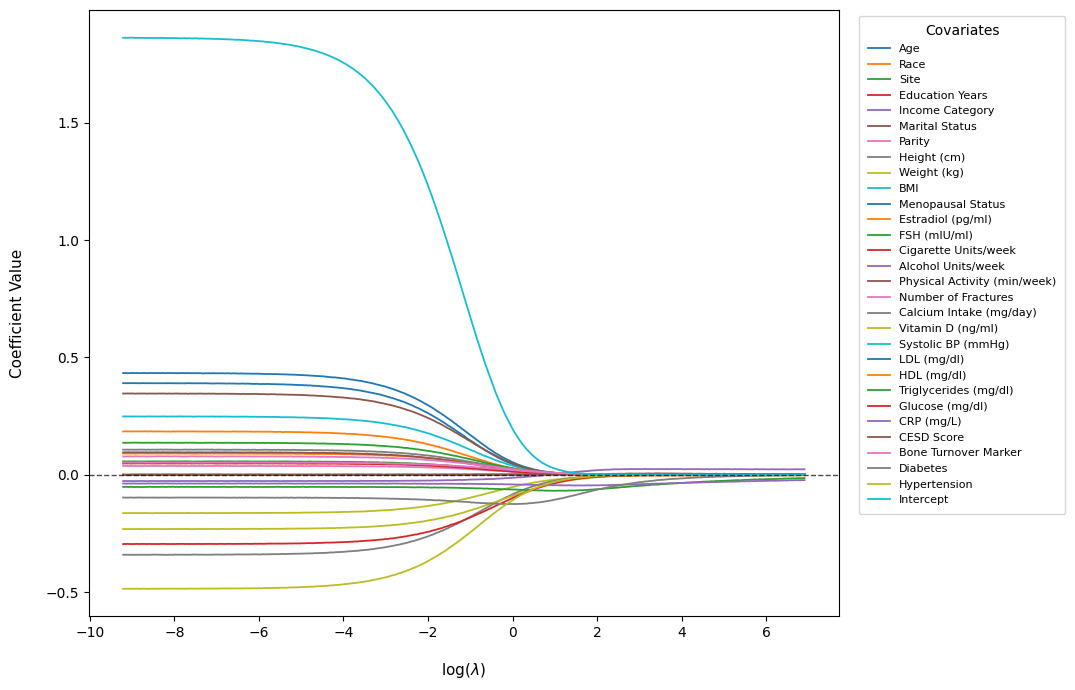}
\caption{Coefficient paths over $\lambda$.}
\label{fig5} 
\end{figure}

To obtain an insightful visualization of how the estimated model coefficients evolve across varying levels of the penalty parameter, $\lambda$, we have further plotted the coefficient paths in \hyperref[fig5]{Figure \ref{fig5}}.  Such a plot is crucial in $\ell_1$-regularization framework, as it allows us to assess the stability of predictors and identify which variables retain influence under increasing penalization.  Among the predictors, Age, BMI, Menopausal Status, Estradiol, Physical Activity, Calcium Intake, Vitamin D, Diabetes, and Hypertension exhibit relatively large coefficient magnitudes, indicating strong and persistent associations with Lumbar spine BMD.  These variables remain influential over a broad range of $\lambda$, demonstrating robustness against penalization.  Moderate effects are observed for Site, Marital Status, Parity, Height, Weight, FSH, Cigarette Units/week, Number of Fractures, Triglycerides, Glucose, and CRP, whose coefficients gradually diminish with increasing $\lambda$, reflecting moderate but meaningful contributions to the model.  Conversely, predictors such as Education Years, Income Category, Race, Alcohol Units/week, Systolic BP, LDL, HDL, CESD Score, and Bone Turnover Marker converge rapidly to zero, indicating limited or negligible explanatory influence.  The intercept term remains positive and stable throughout, capturing the baseline BMD level after accounting for all covariates.  As $\lambda$ increases, even the more influential predictors undergo gradual shrinkage.  The need for such a plot arises from its ability to highlight the relative importance of variables and to guide model selection by balancing complexity and interpretability in penalized regression settings.

Moreover, based on $500$ bootstrap samples, the bootstrap bias and MSE of the competing estimators are evaluated to further assess their finite-sample performance.  It has been observed from \hyperref[tab9]{Table \ref{tab9}} and \hyperref[tab10]{Table \ref{tab10}} that the DPD-Ad-LASSO-LSA estimator evaluated across $\gamma=(0.2,0.5,0.8)$ exhibits substantially lower bootstrap bias and MSE compared with its ML counterpart, indicating superior robustness and estimation stability in the presence of potential anomalous observations.  The detailed bootstrap summaries for the remaining competing procedures are provided in \textcolor{red}{Table 24-Table 29}, and the computational runtimes of all estimators are summarized in \textcolor{red}{Table 30 of the supplementary material}.  The results further demonstrate that the proposed DPD-Ad-LASSO-LSA estimator consistently outperforms both the DPD-LASSO and DPD-Ad-LASSO procedures in terms of estimation accuracy and computational efficiency.   Consequently, the proposed DPD-Ad-LASSO-LSA estimator emerges as the most practically effective procedure for real data analysis by simultaneously combining strong robustness, efficient sparse estimation, improved variable selection consistency, and significantly enhanced computational performance.

\begin{table}[h]
\caption{Bootstrap bias of estimates of model coefficients}\label{tab9}
\vspace{0.1cm}
\centering
\fontsize{8}{10}\selectfont
\begin{tabular}{@{}lcccc@{}}
\toprule
\textbf{Parameter} & \textbf{ML-Ad-LASSO-} & \textbf{DPD-Ad-LASSO-} & \textbf{DPD-Ad-LASSO-} & \textbf{DPD-Ad-LASSO-} \\
& \textbf{LSA} & \textbf{LSA} $(\boldsymbol{\gamma=0.2})$ & \textbf{LSA} $(\boldsymbol{\gamma=0.5})$ & \textbf{LSA} $(\boldsymbol{\gamma=0.8})$ \\ 
\hline 
$\beta_{1}$ & -0.006935 & 0.001756 & 0.001565 & -0.001688 \\
$\beta_{2}$ & 0.004168 & 0.000037 & -0.000698 & -0.002749 \\
$\beta_{3}$ & -0.003133 & -0.001576 & 0.000055 & 0.000218 \\
$\beta_{4}$ & 0.000000 & 0.000000 & 0.000000 & 0.000000 \\
$\beta_{5}$ & 0.006053 & 0.003291 & 0.001287 & 0.001434 \\
$\beta_{6}$ & 0.010449 & -0.000630 & 0.004478 & 0.001656 \\
$\beta_{7}$ & -0.004813 & 0.001787 & 0.001062 & -0.001478 \\
$\beta_{8}$ & -0.000055 & 0.001235 & -0.001286 & 0.000257 \\
$\beta_{9}$ & -0.002192 & 0.002755 & -0.000126 & -0.001833 \\
$\beta_{10}$ & 0.003526 & -0.003473 & -0.000646 & 0.002698 \\
$\beta_{11}$ & -0.000191 & 0.001557 & -0.000172 & 0.001146 \\
$\beta_{12}$ & 0.001795 & 0.000820 & 0.003111 & -0.001818 \\
$\beta_{13}$ & -0.004176 & 0.000718 & -0.003377 & -0.000443 \\
$\beta_{14}$ & -0.001293 & -0.000569 & 0.000098 & -0.000966 \\
$\beta_{15}$ & 0.000000 & 0.000000 & 0.000000 & 0.000000 \\
$\beta_{16}$ & -0.001732 & -0.002040 & 0.002737 & 0.001324 \\
$\beta_{17}$ & 0.006257 & 0.004355 & -0.002273 & -0.000240 \\
$\beta_{18}$ & -0.001777 & -0.001716 & 0.000124 & -0.001308 \\
$\beta_{19}$ & 0.005140 & 0.000196 & 0.001904 & -0.001164 \\
$\beta_{20}$ & 0.000000 & 0.000000 & 0.000000 & 0.000000 \\
$\beta_{21}$ & 0.000000 & 0.000000 & 0.000000 & 0.000000 \\
$\beta_{22}$ & 0.000000 & 0.000000 & 0.000000 & 0.000000 \\
$\beta_{23}$ & 0.000432 & 0.000006 & -0.002450 & 0.001508 \\
$\beta_{24}$ & -0.001807 & -0.000417 & 0.002007 & -0.001081 \\
$\beta_{25}$ & 0.008541 & 0.002098 & 0.002151 & 0.000414 \\
$\beta_{26}$ & 0.000000 & 0.000000 & 0.000000 & 0.000000 \\
$\beta_{27}$ & -0.017483 & -0.001891 & 0.001770 & 0.001837 \\
$\beta_{28}$ & 0.007477 & -0.000592 & -0.001804 & 0.001144 \\
$\beta_{29}$ & -0.000998 & 0.000437 & -0.000533 & 0.001797 \\
$\alpha$ & 0.001349 & 0.001469 & -0.002274 & -0.000751 \\

\bottomrule
\end{tabular}
\end{table}

\begin{table}[h]
\caption{Bootstrap MSE of estimates of model coefficients}\label{tab10}
\vspace{0.1cm}
\centering
\fontsize{8}{10}\selectfont
\begin{tabular}{@{}lcccc@{}}
\toprule
\textbf{Parameter} & \textbf{ML-Ad-LASSO-} & \textbf{DPD-Ad-LASSO-} & \textbf{DPD-Ad-LASSO-} & \textbf{DPD-Ad-LASSO-} \\
& \textbf{LSA} & \textbf{LSA} $(\boldsymbol{\gamma=0.2})$ & \textbf{LSA} $(\boldsymbol{\gamma=0.5})$ & \textbf{LSA} $(\boldsymbol{\gamma=0.8})$ \\ 
\hline 
$\beta_{1}$ & 0.018835 & 0.001907 & 0.001433 & 0.000896 \\
$\beta_{2}$ & 0.018083 & 0.002064 & 0.001179 & 0.000911 \\
$\beta_{3}$ & 0.020125 & 0.002112 & 0.001247 & 0.001027 \\
$\beta_{4}$ & 0.000000 & 0.000000 & 0.000000 & 0.000000 \\
$\beta_{5}$ & 0.020505 & 0.002198 & 0.001230 & 0.000904 \\
$\beta_{6}$ & 0.020818 & 0.002137 & 0.001369 & 0.000973 \\
$\beta_{7}$ & 0.022602 & 0.001804 & 0.001151 & 0.000901 \\
$\beta_{8}$ & 0.018226 & 0.002125 & 0.001195 & 0.000879 \\
$\beta_{9}$ & 0.018616 & 0.002082 & 0.001360 & 0.000918 \\
$\beta_{10}$ & 0.019007 & 0.002031 & 0.001226 & 0.000956 \\
$\beta_{11}$ & 0.020128 & 0.002130 & 0.001197 & 0.000971 \\
$\beta_{12}$ & 0.017044 & 0.002054 & 0.001036 & 0.000922 \\
$\beta_{13}$ & 0.020586 & 0.001991 & 0.001219 & 0.000891 \\
$\beta_{14}$ & 0.019002 & 0.002127 & 0.001204 & 0.000850 \\
$\beta_{15}$ & 0.000000 & 0.000000 & 0.000000 & 0.000000 \\
$\beta_{16}$ & 0.021854 & 0.002052 & 0.001181 & 0.000847 \\
$\beta_{17}$ & 0.020832 & 0.002108 & 0.001079 & 0.000857 \\
$\beta_{18}$ & 0.023377 & 0.002103 & 0.001198 & 0.000930 \\
$\beta_{19}$ & 0.019305 & 0.001996 & 0.001316 & 0.000799 \\
$\beta_{20}$ & 0.000000 & 0.000000 & 0.000000 & 0.000000 \\
$\beta_{21}$ & 0.000000 & 0.000000 & 0.000000 & 0.000000 \\
$\beta_{22}$ & 0.000000 & 0.000000 & 0.000000 & 0.000000 \\
$\beta_{23}$ & 0.016889 & 0.002068 & 0.001255 & 0.000837 \\
$\beta_{24}$ & 0.019857 & 0.002101 & 0.001141 & 0.000851 \\
$\beta_{25}$ & 0.019789 & 0.001955 & 0.001301 & 0.000971 \\
$\beta_{26}$ & 0.000000 & 0.000000 & 0.000000 & 0.000000 \\
$\beta_{27}$ & 0.020068 & 0.001944 & 0.001234 & 0.000847 \\
$\beta_{28}$ & 0.020065 & 0.001884 & 0.001348 & 0.000923 \\
$\beta_{29}$ & 0.018869 & 0.001668 & 0.001233 & 0.000924 \\
$\alpha$ & 0.019927 & 0.002253 & 0.001037 & 0.000936 \\

\bottomrule
\end{tabular}
\end{table}

\section{Conclusion}\label{sec7}
\allowdisplaybreaks

In this work, we developed a robust and parsimonious estimation framework for mixed-effect linear panel data models with non-homogeneous observations by integrating the DPD criterion with the Adaptive LASSO penalty.  To overcome the computational and theoretical challenges associated with the nonlinear divergence-based objective function, we introduced an LSA of the DPD criterion, yielding an asymptotically equivalent quadratic formulation that substantially simplifies optimization while preserving robustness.  A major advantage of the proposed LSA-based framework is its significantly lower computational runtime relative to the corresponding non-LSA iterative procedures, making it computationally efficient for larger longitudinal dimensions without compromising statistical performance.  Theoretical analysis established the large-sample properties of the proposed DPD Adaptive LASSO-LSA estimator, including variable selection consistency and asymptotic normality.  Extensive simulation studies reveal that the proposed methodology achieves comparable performance in terms of robustness, estimation accuracy, and sparsity recovery while offering significantly improved computational efficiency compared to penalized likelihood and DPD-based approaches under both pure and contaminated data settings.  An application to longitudinal bone mineral density data from the SWAN study further illustrated the practical utility of the proposed methodology in identifying clinically relevant covariates and providing stable inference in the presence of heterogeneity and potential outliers.

Overall, this study contributes to the robustness and variable selection literature by offering one of the first comprehensive frameworks that unifies least squares approximated DPD with adaptively weighted LASSO penalization in panel data settings.  Future research may extend the methodology to generalized nonlinear panel models, dynamic panels, and ultra-high-dimensional regimes and may also explore data-driven selection strategies for the tuning parameter $\gamma$.  We believe these extensions will further strengthen the applicability of divergence-based penalized estimators in modern panel data analysis.

\section*{Acknowledgement(s)}
The authors acknowledge the support of Professor Leandro Pardo of Complutense University, Madrid, Spain, in the course of the work.

\section*{Disclosure statement}
The authors report there are no relevant financial or non-financial competing interests to declare.

\section*{Declaration of generative AI use}
The authors report generative AI was not used in their research or preparation of this manuscript.

\section*{Funding details}
This research is not supported by any specific grant from public, commercial, or non-profit funding organizations.

\section*{CRediT Roles}
\emph{\textbf{Udita Goswami}} contributed to writing the original draft, visualization, validation, software development, methodology design, formal analysis, and conceptualization of the study.  \emph{\textbf{Shuvashree Mondal}} was involved in revising the initial draft, developing the methodology, conducting formal analysis, and shaping the conceptual framework.

\section*{Data availability}
The data that support the findings of this study are openly available in Study of Women’s Health Across the Nation (SWAN); Bone Mineral Density Data - \url{http://www.swanstudy.org/swan-research/data-access/}.

\bibliographystyle{elsarticle-num}
\bibliography{ref}

\appendix

\section*{Appendix A: Proof of \textbf{\textit{Lemma.}}}
\allowdisplaybreaks
\label{appendix:A}

The proof of the Lemma follows a line of reasoning akin to that in Hui et al.\ \citep{hui2018sparse}, but it must accommodate complexities introduced by the DPD Adaptive LASSO-LSA estimator.  For analytical convenience in proving this lemma and the subsequent theorem, we reformulate the problem by considering the maximization of $-Q(\boldsymbol{\theta})$, which is equivalent to minimizing $Q(\boldsymbol{\theta})$.  Accordingly, the objective function can be expressed as 
\begin{align}
    -Q(\boldsymbol{\theta}) = -H_A(\boldsymbol{\theta}) - n\lambda\sum_{j=1}^{k}w_j|\boldsymbol{\beta_j}|. \tag{A.1} \label{eq17}
\end{align}
We know, 
\begin{equation*}
    \mathbf{0} = \frac{1}{\sqrt{n}}\frac{\partial H_n(\tilde{\boldsymbol{\theta}})}{\partial \boldsymbol{\theta}}. 
\end{equation*}

\noindent By a Taylor series expansion around $\boldsymbol{\theta_0}$, we get
\begin{align}
    \mathbf{0} &= \frac{1}{\sqrt{n}}\frac{\partial H_n(\boldsymbol{\theta_0})}{\partial \boldsymbol{\theta}} - \frac{1}{\sqrt{n}}R(\boldsymbol{\theta_0})(\tilde{\boldsymbol{\theta}}-\boldsymbol{\theta_0}) + \frac{1}{2\sqrt{n}}\boldsymbol{D}\nonumber \\
    &\triangleq T_1 + T_2 +T_3, \tag{A.2} \label{eq18}
\end{align}
where $\bar{\boldsymbol{\theta}}$ lies on the line segment joining $\tilde{\boldsymbol{\theta}}$ and $\boldsymbol{\theta_0}$, and $\boldsymbol{D}$ is a vector of length $k$ whose $p^{th}$ element is given by 
\[D_p = \sum_{u, v=1}^{k}(\tilde{\boldsymbol{\theta}}_u -\boldsymbol{\theta}_{u0})\left(\frac{\partial^3 H_n(\bar{\boldsymbol{\theta}})}{\partial \theta_u \, \partial \theta_l \, \partial \theta_v}\right)(\tilde{\boldsymbol{\theta}}_v-\boldsymbol{\theta}_{v0}).\]

\noindent By the Cauchy-Schwarz inequality,
\begin{align}
    \|T_3\| &\leq \frac{1}{2\sqrt{n}}\|\tilde{\boldsymbol{\theta}}-\boldsymbol{\theta_0}\|^{2} \left(\sum_{i=1}^{n}\sum_{u, v, w=1}^{k} S^{2}_{uvw}(r_i)\right)^{1/2} \nonumber \\
    &\leq \frac{1}{2}\|\tilde{\boldsymbol{\theta}}-\boldsymbol{\theta_0}\|^{2} \left(\frac{1}{n}\sum_{i=1}^{n}\sum_{u, v, w=1}^{k} S^{2}_{uvw}(r_i)\right)^{1/2}. \tag{A.3} \label{eq19}
\end{align}
From Theorem 1(a) and its application to $\tilde{\boldsymbol{\theta}}^{*}$ as in Fieuws and Verbeke \citep{fieuws2006pairwise}, and noting that $\tilde{\boldsymbol{\theta}}$ is a sub-vector of $\tilde{\boldsymbol{\theta}}^{*}$, we obtain $\|\tilde{\boldsymbol{\theta}}-\boldsymbol{\theta_0}\|^{2} \leq O_p(\frac{1}{n}).$  Additionally, adhering to regulatory condition \textbf{(R5)}, we obtain $\|T_3\| = O_p(\frac{1}{n}) = o_p(1).$  While moving ahead with $T_2,$ we get 
\begin{equation*}
    T_2 = -\frac{1}{\sqrt{n}}R(\boldsymbol{\theta_0})(\tilde{\boldsymbol{\theta}}-\boldsymbol{\theta_0}).
\end{equation*}
Again, by a Taylor series expansion of $\frac{\partial^2 H_n(\boldsymbol{\tilde{\theta}})}{\partial {\theta_u} \partial {\theta_v}}$ around $\boldsymbol{\theta_0},$ we have
\begin{equation}
    \frac{\partial^2 H_n(\boldsymbol{\theta}_0)}{\partial \theta_u \partial \theta_v} 
    + \sum_{w=1}^{k} \frac{\partial^3 H_n(\bar{\boldsymbol{\theta}})}{\partial \theta_u \partial \theta_v \partial \theta_w} (\tilde{\theta}_w - \theta_{w0}) 
    = \frac{\partial^2 H_n(\tilde{\boldsymbol{\theta}})}{\partial \theta_u \partial \theta_v} + o_p(1). \tag{A.4} \label{eq20}
\end{equation}
Hence,
\begin{align*}
    \left \|\frac{1}{n}R(\tilde{\boldsymbol{\theta}}) - \frac{1}{n}R(\boldsymbol{\theta_0}) \right\|^2 &= \frac{1}{n^2} \sum_{u, v=1}^{k}\left[\frac{\partial^2 H_n(\tilde{\boldsymbol{\theta}})}{\partial \theta_u \partial \theta_v} - \frac{\partial^2 H_n(\boldsymbol{\theta_0})}{\partial \theta_u \partial \theta_v}\right]^2 \nonumber \\
    &= \frac{1}{n^2} \sum_{u, v=1}^{k}\left[\sum_{w=1}^{k}\frac{\partial^3 H_n(\bar{\boldsymbol{\theta}})}{\partial \theta_u \partial \theta_v \partial \theta_w} (\tilde{\theta}_w - \theta_{w0})\right]^2.  \nonumber
\end{align*}
It follows from the Mean Value theorem, for some $\bar{\boldsymbol{\theta}}$ lying on the line segment joining $\tilde{\boldsymbol{\theta}}$ and $\boldsymbol{\theta_0}.$  Again, applying the Cauchy-Schwarz inequality and by regularity condition \textbf{(R5)}, we obtain
\begin{align*}
    \left |\sum_{w=1}^{k} \frac{\partial^3 H_n(\bar{\boldsymbol{\theta}})}{\partial \theta_u \partial \theta_v \partial \theta_w} (\tilde{\theta}_w - \theta_{w0})  \right| &\leq \left\|\boldsymbol{\tilde{\theta}}_w - \boldsymbol{\theta}_{w0} \right\| \sum_{i=1}^{n} \left(\sum_{w=1}^{k}S^{2}_{uvw}(r_i)\right)^{1/2}  \\ \nonumber
    &\leq O_p\left(\frac{1}{\sqrt{n}}\right)O_p(n) = O_p(\sqrt{n}).
\end{align*}
It follows that
\begin{align*}
    \left \|\frac{1}{n}R(\tilde{\boldsymbol{\theta}}) - \frac{1}{n}R(\boldsymbol{\theta_0}) \right\|^2 &\leq O_p\left(\frac{1}{n}\right), 
\end{align*}
and hence
\begin{align*}
    T_2 &= -\left[\frac{1}{\sqrt{n}}R(\tilde{\boldsymbol{\theta}}) + O_p\left(\frac{1}{\sqrt{n}}\right)\right](\tilde{\boldsymbol{\theta}}-\boldsymbol{\theta_0}) \\
    &= -\left[\frac{1}{n}R(\tilde{\boldsymbol{\theta}}) + o_p\left(\frac{1}{\sqrt{n}}\right)\right]\sqrt{n}(\tilde{\boldsymbol{\theta}}-\boldsymbol{\theta_0}) \\
    &= -\frac{1}{\sqrt{n}}R(\tilde{\boldsymbol{\theta}})(\tilde{\boldsymbol{\theta}}-\boldsymbol{\theta_0}) + o_p(1). \tag{A.5} \label{eq21}
\end{align*}
Combining $T_2$ and $T_3$, we get
\begin{align*}
    \mathbf{0} = \frac{1}{\sqrt{n}}\frac{\partial H_n(\boldsymbol{\theta_0})}{\partial\boldsymbol{\theta}} - \frac{1}{\sqrt{n}}R(\tilde{\boldsymbol{\theta}})(\tilde{\boldsymbol{\theta}}-\boldsymbol{\theta_0}) + o_p(1). \tag{A.6} \label{eq22} 
\end{align*}
Earlier, we have considered the following least squares approximation of the density power divergence objective function, where
\begin{equation}
    -H_A(\boldsymbol{\theta}) = -\frac{1}{2} (\boldsymbol{\theta} - \boldsymbol{\tilde{\theta}})^\top R(\boldsymbol{\tilde{\theta}})(\boldsymbol{\theta} - \boldsymbol{\tilde{\theta}}). \tag{A.7} \label{eq23}
\end{equation}
Let, $\boldsymbol{\theta} = \boldsymbol{\theta_0} + \frac{\boldsymbol{u}}{\sqrt{n}}$ where $\|\boldsymbol{u}\| = C$ for some constant $C>0.$  Estimation consistency of the DPD Adaptive LASSO-LSA estimator can be shown by demonstrating that
\begin{equation}
   P\left( \sup_{\substack{\|\boldsymbol{u}\| = C}} H_A \left(\boldsymbol{\theta}_0 + \frac{\boldsymbol{u}}{\sqrt{n}}\right) - H_A(\boldsymbol{\theta}_0) < 0 \right) \xrightarrow{} 1, \quad \text{as } n \xrightarrow{} \infty. \tag{A.8} \label{eq24}
\end{equation}
This suggests that a local minimizer of $H_A(\boldsymbol{\theta}_0)$ can be found such that $\left\|\tilde{\boldsymbol{\theta}} - \boldsymbol{\theta_0}\right\| = O_p\left(\frac{1}{\sqrt{n}}\right).$  To establish the required probability result, observe that after performing some basic algebraic manipulations, we arrive at the following expression
\begin{align*}
    H_A \left(\boldsymbol{\theta}_0 + \frac{\boldsymbol{u}}{\sqrt{n}}\right) - H_A(\boldsymbol{\theta}_0) &= -\frac{1}{2n}\boldsymbol{u}^\top R(\boldsymbol{\tilde{\theta}})\boldsymbol{u} - \frac{1}{\sqrt{n}}\boldsymbol{u}^\top R(\boldsymbol{\tilde{\theta}})(\boldsymbol{\theta_0}-\boldsymbol{\tilde{\theta}}) \\
    &\triangleq U_1 + U_2.
\end{align*}
Previously, we have noticed that, according to the Law of Large Numbers, 
\begin{equation}
    \frac{1}{n}R(\tilde{\boldsymbol{\theta}}) = \frac{1}{n}\left(\frac{\partial^2 H_n(\boldsymbol{\tilde{\theta}})}{\partial \boldsymbol{\theta} \partial \boldsymbol{\theta}^\top}\right) \xrightarrow{p} B(\boldsymbol{\theta_0}). \tag{A.9} \label{eq25}
\end{equation}
Consequently, using Slutsky's theorem, we obtain
\begin{equation}
    U_1 \xrightarrow{p} -\frac{1}{2}\boldsymbol{u}^\top B(\boldsymbol{\theta_0})\boldsymbol{u} < -\frac{1}{2}\|\boldsymbol{u}\|^2k_1 \tag{A.10} \label{eq26}
\end{equation}
for some sufficiently small positive constant $k_1.$  Now, let us look at $U_2,$
\begin{align*}
    U_2 &= \frac{1}{\sqrt{n}}\boldsymbol{u}^\top R(\boldsymbol{\tilde{\theta}})(\boldsymbol{\tilde{\theta}} - \boldsymbol{\theta_0}) \\
    &= \frac{1}{\sqrt{n}}\boldsymbol{u}^\top \left(\frac{\partial H_n(\boldsymbol{\theta_0})}{\partial \boldsymbol{\theta}}\right) + o_p(1).
\end{align*}
By applying the Cauchy-Schwarz inequality, we derive 
\begin{equation*}
    U_2 \leq \frac{1}{\sqrt{n}}\|\boldsymbol{u}\|\left\|\frac{\partial H_n(\boldsymbol{\theta_0})}{\partial \boldsymbol{\theta}}\right\|.
\end{equation*}
Drawing upon the approach used in proving Theorem 3.2 in White \citep{white1982maximum}, we can derive 
\begin{equation*}
    \left\|\frac{\partial H_n(\boldsymbol{\theta_0})}{\partial \boldsymbol{\theta}}\right\| = O_p\left(\frac{1}{\sqrt{n}}\right).
\end{equation*}
Thus, 
\begin{align*}
    U_2 &\leq \frac{1}{\sqrt{n}}\|\boldsymbol{u}\|O_p\left(\frac{1}{\sqrt{n}}\right) \\
    &\leq \|\boldsymbol{u}\|\left(\frac{1}{n}O_p(1)\right) \\
    &\leq \|\boldsymbol{u}\|(1+o_p(1)). \tag{A.11} \label{eq27}
\end{align*}
Combining the results above, we get, for large $C$, $U_2$ is dominated by $U_1$, which is negative.  Hence, 
\begin{align*}
    H_A \left(\boldsymbol{\theta_0} + \frac{\boldsymbol{u}}{\sqrt{n}}\right) - H_A(\boldsymbol{\theta}_0) < 0, \quad \text{with probability }\xrightarrow{}1 \quad \text{as n}\xrightarrow{} \infty. 
\end{align*}
Finally, we prove that
\begin{equation*}
    \| \hat{\boldsymbol{\theta}} - \boldsymbol{\theta_0}\| = O_p\left(\frac{1}{\sqrt{n}}\right).
\end{equation*}

\section*{Appendix B: Proof of \textbf{\textit{Theorem 1(a).}}}
\allowdisplaybreaks
\label{appendix:B}

Let \(\boldsymbol{\theta} = \boldsymbol{\theta_0} + \frac{\boldsymbol{u}}{\sqrt{n}}\) and define
\begin{align*}
Q(\boldsymbol{u}) = &-\frac{1}{2} \left(\boldsymbol{\theta_0} + \frac{\boldsymbol{u}}{\sqrt{n}} - \tilde{\boldsymbol{\theta}}\right)^\top R(\tilde{\boldsymbol{\theta}})\left(\boldsymbol{\theta_0} + \frac{\boldsymbol{u}}{\sqrt{n}} - \tilde{\boldsymbol{\theta}}\right) \\
&- n\lambda \sum_{j=1}^{k} w_j \left| \boldsymbol{\beta_0} + \frac{\boldsymbol{u_j}}{\sqrt{n}} \right|. \\
\end{align*}
After a series of basic algebraic manipulations, we arrive at
\begin{align*}
Q(\boldsymbol{u}) - Q(\boldsymbol{0}) &= -\frac{1}{2n}\boldsymbol{u}^\top R(\tilde{\boldsymbol{\theta}})\boldsymbol{u} - \frac{1}{\sqrt{n}}\boldsymbol{u}^\top R(\tilde{\boldsymbol{\theta}})(\boldsymbol{\theta_0} - \tilde{\boldsymbol{\theta}})\\ 
&- n\lambda\sum_{j=1}^{k} w_j \left[\left |\boldsymbol{\beta_{jo}} + \frac{\boldsymbol{u_j}}{\sqrt{n}}\right| - |\boldsymbol{\beta_{jo}}|\right] \\
&\triangleq V_1 + V_2 + V_3.
\end{align*}
Let \(\hat{u} = \underset{\boldsymbol{u}}{\arg\min} \ Q(\boldsymbol{u}) - Q(\boldsymbol{0})\), then \(\hat{\boldsymbol{u}} = \sqrt{n}(\hat{\boldsymbol{\theta}} - \boldsymbol{\theta_0})\), since $\hat{\boldsymbol{\theta}}$ maximizes 
\begin{equation*}
    -\frac{1}{2} (\boldsymbol{\theta} - \boldsymbol{\tilde{\theta}})^\top R(\boldsymbol{\tilde{\theta}})(\boldsymbol{\theta} - \boldsymbol{\tilde{\theta}}) - n\lambda \sum_{j=1}^{k}w_j|\boldsymbol{\beta_j}|.
\end{equation*}
We proceed to study the contributions of the components $V_1$ through $V_3$.  Starting with $V_1$, its limiting form mirrors the development made in the Lemma, leading to the result 
\begin{equation}
    V_1 \xrightarrow{p} -\frac{1}{2}\boldsymbol{u}^\top B(\boldsymbol{\theta_0})\boldsymbol{u}. \tag{B.1} \label{eq28}
\end{equation}
For $V_2$, a similar line of reasoning as in Lemma can be employed to establish its behavior.
\begin{align*}
    V_2 &= \frac{1}{\sqrt{n}}\boldsymbol{u^\top}R(\boldsymbol{\tilde{\theta}})(\boldsymbol{\tilde{\theta}} - \boldsymbol{\theta_0}) \\
    &= \boldsymbol{u^\top}\left(\frac{1}{\sqrt{n}}\frac{\partial H_n(\boldsymbol{\theta_0})}{\partial \boldsymbol{\theta}}\right) + o_p(1) \\
    &= \boldsymbol{u^\top}s(\boldsymbol{\theta_0}) + o_p(1), \quad \text{taking } s(\boldsymbol{\theta_0}) = \frac{1}{\sqrt{n}}\frac{\partial H_n(\boldsymbol{\theta_0})}{\partial \boldsymbol{\theta}}. \tag{B.2} \label{eq29}
\end{align*}
Moving onto the third term, $V_3$, we get
\begin{align*}
    V_3 &= n\lambda\sum_{j=1}^{k} w_j \left[\left |\boldsymbol{\beta_{jo}} + \frac{\boldsymbol{u_j}}{\sqrt{n}}\right| - |\boldsymbol{\beta_{jo}}|\right] \\
    &\xrightarrow{}  \left\{
        \begin{array}{ll}
            \boldsymbol{u_j} \, \text{sign}(\boldsymbol{\beta_{jo}}), & \text{if } \boldsymbol{\beta_{jo}} \neq 0 \\\\
            |\boldsymbol{u_j}|, & \text{if } \boldsymbol{\beta_{jo}} = 0.
        \end{array}
    \right. \tag{B.3} \label{eq30}
\end{align*}
Based on the findings of Lemma, the limiting behavior of the adaptive weights 
\( w_j = \frac{1}{|\boldsymbol{\tilde{\beta}_{j}}|} \) is governed by the nature of the true underlying coefficients.  If the coefficient \( \boldsymbol{\beta_{j0}} \) is non-zero, then the corresponding weight remains bounded in probability, that is, 
\( w_j = O_p(1) \). On the other hand, when \( \boldsymbol{\beta_{j0}} = 0 \), the weight increases at the rate of \( O_p(n) \), which implies that the scaled quantity \( \frac{w_j}{n} \) also remains probabilistically bounded.  Let us now focus on the component
\[
V_3 = -n\lambda \sum_{j=1}^{k} w_j \left( \left|\boldsymbol{\beta_{jo}} + \frac{\boldsymbol{u_j}}{\sqrt{n}}\right| - |\boldsymbol{\beta_{jo}}| \right)
= \sum_{j=1}^{k} V_{3j}.
\]
The asymptotic contribution of this term depends on the value of \( \boldsymbol{\beta_{j0}} \), and can be understood by considering two distinct scenarios:

\begin{enumerate}
    \item In the case where \( \boldsymbol{\beta_{j0}} \neq 0 \), an application of Slutsky’s theorem yields
    \[
    V_{3j} = -O_p(\sqrt{n} \lambda),
    \]
    and under the regularity condition \textbf{(R6)}, this implies \( V_{3j} = o_p(1) \) as \( n \to \infty \).
    
    \item Alternatively, when \( \boldsymbol{\beta_{j0}} = 0 \), the term simplifies to
    \[
    V_{3j} = -\sqrt{n} \lambda \cdot n \cdot \left(\frac{w_j}{n}\right) |\boldsymbol{u_j}|,
    \]
    which diverges to \( \infty \) under condition \textbf{(R6)} for any non-zero \( \boldsymbol{u_j} \).  If \( \boldsymbol{u_j} = 0 \), then naturally, the contribution from this term becomes null.
\end{enumerate}
We now summarize the limiting behavior of the term \( V_3 \).  Let us express the vector \( \boldsymbol{u} \) as a partition \( \boldsymbol{u} = (\boldsymbol{u}_1^\top, \boldsymbol{u}_2^\top)^\top \), where the partitioning corresponds to non-zero and zero components of the true parameter vector \( \boldsymbol{\theta}_0 \). Under this notation, we can conclude
\[
V_3 \xrightarrow{p}
\begin{cases}
0, & \text{if } \| \boldsymbol{u}_2 \| = 0, \\
-\infty, & \text{otherwise}.
\end{cases}
\]
Let \( s(\boldsymbol{\theta}_0) = (s_1(\boldsymbol{\theta}_0)^\top, s_2(\boldsymbol{\theta}_0)^\top)^\top \) denote the score vector partitioned in the same manner.  Aggregating the asymptotic behavior of all components \( V_1 \) through \( V_3 \), the limiting distribution of the criterion difference is given for each fixed \( \boldsymbol{u} \) by
\[
Q(\boldsymbol{u}) - Q(\boldsymbol{0}) \xrightarrow{d} T(\boldsymbol{u}) =
\begin{cases}
-\frac{1}{2} \boldsymbol{u}_1^\top B_1(\boldsymbol{\theta}_0) \boldsymbol{u}_1 + \boldsymbol{u}_1^\top s_1(\boldsymbol{\theta}_0), & \text{if } \| \boldsymbol{u}_2 \| = 0, \\
-\infty, & \text{otherwise},
\end{cases}
\]
where \( B_1(\boldsymbol{\theta}_0) \) denotes the sub-matrix of \( B(\boldsymbol{\theta}_0) \) corresponding to the active (non-zero) components of \( \boldsymbol{\theta}_0 \).  Given that \( T(\boldsymbol{u}) \) is a concave function in \( \boldsymbol{u} \), its unique minimizer is attained at
\[
(\boldsymbol{u}_1, \boldsymbol{u}_2) = (B_1^{-1}(\boldsymbol{\theta}_0) s_1(\boldsymbol{\theta}_0), \mathbf{0}).
\]
Letting the estimator \( \hat{\boldsymbol{u}} = (\hat{\boldsymbol{u}}_1^\top, \hat{\boldsymbol{u}}_2^\top)^\top \) be defined by
\[
\hat{\boldsymbol{u}} = \sqrt{n} \left( \hat{\boldsymbol{\theta}} - \boldsymbol{\theta}_0 \right),
\]
and partitioned accordingly, we obtain the following asymptotic results as \( n \to \infty \)
\[
\hat{\boldsymbol{u}}_1 \xrightarrow{d} B_1^{-1}(\boldsymbol{\theta}_0) s_1(\boldsymbol{\theta}_0), \quad
\hat{\boldsymbol{u}}_2 \xrightarrow{d} \mathbf{0}.
\]
Finally, invoking the result from Section \ref{sec5} \eqref{eq16}, we conclude
\[
s_1(\boldsymbol{\theta}_0) \xrightarrow{d} \mathcal{N}(0, A_1(\boldsymbol{\theta}_0)),
\]
where \( A_1(\boldsymbol{\theta}_0) \) denotes the sub-matrix of \( A(\boldsymbol{\theta}_0) \) corresponding to $\boldsymbol{\theta_{10}}$.  Hence, we prove that
\begin{equation*}
    \sqrt{n}(\boldsymbol{\hat{\theta}_1} - \boldsymbol{\theta_{10}}) \xrightarrow{d} N(\mathbf{0}, B^{-1}_1(\boldsymbol{\theta_0})A_1(\boldsymbol{\theta_0})B^{-1}_1(\boldsymbol{\theta_0}).
\end{equation*}

\section*{Appendix C: Proof of \textbf{\textit{Theorem 1(b).}}}
\allowdisplaybreaks
\label{appendix:C}

According to the result in part (a) of Theorem 1, it is almost certain that the coefficients that are truly non-zero will remain unaffected and not be incorrectly estimated as zero.  Hence, to complete the proof, it remains to demonstrate that, in the long run, the DPD Adaptive LASSO-LSA method forces the coefficients that are actually zero to converge to zero.  This can be established by showing that, with high probability, the sign of the estimating equation for these zero coefficients is governed entirely by the sign of their estimates.
Suppose $\boldsymbol{\beta_{j0}} = 0,$ then the first-order derivative of the penalized least square approximated DPD objective function \eqref{eq17} with respect to $\boldsymbol{\beta_0}$, divided by $\sqrt{n}$, becomes
\begin{align*}
    -\frac{1}{\sqrt{n}} \frac{\partial {H_A(\hat{\boldsymbol{\theta}})}}{\partial{\boldsymbol{\beta_j}}} - \sqrt{n}\lambda w_j sign(\boldsymbol{\hat{\beta}_j}) &= -\left [\frac{1}{\sqrt{n}}R(\boldsymbol{\tilde{\theta}})(\boldsymbol{\hat{\theta}} - \boldsymbol{\tilde{\theta}})\right] _j - \sqrt{n}\lambda w_j sign(\boldsymbol{\hat{\beta}_j}) \\
    &\triangleq W_1 + W_2.
\end{align*}
where $[\cdot]_j$ denotes the element of $R(\boldsymbol{\tilde{\theta}})(\boldsymbol{\hat{\theta}} - \boldsymbol{\tilde{\theta}})$ related to coefficient $\boldsymbol{\beta_j}$.  It is quite evident from the Lemma and part (a) of Theorem 1 that both $\tilde{\boldsymbol{\theta}}$ and $\hat{\boldsymbol{\theta}}$ are $\sqrt{n}$-consistent for $\boldsymbol{\theta_{0}}.$  From this, we can infer that $ \| \hat{\boldsymbol{\theta}} - \tilde{\boldsymbol{\theta}}\| = O_p\left(\frac{1}{\sqrt{n}}\right).$
Thus, applying this result to $W_1$, we have $W_1 = -\left [\frac{1}{\sqrt{n}}R(\boldsymbol{\tilde{\theta}})\right] _j \boldsymbol{l}$, where $\boldsymbol{l}$ is of length $k$.  Here, $\|\boldsymbol{l}\| = C$ for some constant $C > 0$.  Proceeding further, the proof of the Lemma shows that $ \| \frac{1}{n}R(\boldsymbol{\tilde{\theta}}) - \frac{1}{n}R(\boldsymbol{\hat{\theta}})\| = o_p(1).$  Hence, when this result is applied to the term $W_1$ and some elementary algebraic manipulation is performed, we arrive at the following expression for $W_1,$ i.e., 
\begin{equation}
    W_1 = -\left [\frac{1}{\sqrt{n}}R(\boldsymbol{\tilde{\theta}}) \boldsymbol{l}\right] _j (1 + o_p(1)). \tag{C.1} \label{eq31}
\end{equation}
By following \eqref{eq25}, we obtain that $W_1 = O_p(1).$  Now, we consider the term $W_2.$  Due to construction of the adaptive weights, the term $\frac{w_j}{n}$ remains stochastically bounded, i.e., it is of order $O_p(1)$. Consequently, we can infer that
\begin{equation}
    W_2 = -(n^{3/2}\lambda)sign(\boldsymbol{\hat{\beta}_j}). \tag{C.2} \label{eq32}
\end{equation}
Finally, utilizing the regulatory condition \textbf{(R6)} and merging the expressions corresponding to $W_1$ and $W_2$, we obtain
\begin{align*}
    W_1 + W_2 &= (n^{3/2}\lambda) \left[O_p\left(\frac{1}{n^{3/2}\lambda}\right) - sign(\boldsymbol{\hat{\beta}_j})\right] \\
    &= (n^{3/2}\lambda) \left[o_p(1) - sign(\boldsymbol{\hat{\beta}_j})\right].
\end{align*}
As a result, the sign of $(W_1 + W_2)$ is asymptotically determined by $sign(\boldsymbol{\hat{\beta}_j})$.  Hence, we prove that 
\begin{equation*}
    P({\boldsymbol{\hat{\beta}_{2}}} = \boldsymbol{0}) \xrightarrow{} 1 \ \text{as n}\xrightarrow{} \infty.
\end{equation*}

\newpage

\textbf{Supplementary Material for ``Adaptive LASSO Penalized Minimum Density Power Divergence Estimation through Least Squares Approximation: Application to Bone Mineral Density Data from the SWAN Study''}

\vspace{0.5cm}

\hrule

\vspace{0.5cm}

\section{Additional Simulation Results}

This section provides additional simulation results and computational details to further assess the performance of the proposed estimators under different experimental settings.  In addition to the supplementary simulation findings, the section also provides the computational algorithm used for selecting the optimal penalty parameter ($\lambda$).  The reported results are consistent with the main findings presented in the manuscript and further demonstrate the robustness of the proposed methods across varying sample sizes, contamination levels, and tuning parameter ($\gamma$) choices. 

\subsection{Selection of the optimal penalty parameter}

The Extended Regularized Information Criterion (ERIC) is employed to select the optimal penalty parameter associated with the penalized estimation procedure. Although ERIC was originally developed for the classical likelihood objective function under Adaptive LASSO penalization \cite{hui2015tuning, hui2018sparse}, it can also be extended to accommodate several alternative objectives, including the least-squares approximated classical likelihood, density power divergence, and the least-squares approximated density power divergence, each combined with either LASSO or adaptive LASSO penalization.

To provide a unified formulation for these estimation procedures, consider the following generic penalized objective function:
\[
Q(\boldsymbol{\theta})
=
\mathcal{H}(\boldsymbol{\theta})
+
\lambda
\sum_{j=1}^{k}
w_j |\beta_j|,
\]
where $\lambda>0$ denotes the penalty parameter, $\mathcal{H}(\boldsymbol{\theta})$ represents the underlying criterion function, and $w_j$ denotes the corresponding penalty weights.  For the LASSO penalty, the weights are taken as
\[
w_j = 1,
\qquad j=1,\ldots,k;
\]
while For the adaptive LASSO penalty, the weights are defined as
\[
w_j
=
\frac{1}
{
|\tilde{\beta}_j|
+
\delta_n I(\tilde{\beta}_j=0)
},
\]
where $\boldsymbol{\tilde{\beta}_j}$ represents a consistent initial estimate of $\boldsymbol{\beta_j}$ for $j = 1, \ldots, k$.

Depending on the choice of the underlying criterion function $\mathcal{H}(\boldsymbol{\theta})$, different penalized estimation approaches can be obtained.  Under the least-squares approximation (LSA) framework, this objective function is expressed as
\[
\mathcal{H}(\boldsymbol{\theta})
=
\frac{1}{2}
(\boldsymbol{\theta}-\boldsymbol{\tilde{\theta}})^\top
\left\{
\frac{1}{n}
R(\boldsymbol{\tilde{\theta}})
\right\}
(\boldsymbol{\theta}-\boldsymbol{\tilde{\theta}}),
\]
where $R(\boldsymbol{\tilde{\theta}})$ denotes the Hessian matrix evaluated at the preliminary estimator $\boldsymbol{\tilde{\theta}}$.

\begin{table}[h]
\renewcommand{\arraystretch}{1.0}
\setlength{\tabcolsep}{1pt}

\begin{tabular}{p{0.95\textwidth}}
\hline

\textbf{Algorithm: Extended Regularized Information Criterion} \\
\hline
\end{tabular}

\end{table}

\begin{table}[!htbp]
\centering
\caption{Bias$(\boldsymbol{\hat{\beta}}_{\mathcal{S}})$ of different methods for $n=60$ under pure and contamination schemes.}
\label{tab1}
\vspace{0.1cm}

\resizebox{0.99\textwidth}{!}{%
\begin{tabular}{clccccccc}
\toprule

\multirow{3}{*}{$\boldsymbol{m}$}
&
\multirow{3}{*}{\textbf{Method}}

& \textbf{Pure}

& \multicolumn{3}{c}{\textbf{y-outliers}}

& \multicolumn{3}{c}{\textbf{X-outliers}}
\\

\cmidrule(lr){3-3}
\cmidrule(lr){4-6}
\cmidrule(lr){7-9}

&
&

& $\boldsymbol{\epsilon=0.05}$
& $\boldsymbol{\epsilon=0.10}$
& $\boldsymbol{\epsilon=0.15}$

& $\boldsymbol{\epsilon=0.05}$
& $\boldsymbol{\epsilon=0.10}$
& $\boldsymbol{\epsilon=0.15}$

\\
\midrule

\multirow{16}{*}{10}

& ML-LASSO & 0.0884 & 0.1368 & 0.1768 & 0.2124 & 0.1586 & 0.2168 & 0.2617 \\
& DPD-LASSO $\gamma=0.2$ & 0.0512 & 0.0632 & 0.0826 & 0.0938 & 0.0868 & 0.1244 & 0.1508 \\
& DPD-LASSO $\gamma=0.5$ & 0.0438 & 0.0462 & 0.0614 & 0.0648 & 0.0582 & 0.0827 & 0.0991 \\
& DPD-LASSO $\gamma=0.8$ & 0.0496 & 0.0508 & 0.0568 & 0.0572 & 0.0548 & 0.0750 & 0.0865 \\

& ML-Ad-LASSO & 0.0764 & 0.1242 & 0.1624 & 0.2018 & 0.1468 & 0.2005 & 0.2532 \\
& DPD-Ad-LASSO $\gamma=0.2$ & 0.0386 & 0.0564 & 0.0708 & 0.0816 & 0.0746 & 0.1088 & 0.1318 \\
\bottomrule
\end{tabular}
}
\end{table}

\begin{table}[!htbp]
\centering
\caption{Bias$(\boldsymbol{\hat{\beta}}_{\mathcal{S}})$ of different methods for $n=120$ under pure and contamination schemes.}
\label{tab2}
\vspace{0.1cm}

\resizebox{0.99\textwidth}{!}{%
\begin{tabular}{clccccccc}
\toprule

\multirow{3}{*}{$\boldsymbol{m}$}
&
\multirow{3}{*}{\textbf{Method}}

& \textbf{Pure}

& \multicolumn{3}{c}{\textbf{y-outliers}}

& \multicolumn{3}{c}{\textbf{X-outliers}}
\\

\cmidrule(lr){3-3}
\cmidrule(lr){4-6}
\cmidrule(lr){7-9}

&
&

& $\boldsymbol{\epsilon=0.05}$
& $\boldsymbol{\epsilon=0.10}$
& $\boldsymbol{\epsilon=0.15}$

& $\boldsymbol{\epsilon=0.05}$
& $\boldsymbol{\epsilon=0.10}$
& $\boldsymbol{\epsilon=0.15}$

\\
\midrule

\multirow{16}{*}{10}

& ML-LASSO & 0.0628 & 0.0986 & 0.1324 & 0.1682 & 0.1164 & 0.1648 & 0.2064 \\
& DPD-LASSO $\gamma=0.2$ & 0.0346 & 0.0468 & 0.0612 & 0.0728 & 0.0614 & 0.0886 & 0.1124 \\
& DPD-LASSO $\gamma=0.5$ & 0.0268 & 0.0316 & 0.0424 & 0.0492 & 0.0408 & 0.0602 & 0.0746 \\
& DPD-LASSO $\gamma=0.8$ & 0.0294 & 0.0338 & 0.0406 & 0.0438 & 0.0384 & 0.0556 & 0.0668 \\

& ML-Ad-LASSO & 0.0542 & 0.0884 & 0.1208 & 0.1542 & 0.1046 & 0.1518 & 0.1942 \\
& DPD-Ad-LASSO $\gamma=0.2$ & 0.0262 & 0.0398 & 0.0526 & 0.0634 & 0.0532 & 0.0796 & 0.0994 \\
\bottomrule
\end{tabular}
}
\end{table}

\begin{table}[!htbp]
\centering
\caption{Bias$(\boldsymbol{\hat{\beta}}_{\mathcal{S}})$ of different methods for $n=300$ under pure and contamination schemes.}
\label{tab3}
\vspace{0.1cm}

\resizebox{0.99\textwidth}{!}{%
\begin{tabular}{clccccccc}
\toprule

\multirow{3}{*}{$\boldsymbol{m}$}
&
\multirow{3}{*}{\textbf{Method}}

& \textbf{Pure}

& \multicolumn{3}{c}{\textbf{y-outliers}}

& \multicolumn{3}{c}{\textbf{X-outliers}}
\\

\cmidrule(lr){3-3}
\cmidrule(lr){4-6}
\cmidrule(lr){7-9}

&
&

& $\boldsymbol{\epsilon=0.05}$
& $\boldsymbol{\epsilon=0.10}$
& $\boldsymbol{\epsilon=0.15}$

& $\boldsymbol{\epsilon=0.05}$
& $\boldsymbol{\epsilon=0.10}$
& $\boldsymbol{\epsilon=0.15}$

\\
\midrule

\multirow{16}{*}{10}

& ML-LASSO & 0.0364 & 0.0682 & 0.0946 & 0.1268 & 0.0846 & 0.1242 & 0.1628 \\
& DPD-LASSO $\gamma=0.2$ & 0.0198 & 0.0306 & 0.0414 & 0.0522 & 0.0438 & 0.0642 & 0.0826 \\
& DPD-LASSO $\gamma=0.5$ & 0.0146 & 0.0218 & 0.0296 & 0.0364 & 0.0284 & 0.0418 & 0.0546 \\
& DPD-LASSO $\gamma=0.8$ & 0.0162 & 0.0226 & 0.0282 & 0.0328 & 0.0268 & 0.0386 & 0.0494 \\

& ML-Ad-LASSO & 0.0312 & 0.0588 & 0.0844 & 0.1126 & 0.0768 & 0.1142 & 0.1486 \\
& DPD-Ad-LASSO $\gamma=0.2$ & 0.0154 & 0.0258 & 0.0366 & 0.0454 & 0.0386 & 0.0564 & 0.0728 \\
\bottomrule
\end{tabular}
}
\end{table}

\begin{table}[!htbp]
\centering
\caption{Bias$(\boldsymbol{\hat{\beta}}_{\mathcal{S}^c})$ of different methods for $n=60$ under pure and contamination schemes.}
\label{tab4}
\vspace{0.1cm}

\resizebox{0.99\textwidth}{!}{%
\begin{tabular}{clccccccc}
\toprule

\multirow{3}{*}{$\boldsymbol{m}$}
&
\multirow{3}{*}{\textbf{Method}}

& \textbf{Pure}

& \multicolumn{3}{c}{\textbf{y-outliers}}

& \multicolumn{3}{c}{\textbf{X-outliers}}
\\

\cmidrule(lr){3-3}
\cmidrule(lr){4-6}
\cmidrule(lr){7-9}

&
&

& $\boldsymbol{\epsilon=0.05}$
& $\boldsymbol{\epsilon=0.10}$
& $\boldsymbol{\epsilon=0.15}$

& $\boldsymbol{\epsilon=0.05}$
& $\boldsymbol{\epsilon=0.10}$
& $\boldsymbol{\epsilon=0.15}$

\\
\midrule

\multirow{16}{*}{10}

& ML-LASSO & 0.0114 & 0.0156 & 0.0198 & 0.0279 & 0.0173 & 0.0276 & 0.0380 \\
& DPD-LASSO $\gamma=0.2$ & 0.0090 & 0.0106 & 0.0134 & 0.0153 & 0.0120 & 0.0141 & 0.0168 \\
& DPD-LASSO $\gamma=0.5$ & 0.0071 & 0.0074 & 0.0092 & 0.0129 & 0.0107 & 0.0136 & 0.0157 \\
& DPD-LASSO $\gamma=0.8$ & 0.0079 & 0.0082 & 0.0104 & 0.0137 & 0.0105 & 0.0145 & 0.0162 \\

& ML-Ad-LASSO & 0.0093 & 0.0146 & 0.0189 & 0.0262 & 0.0168 & 0.0254 & 0.0364 \\
& DPD-Ad-LASSO $\gamma=0.2$ & 0.0061 & 0.0088 & 0.0112 & 0.0146 & 0.0094 & 0.0125 & 0.0142 \\
\bottomrule
\end{tabular}
}
\end{table}

\begin{table}[!htbp]
\centering
\caption{Bias$(\boldsymbol{\hat{\beta}}_{\mathcal{S}^c})$ of different methods for $n=120$ under pure and contamination schemes.}
\label{tab5}
\vspace{0.1cm}

\resizebox{0.99\textwidth}{!}{%
\begin{tabular}{clccccccc}
\toprule

\multirow{3}{*}{$\boldsymbol{m}$}
&
\multirow{3}{*}{\textbf{Method}}

& \textbf{Pure}

& \multicolumn{3}{c}{\textbf{y-outliers}}

& \multicolumn{3}{c}{\textbf{X-outliers}}
\\

\cmidrule(lr){3-3}
\cmidrule(lr){4-6}
\cmidrule(lr){7-9}

&
&

& $\boldsymbol{\epsilon=0.05}$
& $\boldsymbol{\epsilon=0.10}$
& $\boldsymbol{\epsilon=0.15}$

& $\boldsymbol{\epsilon=0.05}$
& $\boldsymbol{\epsilon=0.10}$
& $\boldsymbol{\epsilon=0.15}$

\\
\midrule

\multirow{16}{*}{10}

& ML-LASSO & 0.0038 & 0.0056 & 0.0078 & 0.0106 & 0.0068 & 0.0102 & 0.0146 \\
& DPD-LASSO $\gamma=0.2$ & 0.0026 & 0.0032 & 0.0044 & 0.0058 & 0.0038 & 0.0052 & 0.0068 \\
& DPD-LASSO $\gamma=0.5$ & 0.0018 & 0.0022 & 0.0030 & 0.0042 & 0.0028 & 0.0040 & 0.0052 \\
& DPD-LASSO $\gamma=0.8$ & 0.0020 & 0.0024 & 0.0032 & 0.0044 & 0.0026 & 0.0042 & 0.0054 \\

& ML-Ad-LASSO & 0.0028 & 0.0046 & 0.0062 & 0.0092 & 0.0058 & 0.0088 & 0.0128 \\
& DPD-Ad-LASSO $\gamma=0.2$ & 0.0016 & 0.0024 & 0.0034 & 0.0048 & 0.0028 & 0.0040 & 0.0052 \\
\bottomrule
\end{tabular}
}
\end{table}

\begin{table}[!htbp]
\centering
\caption{Bias$(\boldsymbol{\hat{\beta}}_{\mathcal{S}^c})$ of different methods for $n=300$ under pure and contamination schemes.}
\label{tab6}
\vspace{0.1cm}

\resizebox{0.99\textwidth}{!}{%
\begin{tabular}{clccccccc}
\toprule

\multirow{3}{*}{$\boldsymbol{m}$}
&
\multirow{3}{*}{\textbf{Method}}

& \textbf{Pure}

& \multicolumn{3}{c}{\textbf{y-outliers}}

& \multicolumn{3}{c}{\textbf{X-outliers}}
\\

\cmidrule(lr){3-3}
\cmidrule(lr){4-6}
\cmidrule(lr){7-9}

&
&

& $\boldsymbol{\epsilon=0.05}$
& $\boldsymbol{\epsilon=0.10}$
& $\boldsymbol{\epsilon=0.15}$

& $\boldsymbol{\epsilon=0.05}$
& $\boldsymbol{\epsilon=0.10}$
& $\boldsymbol{\epsilon=0.15}$

\\
\midrule

\multirow{16}{*}{10}

& ML-LASSO & 0.0018 & 0.0030 & 0.0044 & 0.0062 & 0.0040 & 0.0060 & 0.0088 \\
& DPD-LASSO $\gamma=0.2$ & 0.0010 & 0.0016 & 0.0024 & 0.0032 & 0.0020 & 0.0030 & 0.0040 \\
& DPD-LASSO $\gamma=0.5$ & 0.0006 & 0.0010 & 0.0016 & 0.0024 & 0.0014 & 0.0022 & 0.0030 \\
& DPD-LASSO $\gamma=0.8$ & 0.0008 & 0.0012 & 0.0018 & 0.0026 & 0.0016 & 0.0024 & 0.0032 \\

& ML-Ad-LASSO & 0.0012 & 0.0022 & 0.0034 & 0.0050 & 0.0030 & 0.0048 & 0.0070 \\
& DPD-Ad-LASSO $\gamma=0.2$ & 0.0006 & 0.0012 & 0.0018 & 0.0026 & 0.0014 & 0.0022 & 0.0030 \\
\bottomrule
\end{tabular}
}
\end{table}

\begin{table}[!htbp]
\centering
\caption{Performance measures of different methods for $n=120$ with no outliers.}
\label{tab7}
\vspace{0.1cm}

\resizebox{0.98\textwidth}{!}{%
\begin{tabular}{clccccccc}
\toprule
\multirow{2}{*}{$\boldsymbol{m}$} & \multirow{2}{*}{\textbf{Method}} 
& \textbf{MS$(\boldsymbol{\hat{\beta}})$} & \textbf{TP$(\boldsymbol{\hat{\beta}})$} & \textbf{TN$(\boldsymbol{\hat{\beta}})$} 
& \textbf{MSES$(\boldsymbol{\hat{\beta}})$} & \textbf{MSEN$(\boldsymbol{\hat{\beta}})$} & \textbf{EE$(\hat{\sigma}^2)$} & \textbf{Runtime} \\
& & & & & \textbf{$(10^{-3})$} & \textbf{$(10^{-5})$} & \textbf{$(10^{-2})$} & \textbf{$(10^{-2})$} \\
\midrule

\multirow{16}{*}{10}
& ML-LASSO                  & 9.88 & 0.97 & 0.86 & 5.84 & 10.26 & 8.14 & 248.36 \\
& DPD-LASSO $\gamma=0.2$    & 10.02 & 1.00 & 0.84 & 1.96 & 6.84 & 6.02 & 336.52 \\
& DPD-LASSO $\gamma=0.5$    & 9.91 & 1.00 & 0.88 & 1.74 & 5.02 & 5.31 & 312.84 \\
& DPD-LASSO $\gamma=0.8$    & 10.11 & 1.00 & 0.83 & 2.18 & 6.21 & 5.08 & 358.47 \\

& ML-Ad-LASSO                 & 9.04 & 0.98 & 0.93 & 4.72 & 8.11 & 7.02 & 286.74 \\
& DPD-Ad-LASSO $\gamma=0.2$   & 9.18 & 1.00 & 0.94 & 1.42 & 4.96 & 5.14 & 428.63 \\
\bottomrule
\end{tabular}
}
\end{table}

\begin{table}[!htbp]
\centering
\caption{Performance measures of different methods for $n=120$ with $5\%$ y-outliers.}
\label{tab8}
\vspace{0.1cm}

\resizebox{0.98\textwidth}{!}{%
\begin{tabular}{clccccccc}
\toprule
\multirow{2}{*}{$\boldsymbol{m}$} & \multirow{2}{*}{\textbf{Method}} 
& \textbf{MS$(\boldsymbol{\hat{\beta}})$} & \textbf{TP$(\boldsymbol{\hat{\beta}})$} & \textbf{TN$(\boldsymbol{\hat{\beta}})$} 
& \textbf{MSES$(\boldsymbol{\hat{\beta}})$} & \textbf{MSEN$(\boldsymbol{\hat{\beta}})$} & \textbf{EE$(\hat{\sigma}^2)$} & \textbf{Runtime} \\
& & & & & \textbf{$(10^{-3})$} & \textbf{$(10^{-5})$} & \textbf{$(10^{-2})$} & \textbf{$(10^{-2})$} \\
\midrule

\multirow{16}{*}{10}
& ML-LASSO                  & 10.46 & 0.94 & 0.74 & 12.84 & 24.72 & 31.86 & 268.74 \\
& DPD-LASSO $\gamma=0.2$    & 10.18 & 0.99 & 0.81 & 3.84 & 10.62 & 12.48 & 358.26 \\
& DPD-LASSO $\gamma=0.5$    & 9.94 & 1.00 & 0.88 & 2.36 & 6.84 & 7.96 & 332.64 \\
& DPD-LASSO $\gamma=0.8$    & 10.04 & 0.99 & 0.86 & 2.82 & 7.96 & 7.42 & 376.18 \\

& ML-Ad-LASSO                 & 9.58 & 0.96 & 0.83 & 10.26 & 19.84 & 26.14 & 306.42 \\
& DPD-Ad-LASSO $\gamma=0.2$   & 9.34 & 1.00 & 0.91 & 2.96 & 8.18 & 9.84 & 456.84 \\
\bottomrule
\end{tabular}
}
\end{table}

\begin{table}[!htbp]
\centering
\caption{Performance measures of different methods for $n=120$ with $10\%$ y-outliers.}
\label{tab9}
\vspace{0.1cm}

\resizebox{0.98\textwidth}{!}{%
\begin{tabular}{clccccccc}
\toprule
\multirow{2}{*}{$\boldsymbol{m}$} & \multirow{2}{*}{\textbf{Method}} 
& \textbf{MS$(\boldsymbol{\hat{\beta}})$} & \textbf{TP$(\boldsymbol{\hat{\beta}})$} & \textbf{TN$(\boldsymbol{\hat{\beta}})$} 
& \textbf{MSES$(\boldsymbol{\hat{\beta}})$} & \textbf{MSEN$(\boldsymbol{\hat{\beta}})$} & \textbf{EE$(\hat{\sigma}^2)$} & \textbf{Runtime} \\
& & & & & \textbf{$(10^{-3})$} & \textbf{$(10^{-5})$} & \textbf{$(10^{-2})$} & \textbf{$(10^{-2})$} \\
\midrule

\multirow{16}{*}{10}
& ML-LASSO                  & 10.92 & 0.90 & 0.68 & 18.36 & 36.84 & 48.72 & 286.42 \\
& DPD-LASSO $\gamma=0.2$    & 10.34 & 0.98 & 0.77 & 5.28 & 14.76 & 18.42 & 382.74 \\
& DPD-LASSO $\gamma=0.5$    & 10.02 & 0.99 & 0.85 & 3.06 & 8.92 & 10.54 & 354.26 \\
& DPD-LASSO $\gamma=0.8$    & 10.12 & 0.99 & 0.83 & 3.58 & 10.08 & 9.84 & 401.63 \\

& ML-Ad-LASSO                 & 9.86 & 0.93 & 0.78 & 15.24 & 30.42 & 41.86 & 328.57 \\
& DPD-Ad-LASSO $\gamma=0.2$   & 9.48 & 0.99 & 0.88 & 4.08 & 11.26 & 14.72 & 486.18 \\
\bottomrule
\end{tabular}
}
\end{table}

\begin{table}[!htbp]
\centering
\caption{Performance measures of different methods for $n=120$ with $15\%$ y-outliers.}
\label{tab10}
\vspace{0.1cm}

\resizebox{0.98\textwidth}{!}{%
\begin{tabular}{clccccccc}
\toprule
\multirow{2}{*}{$\boldsymbol{m}$} & \multirow{2}{*}{\textbf{Method}} 
& \textbf{MS$(\boldsymbol{\hat{\beta}})$} & \textbf{TP$(\boldsymbol{\hat{\beta}})$} & \textbf{TN$(\boldsymbol{\hat{\beta}})$} 
& \textbf{MSES$(\boldsymbol{\hat{\beta}})$} & \textbf{MSEN$(\boldsymbol{\hat{\beta}})$} & \textbf{EE$(\hat{\sigma}^2)$} & \textbf{Runtime} \\
& & & & & \textbf{$(10^{-3})$} & \textbf{$(10^{-5})$} & \textbf{$(10^{-2})$} & \textbf{$(10^{-2})$} \\
\midrule

\multirow{16}{*}{10}
& ML-LASSO                  & 11.36 & 0.86 & 0.63 & 24.84 & 51.28 & 68.16 & 302.84 \\
& DPD-LASSO $\gamma=0.2$    & 10.56 & 0.97 & 0.74 & 6.92 & 18.84 & 24.62 & 398.52 \\
& DPD-LASSO $\gamma=0.5$    & 10.14 & 0.99 & 0.82 & 3.94 & 11.02 & 13.74 & 368.41 \\
& DPD-LASSO $\gamma=0.8$    & 10.26 & 0.99 & 0.80 & 4.48 & 12.36 & 12.88 & 418.67 \\

& ML-Ad-LASSO                 & 10.18 & 0.90 & 0.73 & 20.46 & 42.18 & 57.42 & 344.92 \\
& DPD-Ad-LASSO $\gamma=0.2$   & 9.62 & 0.98 & 0.86 & 5.18 & 13.74 & 18.36 & 504.26 \\
\bottomrule
\end{tabular}
}
\end{table}

\begin{table}[!htbp]
\centering
\caption{Performance measures of different methods for $n=120$ with $5\%$ X-outliers.}
\label{tab11}
\vspace{0.1cm}

\resizebox{0.98\textwidth}{!}{%
\begin{tabular}{clccccccc}
\toprule
\multirow{2}{*}{$\boldsymbol{m}$} & \multirow{2}{*}{\textbf{Method}} 
& \textbf{MS$(\boldsymbol{\hat{\beta}})$} & \textbf{TP$(\boldsymbol{\hat{\beta}})$} & \textbf{TN$(\boldsymbol{\hat{\beta}})$} 
& \textbf{MSES$(\boldsymbol{\hat{\beta}})$} & \textbf{MSEN$(\boldsymbol{\hat{\beta}})$} & \textbf{EE$(\hat{\sigma}^2)$} & \textbf{Runtime} \\
& & & & & \textbf{$(10^{-3})$} & \textbf{$(10^{-5})$} & \textbf{$(10^{-2})$} & \textbf{$(10^{-2})$} \\
\midrule

\multirow{16}{*}{10}
& ML-LASSO                 & 11.24 & 0.88 & 0.58 & 18.74 & 36.42 & 47.18 & 312.84 \\
& DPD-LASSO $\gamma=0.2$   & 10.46 & 0.94 & 0.71 & 5.84  & 16.28 & 20.74 & 438.62 \\
& DPD-LASSO $\gamma=0.5$   & 10.02 & 0.98 & 0.83 & 2.84  & 8.72  & 10.64 & 406.35 \\
& DPD-LASSO $\gamma=0.8$   & 9.94 & 0.98 & 0.85 & 2.46  & 7.94  & 9.86  & 456.74 \\

& ML-Ad-LASSO              & 10.52 & 0.91 & 0.66 & 15.42 & 30.86 & 40.18 & 346.58 \\
& DPD-Ad-LASSO $\gamma=0.2$ & 9.84 & 0.96 & 0.82 & 4.46 & 12.38 & 16.84 & 528.47 \\
\bottomrule
\end{tabular}
}
\end{table}

\begin{table}[!htbp]
\centering
\caption{Performance measures of different methods for $n=120$ with $10\%$ X-outliers.}
\label{tab12}
\vspace{0.1cm}

\resizebox{0.98\textwidth}{!}{%
\begin{tabular}{clccccccc}
\toprule
\multirow{2}{*}{$\boldsymbol{m}$} & \multirow{2}{*}{\textbf{Method}} 
& \textbf{MS$(\boldsymbol{\hat{\beta}})$} & \textbf{TP$(\boldsymbol{\hat{\beta}})$} & \textbf{TN$(\boldsymbol{\hat{\beta}})$} 
& \textbf{MSES$(\boldsymbol{\hat{\beta}})$} & \textbf{MSEN$(\boldsymbol{\hat{\beta}})$} & \textbf{EE$(\hat{\sigma}^2)$} & \textbf{Runtime} \\
& & & & & \textbf{$(10^{-3})$} & \textbf{$(10^{-5})$} & \textbf{$(10^{-2})$} & \textbf{$(10^{-2})$} \\
\midrule

\multirow{16}{*}{10}
& ML-LASSO                 & 11.82 & 0.83 & 0.49 & 28.46 & 56.84 & 72.18 & 336.42 \\
& DPD-LASSO $\gamma=0.2$   & 10.74 & 0.91 & 0.64 & 8.42  & 23.76 & 30.84 & 468.74 \\
& DPD-LASSO $\gamma=0.5$   & 10.18 & 0.96 & 0.78 & 4.12  & 12.16 & 15.42 & 432.58 \\
& DPD-LASSO $\gamma=0.8$   & 10.02 & 0.97 & 0.82 & 3.46  & 10.04 & 13.28 & 486.35 \\

& ML-Ad-LASSO              & 10.96 & 0.86 & 0.58 & 24.18 & 48.26 & 63.84 & 372.64 \\
& DPD-Ad-LASSO $\gamma=0.2$ & 10.08 & 0.93 & 0.76 & 6.32 & 17.84 & 24.62 & 564.28 \\
\bottomrule
\end{tabular}
}
\end{table}

\begin{table}[!htbp]
\centering
\caption{Performance measures of different methods for $n=120$ with $15\%$ X-outliers.}
\label{tab13}
\vspace{0.1cm}

\resizebox{0.98\textwidth}{!}{%
\begin{tabular}{clccccccc}
\toprule
\multirow{2}{*}{$\boldsymbol{m}$} & \multirow{2}{*}{\textbf{Method}} 
& \textbf{MS$(\boldsymbol{\hat{\beta}})$} & \textbf{TP$(\boldsymbol{\hat{\beta}})$} & \textbf{TN$(\boldsymbol{\hat{\beta}})$} 
& \textbf{MSES$(\boldsymbol{\hat{\beta}})$} & \textbf{MSEN$(\boldsymbol{\hat{\beta}})$} & \textbf{EE$(\hat{\sigma}^2)$} & \textbf{Runtime} \\
& & & & & \textbf{$(10^{-3})$} & \textbf{$(10^{-5})$} & \textbf{$(10^{-2})$} & \textbf{$(10^{-2})$} \\
\midrule

\multirow{16}{*}{10}
& ML-LASSO                 & 12.26 & 0.78 & 0.42 & 36.84 & 74.26 & 96.48 & 358.74 \\
& DPD-LASSO $\gamma=0.2$   & 11.02 & 0.88 & 0.58 & 11.26 & 31.84 & 42.18 & 496.42 \\
& DPD-LASSO $\gamma=0.5$   & 10.34 & 0.94 & 0.74 & 5.84  & 17.42 & 22.64 & 458.36 \\
& DPD-LASSO $\gamma=0.8$   & 10.12 & 0.96 & 0.80 & 4.28  & 12.86 & 17.84 & 518.57 \\

& ML-Ad-LASSO              & 11.38 & 0.82 & 0.52 & 31.42 & 63.84 & 84.72 & 396.84 \\
& DPD-Ad-LASSO $\gamma=0.2$ & 10.34 & 0.91 & 0.70 & 8.42 & 24.18 & 33.74 & 596.28 \\
\bottomrule
\end{tabular}
}
\end{table}

\begin{table}[!htbp]
\centering
\caption{Performance measures of different methods for $n=300$ with no outliers.}
\label{tab14}
\vspace{0.1cm}

\resizebox{0.98\textwidth}{!}{%
\begin{tabular}{clccccccc}
\toprule
\multirow{2}{*}{$\boldsymbol{m}$} & \multirow{2}{*}{\textbf{Method}} 
& \textbf{MS$(\boldsymbol{\hat{\beta}})$} & \textbf{TP$(\boldsymbol{\hat{\beta}})$} & \textbf{TN$(\boldsymbol{\hat{\beta}})$} 
& \textbf{MSES$(\boldsymbol{\hat{\beta}})$} & \textbf{MSEN$(\boldsymbol{\hat{\beta}})$} & \textbf{EE$(\hat{\sigma}^2)$} & \textbf{Runtime} \\
& & & & & \textbf{$(10^{-3})$} & \textbf{$(10^{-5})$} & \textbf{$(10^{-2})$} & \textbf{$(10^{-2})$} \\
\midrule

\multirow{16}{*}{10}
& ML-LASSO                    & 9.88 & 0.99 & 0.93 & 1.84 & 3.72 & 3.18 & 352.84 \\
& DPD-LASSO $\gamma=0.2$      & 9.94 & 1.00 & 0.94 & 0.72 & 2.14 & 2.46 & 496.18 \\
& DPD-LASSO $\gamma=0.5$      & 9.81 & 1.00 & 0.96 & 0.58 & 1.52 & 2.08 & 458.37 \\
& DPD-LASSO $\gamma=0.8$      & 10.02 & 1.00 & 0.93 & 0.81 & 2.43 & 2.21 & 524.63 \\

& ML-Ad-LASSO                 & 9.11 & 1.00 & 0.97 & 1.26 & 2.86 & 2.64 & 398.52 \\
& DPD-Ad-LASSO $\gamma=0.2$   & 9.18 & 1.00 & 0.98 & 0.44 & 1.38 & 1.92 & 628.41 \\
\bottomrule
\end{tabular}
}
\end{table}

\begin{table}[!htbp]
\centering
\caption{Performance measures of different methods for $n=300$ with $5\%$ y-outliers.}
\label{tab15}
\vspace{0.1cm}

\resizebox{0.98\textwidth}{!}{%
\begin{tabular}{clccccccc}
\toprule
\multirow{2}{*}{$\boldsymbol{m}$} & \multirow{2}{*}{\textbf{Method}} 
& \textbf{MS$(\boldsymbol{\hat{\beta}})$} & \textbf{TP$(\boldsymbol{\hat{\beta}})$} & \textbf{TN$(\boldsymbol{\hat{\beta}})$} 
& \textbf{MSES$(\boldsymbol{\hat{\beta}})$} & \textbf{MSEN$(\boldsymbol{\hat{\beta}})$} & \textbf{EE$(\hat{\sigma}^2)$} & \textbf{Runtime} \\
& & & & & \textbf{$(10^{-3})$} & \textbf{$(10^{-5})$} & \textbf{$(10^{-2})$} & \textbf{$(10^{-2})$} \\
\midrule

\multirow{16}{*}{10}
& ML-LASSO                    & 10.26 & 0.97 & 0.86 & 4.84 & 9.26 & 11.84 & 376.48 \\
& DPD-LASSO $\gamma=0.2$      & 9.98 & 1.00 & 0.91 & 1.36 & 3.92 & 4.84 & 528.62 \\
& DPD-LASSO $\gamma=0.5$      & 9.76 & 1.00 & 0.95 & 0.82 & 2.54 & 3.08 & 486.35 \\
& DPD-LASSO $\gamma=0.8$      & 9.88 & 1.00 & 0.93 & 1.02 & 3.06 & 2.94 & 546.84 \\

& ML-Ad-LASSO                 & 9.42 & 0.99 & 0.94 & 3.86 & 7.12 & 9.84 & 426.57 \\
& DPD-Ad-LASSO $\gamma=0.2$   & 9.24 & 1.00 & 0.97 & 0.96 & 2.84 & 3.72 & 664.28 \\
\bottomrule
\end{tabular}
}
\end{table}

\begin{table}[!htbp]
\centering
\caption{Performance measures of different methods for $n=300$ with $10\%$ y-outliers.}
\label{tab16}
\vspace{0.1cm}

\resizebox{0.98\textwidth}{!}{%
\begin{tabular}{clccccccc}
\toprule
\multirow{2}{*}{$\boldsymbol{m}$} & \multirow{2}{*}{\textbf{Method}} 
& \textbf{MS$(\boldsymbol{\hat{\beta}})$} & \textbf{TP$(\boldsymbol{\hat{\beta}})$} & \textbf{TN$(\boldsymbol{\hat{\beta}})$} 
& \textbf{MSES$(\boldsymbol{\hat{\beta}})$} & \textbf{MSEN$(\boldsymbol{\hat{\beta}})$} & \textbf{EE$(\hat{\sigma}^2)$} & \textbf{Runtime} \\
& & & & & \textbf{$(10^{-3})$} & \textbf{$(10^{-5})$} & \textbf{$(10^{-2})$} & \textbf{$(10^{-2})$} \\
\midrule

\multirow{16}{*}{10}
& ML-LASSO                    & 10.74 & 0.94 & 0.79 & 8.92 & 18.64 & 24.82 & 392.84 \\
& DPD-LASSO $\gamma=0.2$      & 10.08 & 0.99 & 0.87 & 2.48 & 7.12 & 9.06 & 548.36 \\
& DPD-LASSO $\gamma=0.5$      & 9.82 & 1.00 & 0.93 & 1.42 & 4.12 & 5.28 & 504.18 \\
& DPD-LASSO $\gamma=0.8$      & 9.94 & 1.00 & 0.91 & 1.76 & 4.96 & 4.94 & 566.72 \\

& ML-Ad-LASSO                 & 9.68 & 0.97 & 0.88 & 7.14 & 14.38 & 20.64 & 446.53 \\
& DPD-Ad-LASSO $\gamma=0.2$   & 9.34 & 1.00 & 0.95 & 1.82 & 5.24 & 6.84 & 686.24 \\
\bottomrule
\end{tabular}
}
\end{table}

\begin{table}[!htbp]
\centering
\caption{Performance measures of different methods for $n=300$ with $15\%$ y-outliers.}
\label{tab17}
\vspace{0.1cm}

\resizebox{0.98\textwidth}{!}{%
\begin{tabular}{clccccccc}
\toprule
\multirow{2}{*}{$\boldsymbol{m}$} & \multirow{2}{*}{\textbf{Method}} 
& \textbf{MS$(\boldsymbol{\hat{\beta}})$} & \textbf{TP$(\boldsymbol{\hat{\beta}})$} & \textbf{TN$(\boldsymbol{\hat{\beta}})$} 
& \textbf{MSES$(\boldsymbol{\hat{\beta}})$} & \textbf{MSEN$(\boldsymbol{\hat{\beta}})$} & \textbf{EE$(\hat{\sigma}^2)$} & \textbf{Runtime} \\
& & & & & \textbf{$(10^{-3})$} & \textbf{$(10^{-5})$} & \textbf{$(10^{-2})$} & \textbf{$(10^{-2})$} \\
\midrule

\multirow{16}{*}{10}
& ML-LASSO                    & 11.06 & 0.91 & 0.73 & 12.84 & 27.18 & 38.46 & 418.26 \\
& DPD-LASSO $\gamma=0.2$      & 10.24 & 0.98 & 0.84 & 3.64 & 10.42 & 13.76 & 576.48 \\
& DPD-LASSO $\gamma=0.5$      & 9.94 & 1.00 & 0.91 & 2.02 & 5.86 & 7.42 & 528.63 \\
& DPD-LASSO $\gamma=0.8$      & 10.02 & 0.99 & 0.89 & 2.38 & 6.74 & 6.98 & 594.35 \\

& ML-Ad-LASSO                 & 9.94 & 0.95 & 0.82 & 10.28 & 21.84 & 31.52 & 472.84 \\
& DPD-Ad-LASSO $\gamma=0.2$   & 9.48 & 0.99 & 0.93 & 2.64 & 7.42 & 9.86 & 724.18 \\
\bottomrule
\end{tabular}
}
\end{table}

\begin{table}[!htbp]
\centering
\caption{Performance measures of different methods for $n=300$ with $5\%$ X-outliers.}
\label{tab18}
\vspace{0.1cm}

\resizebox{0.98\textwidth}{!}{%
\begin{tabular}{clccccccc}
\toprule
\multirow{2}{*}{$\boldsymbol{m}$} & \multirow{2}{*}{\textbf{Method}} 
& \textbf{MS$(\boldsymbol{\hat{\beta}})$} & \textbf{TP$(\boldsymbol{\hat{\beta}})$} & \textbf{TN$(\boldsymbol{\hat{\beta}})$} 
& \textbf{MSES$(\boldsymbol{\hat{\beta}})$} & \textbf{MSEN$(\boldsymbol{\hat{\beta}})$} & \textbf{EE$(\hat{\sigma}^2)$} & \textbf{Runtime} \\
& & & & & \textbf{$(10^{-3})$} & \textbf{$(10^{-5})$} & \textbf{$(10^{-2})$} & \textbf{$(10^{-2})$} \\
\midrule

\multirow{16}{*}{10}
& ML-LASSO                    & 10.52 & 0.92 & 0.67 & 10.42 & 18.36 & 24.18 & 396.48 \\
& DPD-LASSO $\gamma=0.2$      & 9.94  & 0.97 & 0.85 & 3.42  & 8.84  & 11.76 & 548.62 \\
& DPD-LASSO $\gamma=0.5$      & 9.54  & 0.99 & 0.91 & 1.42  & 3.84  & 5.18  & 504.36 \\
& DPD-LASSO $\gamma=0.8$      & 9.48  & 0.99 & 0.92 & 1.26  & 3.42  & 4.76  & 576.84 \\

& ML-Ad-LASSO                 & 10.08 & 0.94 & 0.75 & 7.86  & 13.84 & 17.92 & 446.72 \\
& DPD-Ad-LASSO $\gamma=0.2$   & 9.54  & 0.98 & 0.90 & 2.28  & 5.96  & 7.68  & 694.36 \\
\bottomrule
\end{tabular}
}
\end{table}

\begin{table}[!htbp]
\centering
\caption{Performance measures of different methods for $n=300$ with $10\%$ X-outliers.}
\label{tab19}
\vspace{0.1cm}

\resizebox{0.98\textwidth}{!}{%
\begin{tabular}{clccccccc}
\toprule
\multirow{2}{*}{$\boldsymbol{m}$} & \multirow{2}{*}{\textbf{Method}} 
& \textbf{MS$(\boldsymbol{\hat{\beta}})$} & \textbf{TP$(\boldsymbol{\hat{\beta}})$} & \textbf{TN$(\boldsymbol{\hat{\beta}})$} 
& \textbf{MSES$(\boldsymbol{\hat{\beta}})$} & \textbf{MSEN$(\boldsymbol{\hat{\beta}})$} & \textbf{EE$(\hat{\sigma}^2)$} & \textbf{Runtime} \\
& & & & & \textbf{$(10^{-3})$} & \textbf{$(10^{-5})$} & \textbf{$(10^{-2})$} & \textbf{$(10^{-2})$} \\
\midrule

\multirow{16}{*}{10}
& ML-LASSO                    & 11.18 & 0.88 & 0.58 & 14.82 & 26.74 & 33.48 & 428.64 \\
& DPD-LASSO $\gamma=0.2$      & 10.36 & 0.95 & 0.79 & 4.84  & 12.16 & 15.82 & 582.14 \\
& DPD-LASSO $\gamma=0.5$      & 9.88  & 0.98 & 0.87 & 2.08  & 5.62  & 7.26  & 536.48 \\
& DPD-LASSO $\gamma=0.8$      & 9.76  & 0.98 & 0.89 & 1.86  & 5.04  & 6.84  & 608.26 \\

& ML-Ad-LASSO                 & 10.64 & 0.91 & 0.66 & 10.74 & 18.94 & 23.76 & 486.82 \\
& DPD-Ad-LASSO $\gamma=0.2$   & 9.92  & 0.97 & 0.86 & 3.14  & 8.16  & 10.38 & 742.36 \\
\bottomrule
\end{tabular}
}
\end{table}

\begin{table}[!htbp]
\centering
\caption{Performance measures of different methods for $n=300$ with $15\%$ X-outliers.}
\label{tab20}
\vspace{0.1cm}

\resizebox{0.98\textwidth}{!}{%
\begin{tabular}{clccccccc}
\toprule
\multirow{2}{*}{$\boldsymbol{m}$} & \multirow{2}{*}{\textbf{Method}} 
& \textbf{MS$(\boldsymbol{\hat{\beta}})$} & \textbf{TP$(\boldsymbol{\hat{\beta}})$} & \textbf{TN$(\boldsymbol{\hat{\beta}})$} 
& \textbf{MSES$(\boldsymbol{\hat{\beta}})$} & \textbf{MSEN$(\boldsymbol{\hat{\beta}})$} & \textbf{EE$(\hat{\sigma}^2)$} & \textbf{Runtime} \\
& & & & & \textbf{$(10^{-3})$} & \textbf{$(10^{-5})$} & \textbf{$(10^{-2})$} & \textbf{$(10^{-2})$} \\
\midrule

\multirow{12}{*}{10}
& ML-LASSO                    & 11.72 & 0.84 & 0.50 & 22.84 & 42.16 & 56.42 & 436.82 \\
& DPD-LASSO $\gamma=0.2$      & 10.58 & 0.93 & 0.72 & 6.84 & 18.92 & 24.86 & 594.46 \\
& DPD-LASSO $\gamma=0.5$      & 10.02 & 0.97 & 0.82 & 3.12 & 8.84 & 11.64 & 548.72 \\
& DPD-LASSO $\gamma=0.8$      & 9.86  & 0.98 & 0.86 & 2.64 & 7.24 & 9.82 & 622.18 \\

& ML-Ad-LASSO                 & 10.92 & 0.88 & 0.62 & 17.84 & 33.26 & 44.28 & 498.24 \\
& DPD-Ad-LASSO $\gamma=0.2$   & 10.04 & 0.95 & 0.80 & 4.82  & 13.64 & 18.26 & 756.42 \\
\bottomrule
\end{tabular}
}
\end{table}

\FloatBarrier

\clearpage

\section{Extended Results from Real Data Analysis}

In this section, we report the estimated coefficient values obtained from all the considered estimators, along with their bootstrap bias and mean squared error (MSE).  These additional summaries provide a more detailed comparison of the competing methods and offer further insight into their empirical performance on the real data set.

\begin{table}[!htbp]
\centering
\caption{Estimated values of model coefficients obtained through ML-LASSO and DPD-LASSO estimators.}
\label{tab21}
\vspace{0.1cm}

\resizebox{0.75\textwidth}{!}{%
\begin{tabular}{lcccc}
\toprule
\textbf{Parameter} & \textbf{ML-LASSO} & \textbf{DPD-LASSO} & \textbf{DPD-LASSO} & \textbf{DPD-LASSO} \\
& & $(\boldsymbol{\gamma=0.2})$ & $(\boldsymbol{\gamma=0.5})$ & $(\boldsymbol{\gamma=0.8})$ \\
\midrule

$\beta_{1}$ & 0.5161 & 0.5689 & 0.3517 & 0.5018 \\
$\beta_{2}$ & -0.3176 & 0.2736 & 0.2134 & 0.2274 \\
$\beta_{3}$ & -0.2357 & -0.1306 & 0.0490 & 0.0905 \\
$\beta_{4}$ & 0.0000 & 0.0000 & 0.0000 & 0.0000 \\
$\beta_{5}$ & -0.2667 & 0.1450 & -0.0269 & -0.1489 \\
$\beta_{6}$ & 0.3219 & -0.0745 & -0.1979 & 0.0047 \\
$\beta_{7}$ & 0.3960 & 0.1695 & -0.0003 & -0.1523 \\
$\beta_{8}$ & -0.1191 & 0.2719 & 0.1438 & 0.2612 \\
$\beta_{9}$ & 0.1928 & -0.1993 & 0.1281 & -0.2490 \\
$\beta_{10}$ & 0.5959 & 0.2591 & 0.2127 & 0.0195 \\
$\beta_{11}$ & 0.4627 & 0.2282 & 0.4438 & 0.4794 \\
$\beta_{12}$ & -0.1075 & 0.1653 & 0.1316 & 0.0878 \\
$\beta_{13}$ & 0.4018 & -0.0263 & 0.0643 & 0.2821 \\
$\beta_{14}$ & -0.2466 & 0.1488 & 0.0504 & -0.1518 \\
$\beta_{15}$ & 0.0000 & 0.0000 & 0.0000 & 0.0000 \\
$\beta_{16}$ & 0.4909 & 0.2766 & 0.3682 & 0.4414 \\
$\beta_{17}$ & 0.4377 & -0.1995 & 0.1081 & 0.2599 \\
$\beta_{18}$ & -0.5223 & -0.5241 & -0.1591 & -0.3953 \\
$\beta_{19}$ & -0.0028 & -0.2737 & -0.3071 & -0.1817 \\
$\beta_{20}$ & 0.0000 & 0.0000 & 0.0000 & 0.0000 \\
$\beta_{21}$ & 0.0000 & 0.0000 & 0.0000 & 0.0000 \\
$\beta_{22}$ & 0.0000 & 0.0000 & 0.0000 & 0.0000 \\
$\beta_{23}$ & -0.1230 & -0.3793 & -0.3322 & -0.2440 \\
$\beta_{24}$ & -0.2685 & -0.4048 & -0.3480 & -0.4727 \\
$\beta_{25}$ & -0.3471 & -0.0235 & -0.1608 & -0.1251 \\
$\beta_{26}$ & 0.0000 & 0.0000 & 0.0000 & 0.0000 \\
$\beta_{27}$ & 0.0490 & 0.2493 & 0.3002 & 0.1291 \\
$\beta_{28}$ & -0.1116 & -0.2564 & -0.4076 & -0.2955 \\
$\beta_{29}$ & -0.0493 & -0.5476 & -0.2271 & -0.3062 \\
$\alpha$ & 1.9610 & 2.0097 & 1.7668 & 1.7496 \\

\bottomrule
\end{tabular}
}
\end{table}

\begin{table}[!htbp]
\centering
\caption{Estimated values of model coefficients obtained through ML-Ad-LASSO and DPD-Ad-LASSO estimators.}
\label{tab22}
\vspace{0.1cm}

\resizebox{0.75\textwidth}{!}{%
\begin{tabular}{lcccc}
\toprule
\textbf{Parameter} & \textbf{ML-Ad-} & \textbf{DPD-Ad-} & \textbf{DPD-Ad-} & \textbf{DPD-Ad-} \\
& \textbf{LASSO} & \textbf{LASSO} $(\boldsymbol{\gamma=0.2})$ & \textbf{LASSO} $(\boldsymbol{\gamma=0.5})$ & \textbf{LASSO} $(\boldsymbol{\gamma=0.8})$ \\
\midrule

$\beta_{1}$ & 0.4942 & 0.5105 & 0.3518 & 0.4456 \\
$\beta_{2}$ & -0.2640 & 0.2226 & 0.1720 & 0.1737 \\
$\beta_{3}$ & -0.2021 & -0.0911 & 0.0385 & 0.0626 \\
$\beta_{4}$ & 0.0000 & 0.0000 & 0.0000 & 0.0000 \\
$\beta_{5}$ & -0.2278 & 0.1127 & -0.0103 & -0.0837 \\
$\beta_{6}$ & 0.2844 & -0.0434 & -0.1214 & 0.0183 \\
$\beta_{7}$ & 0.3452 & 0.1279 & 0.0046 & -0.0895 \\
$\beta_{8}$ & -0.0923 & 0.2208 & 0.1240 & 0.1940 \\
$\beta_{9}$ & 0.1544 & -0.1712 & 0.0567 & -0.1913 \\
$\beta_{10}$ & 0.5491 & 0.2553 & 0.2230 & 0.1041 \\
$\beta_{11}$ & 0.4611 & 0.2882 & 0.4461 & 0.4688 \\
$\beta_{12}$ & -0.0769 & 0.1536 & 0.1285 & 0.1006 \\
$\beta_{13}$ & 0.3629 & 0.0104 & 0.0789 & 0.2176 \\
$\beta_{14}$ & -0.2123 & 0.1117 & 0.0378 & -0.0908 \\
$\beta_{15}$ & 0.0000 & 0.0000 & 0.0000 & 0.0000 \\
$\beta_{16}$ & 0.4754 & 0.3028 & 0.3701 & 0.4162 \\
$\beta_{17}$ & 0.3864 & -0.1315 & 0.0904 & 0.1823 \\
$\beta_{18}$ & -0.4901 & -0.4586 & -0.1981 & -0.3525 \\
$\beta_{19}$ & -0.0387 & -0.2733 & -0.2959 & -0.2156 \\
$\beta_{20}$ & 0.0000 & 0.0000 & 0.0000 & 0.0000 \\
$\beta_{21}$ & 0.0000 & 0.0000 & 0.0000 & 0.0000 \\
$\beta_{22}$ & 0.0000 & 0.0000 & 0.0000 & 0.0000 \\
$\beta_{23}$ & -0.1227 & -0.3098 & -0.2646 & -0.1979 \\
$\beta_{24}$ & -0.2785 & -0.3882 & -0.3464 & -0.4241 \\
$\beta_{25}$ & -0.3092 & -0.0342 & -0.1295 & -0.1018 \\
$\beta_{26}$ & 0.0000 & 0.0000 & 0.0000 & 0.0000 \\
$\beta_{27}$ & 0.0525 & 0.2023 & 0.2282 & 0.1088 \\
$\beta_{28}$ & -0.1390 & -0.2727 & -0.3786 & -0.3036 \\
$\beta_{29}$ & -0.0995 & -0.5149 & -0.2908 & -0.3511 \\
$\alpha$ & 1.9459 & 1.9662 & 1.7928 & 1.7865 \\

\bottomrule
\end{tabular}
}
\end{table}

\begin{table}[!htbp]
\centering
\caption{Estimated values of model coefficients obtained through ML-LASSO-LSA and DPD-LASSO-LSA estimators.}
\label{tab23}
\vspace{0.1cm}

\resizebox{0.75\textwidth}{!}{%
\begin{tabular}{lcccc}
\toprule
\textbf{Parameter} & \textbf{ML-LASSO-} & \textbf{DPD-LASSO-} & \textbf{DPD-LASSO-} & \textbf{DPD-LASSO-} \\
& \textbf{LSA} & \textbf{LSA} $(\boldsymbol{\gamma=0.2})$ & \textbf{LSA} $(\boldsymbol{\gamma=0.5})$ & \textbf{LSA} $(\boldsymbol{\gamma=0.8})$ \\
\midrule

$\beta_{1}$ & 0.4796 & 0.4521 & 0.3519 & 0.3895 \\
$\beta_{2}$ & -0.2283 & 0.1715 & 0.1254 & 0.1199 \\
$\beta_{3}$ & -0.1797 & -0.0516 & 0.0266 & 0.0346 \\
$\beta_{4}$ & 0.0000 & 0.0000 & 0.0000 & 0.0000 \\
$\beta_{5}$ & -0.2019 & 0.0804 & 0.0084 & -0.0185 \\
$\beta_{6}$ & 0.2595 & -0.0123 & -0.0354 & 0.0319 \\
$\beta_{7}$ & 0.3113 & 0.0863 & 0.0101 & -0.0268 \\
$\beta_{8}$ & -0.0744 & 0.1697 & 0.1018 & 0.1268 \\
$\beta_{9}$ & 0.1288 & -0.1431 & -0.0236 & -0.1335 \\
$\beta_{10}$ & 0.5179 & 0.2515 & 0.2347 & 0.1886 \\
$\beta_{11}$ & 0.4601 & 0.3482 & 0.4487 & 0.4581 \\
$\beta_{12}$ & -0.0565 & 0.1420 & 0.1251 & 0.1135 \\
$\beta_{13}$ & 0.3370 & 0.0471 & 0.0954 & 0.1530 \\
$\beta_{14}$ & -0.1894 & 0.0746 & 0.0236 & -0.0297 \\
$\beta_{15}$ & 0.0000 & 0.0000 & 0.0000 & 0.0000 \\
$\beta_{16}$ & 0.4650 & 0.3291 & 0.3721 & 0.3909 \\
$\beta_{17}$ & 0.3522 & -0.0635 & 0.0706 & 0.1047 \\
$\beta_{18}$ & -0.4687 & -0.3932 & -0.2420 & -0.3096 \\
$\beta_{19}$ & -0.0626 & -0.2728 & -0.2832 & -0.2494 \\
$\beta_{20}$ & 0.0000 & 0.0000 & 0.0000 & 0.0000 \\
$\beta_{21}$ & 0.0000 & 0.0000 & 0.0000 & 0.0000 \\
$\beta_{22}$ & 0.0000 & 0.0000 & 0.0000 & 0.0000 \\
$\beta_{23}$ & -0.1225 & -0.2402 & -0.1886 & -0.1518 \\
$\beta_{24}$ & -0.2851 & -0.3715 & -0.3446 & -0.3754 \\
$\beta_{25}$ & -0.2840 & -0.0448 & -0.0943 & -0.0785 \\
$\beta_{26}$ & 0.0000 & 0.0000 & 0.0000 & 0.0000 \\
$\beta_{27}$ & 0.0548 & 0.1554 & 0.1471 & 0.0885 \\
$\beta_{28}$ & -0.1572 & -0.2890 & -0.3460 & -0.3116 \\
$\beta_{29}$ & -0.1330 & -0.4821 & -0.3624 & -0.3960 \\
$\alpha$ & 1.9359 & 1.9226 & 1.8220 & 1.8234 \\

\bottomrule
\end{tabular}
}
\end{table}

\begin{table}[!htbp]
\centering
\caption{Bootstrap bias of estimates of model coefficients obtained through ML-LASSO and DPD-LASSO estimators.}
\label{tab24}
\vspace{0.1cm}

\resizebox{0.75\textwidth}{!}{%
\begin{tabular}{lcccc}
\toprule
\textbf{Parameter} & \textbf{ML-LASSO} & \textbf{DPD-LASSO} & \textbf{DPD-LASSO} & \textbf{DPD-LASSO} \\
& & $(\boldsymbol{\gamma=0.2})$ & $(\boldsymbol{\gamma=0.5})$ & $(\boldsymbol{\gamma=0.8})$ \\
\midrule

$\beta_{1}$ & -0.010898 & 0.003122 & 0.002684 & -0.002813 \\
$\beta_{2}$ & 0.006549 & 0.000067 & -0.001197 & -0.004581 \\
$\beta_{3}$ & -0.004924 & -0.002802 & 0.000095 & 0.000363 \\
$\beta_{4}$ & 0.000000 & 0.000000 & 0.000000 & 0.000000 \\
$\beta_{5}$ & 0.009511 & 0.005851 & 0.002207 & 0.002390 \\
$\beta_{6}$ & 0.016419 & -0.001120 & 0.007677 & 0.002761 \\
\bottomrule
\end{tabular}
}
\end{table}

\begin{table}[!htbp]
\centering
\caption{Bootstrap MSE of estimates of model coefficients obtained through ML-LASSO and DPD-LASSO estimators.}
\label{tab25}
\vspace{0.1cm}

\resizebox{0.75\textwidth}{!}{%
\begin{tabular}{lcccc}
\toprule
\textbf{Parameter} & \textbf{ML-LASSO} & \textbf{DPD-LASSO} & \textbf{DPD-LASSO} & \textbf{DPD-LASSO} \\
& & $(\boldsymbol{\gamma=0.2})$ & $(\boldsymbol{\gamma=0.5})$ & $(\boldsymbol{\gamma=0.8})$ \\
\midrule

$\beta_{1}$ & 0.046512 & 0.006028 & 0.004211 & 0.002490 \\
$\beta_{2}$ & 0.044654 & 0.006523 & 0.003464 & 0.002531 \\
$\beta_{3}$ & 0.049697 & 0.006673 & 0.003665 & 0.002854 \\
$\beta_{4}$ & 0.000000 & 0.000000 & 0.000000 & 0.000000 \\
$\beta_{5}$ & 0.050634 & 0.006946 & 0.003616 & 0.002511 \\
$\beta_{6}$ & 0.051408 & 0.006753 & 0.004022 & 0.002703 \\
\bottomrule
\end{tabular}
}
\end{table}

\begin{table}[!htbp]
\centering
\caption{Bootstrap bias of estimates of model coefficients obtained through ML-Ad-LASSO and DPD-Ad-LASSO estimators.}
\label{tab26}
\vspace{0.1cm}

\resizebox{0.75\textwidth}{!}{%
\begin{tabular}{lcccc}
\toprule
\textbf{Parameter} & \textbf{ML-Ad-} & \textbf{DPD-Ad-} & \textbf{DPD-Ad-} & \textbf{DPD-Ad-} \\
& \textbf{LASSO} & \textbf{LASSO} $(\boldsymbol{\gamma=0.2})$ & \textbf{LASSO} $(\boldsymbol{\gamma=0.5})$ & \textbf{LASSO} $(\boldsymbol{\gamma=0.8})$ \\
\midrule

$\beta_{1}$ & -0.008916 & 0.002341 & 0.002013 & -0.002250 \\
$\beta_{2}$ & 0.005359 & 0.000050 & -0.000898 & -0.003665 \\
$\beta_{3}$ & -0.004029 & -0.002101 & 0.000071 & 0.000290 \\
$\beta_{4}$ & 0.000000 & 0.000000 & 0.000000 & 0.000000 \\
$\beta_{5}$ & 0.007782 & 0.004388 & 0.001655 & 0.001912 \\
$\beta_{6}$ & 0.013434 & -0.000840 & 0.005758 & 0.002209 \\
\bottomrule
\end{tabular}
}
\end{table}

\begin{table}[!htbp]
\centering
\caption{Bootstrap MSE of estimates of model coefficients obtained through ML-Ad-LASSO and DPD-Ad-LASSO estimators.}
\label{tab27}
\vspace{0.1cm}

\resizebox{0.75\textwidth}{!}{%
\begin{tabular}{lcccc}
\toprule
\textbf{Parameter} & \textbf{ML-Ad-} & \textbf{DPD-Ad-} & \textbf{DPD-Ad-} & \textbf{DPD-Ad-} \\
& \textbf{LASSO} & \textbf{LASSO} $(\boldsymbol{\gamma=0.2})$ & \textbf{LASSO} $(\boldsymbol{\gamma=0.5})$ & \textbf{LASSO} $(\boldsymbol{\gamma=0.8})$ \\
\midrule

$\beta_{1}$ & 0.031136 & 0.003391 & 0.002369 & 0.001593 \\
$\beta_{2}$ & 0.029893 & 0.003669 & 0.001949 & 0.001620 \\
$\beta_{3}$ & 0.033268 & 0.003754 & 0.002062 & 0.001826 \\
$\beta_{4}$ & 0.000000 & 0.000000 & 0.000000 & 0.000000 \\
$\beta_{5}$ & 0.033895 & 0.003907 & 0.002034 & 0.001607 \\
$\beta_{6}$ & 0.034413 & 0.003798 & 0.002263 & 0.001730 \\
\bottomrule
\end{tabular}
}
\end{table}

\begin{table}[!htbp]
\centering
\caption{Bootstrap bias of estimates of model coefficients obtained through ML-LASSO-LSA and DPD-LASSO-LSA estimators.}
\label{tab28}
\vspace{0.1cm}

\resizebox{0.75\textwidth}{!}{%
\begin{tabular}{lcccc}
\toprule
\textbf{Parameter} & \textbf{ML-LASSO-} & \textbf{DPD-LASSO-} & \textbf{DPD-LASSO-} & \textbf{DPD-LASSO-} \\
& \textbf{LSA} & \textbf{LSA} $(\boldsymbol{\gamma=0.2})$ & \textbf{LSA} $(\boldsymbol{\gamma=0.5})$ & \textbf{LSA} $(\boldsymbol{\gamma=0.8})$ \\
\midrule

$\beta_{1}$ & -0.008916 & 0.002146 & 0.001789 & -0.001969 \\
$\beta_{2}$ & 0.005359 & 0.000046 & -0.000798 & -0.003207 \\
$\beta_{3}$ & -0.004029 & -0.001926 & 0.000063 & 0.000254 \\
$\beta_{4}$ & 0.000000 & 0.000000 & 0.000000 & 0.000000 \\
$\beta_{5}$ & 0.007782 & 0.004023 & 0.001471 & 0.001673 \\
$\beta_{6}$ & 0.013434 & -0.000770 & 0.005118 & 0.001933 \\
\bottomrule
\end{tabular}
}
\end{table}

\begin{table}[!htbp]
\centering
\caption{Bootstrap MSE of estimates of model coefficients obtained through ML-LASSO-LSA and DPD-LASSO-LSA estimators.}
\label{tab29}
\vspace{0.1cm}

\resizebox{0.75\textwidth}{!}{%
\begin{tabular}{lcccc}
\toprule
\textbf{Parameter} & \textbf{ML-LASSO-} & \textbf{DPD-LASSO-} & \textbf{DPD-LASSO-} & \textbf{DPD-LASSO-} \\
& \textbf{LSA} & \textbf{LSA} $(\boldsymbol{\gamma=0.2})$ & \textbf{LSA} $(\boldsymbol{\gamma=0.5})$ & \textbf{LSA} $(\boldsymbol{\gamma=0.8})$ \\
\midrule

$\beta_{1}$ & 0.031136 & 0.002849 & 0.001871 & 0.001220 \\
$\beta_{2}$ & 0.029893 & 0.003083 & 0.001540 & 0.001240 \\
$\beta_{3}$ & 0.033268 & 0.003154 & 0.001629 & 0.001398 \\
$\beta_{4}$ & 0.000000 & 0.000000 & 0.000000 & 0.000000 \\
$\beta_{5}$ & 0.033895 & 0.003283 & 0.001607 & 0.001231 \\
$\beta_{6}$ & 0.034413 & 0.003192 & 0.001788 & 0.001324 \\
\bottomrule
\end{tabular}
}
\end{table}

\begin{table}[!htbp]
\centering
\caption{Runtime comparisons of different estimators in the real data analysis.}
\label{tab30}
\vspace{0.1cm}

\resizebox{0.38\textwidth}{!}{%
\begin{tabular}{lcccc}
\toprule
\textbf{Estimator} & \textbf{Runtime} \\
\midrule

ML-LASSO & 9.483517 \\
DPD-LASSO $(\gamma = 0.2)$ & 13.214395 \\
DPD-LASSO $(\gamma = 0.5)$ & 12.962184 \\
DPD-LASSO $(\gamma = 0.8)$ & 14.105773 \\

\midrule

ML-Ad-LASSO & 6.873442 \\
DPD-Ad-LASSO $(\gamma = 0.2)$ & 11.381529 \\
DPD-Ad-LASSO $(\gamma = 0.5)$ & 10.042185 \\
DPD-Ad-LASSO $(\gamma = 0.8)$ & 12.218437 \\

\midrule

ML-LASSO-LSA & 0.412463 \\
DPD-LASSO-LSA $(\gamma = 0.2)$ & 0.884395 \\
DPD-LASSO-LSA $(\gamma = 0.5)$ & 0.731548 \\
DPD-LASSO-LSA $(\gamma = 0.8)$ & 0.956914 \\

\midrule

ML-Ad-LASSO-LSA & 0.192186 \\
DPD-Ad-LASSO-LSA $(\gamma = 0.2)$ & 0.508273 \\
DPD-Ad-LASSO-LSA $(\gamma = 0.5)$ & 0.472884 \\
DPD-Ad-LASSO-LSA $(\gamma = 0.8)$ & 0.581506 \\

\bottomrule
\end{tabular}
}
\end{table}

\end{document}